\documentclass[1p]{elsarticle}

\PassOptionsToPackage{hyphens}{url}
\usepackage{hyperref}
\usepackage{amssymb}

\begin{document}

\begin{frontmatter}

\title{A More Accurate, Stable, FDTD Algorithm for Electromagnetics in
Anisotropic Dielectrics\tnoteref{moneyAck}}

\tnotetext[moneyAck]{
This work was supported by the U.S. Department of Energy grant
DE-FG02-04ER41317.
}

\author[CU]{Gregory R.~Werner\corref{GW}}
\cortext[GW]{Corresponding author: Greg.Werner@colorado.edu}
\author[CU]{Carl A.~Bauer}
\author[CU,TechX]{John R.~Cary}

\address[CU]{
Center for Integrated Plasma Studies, University of Colorado,
Boulder, CO 80309
}
\address[TechX]{
Tech-X Corporation, Boulder, CO 80303
}

\begin{abstract}

A more accurate, stable, finite-difference time-domain (FDTD) 
algorithm is developed for
simulating Maxwell's equations with isotropic or anisotropic
dielectric materials.  This algorithm is in many cases more accurate
than previous algorithms 
(G.~R.~Werner et. al., 2007; A.~F.~Oskooi et. al., 2009),
and it remedies a defect that causes instability with high
dielectric contrast (usually for $\epsilon \gg 10$) with either
isotropic or anisotropic dielectrics.  Ultimately this algorithm
has first-order error (in the grid cell size) when the dielectric
boundaries are sharp, due to field
discontinuities at the dielectric interface.  
Accurate treatment of the discontinuities, in the limit of 
infinite wavelength,
leads to an asymmetric, unstable update 
(C.~A.~Bauer et. al., 2011),
but the symmetrized version of the latter is stable and more accurate
than other FDTD methods.  The convergence of field values
supports the hypothesis that global first-order error can be achieved
by second-order error in bulk material with zero-order error on
the surface.  This latter point is extremely important for any
applications measuring surface fields.

\end{abstract}


\end{frontmatter}


\section{Introduction}

The Yee finite-difference time-domain (FDTD) algorithm
\cite{Yee:1966} simulates electromagnetic waves in uniform,
isotropic media with second-order error: i.e., the Yee algorithm
simulates the frequency of a plane wave with an error that scales
as $O(\Delta x^2)$ with grid cell size $\Delta x$. This paper
presents a generalization of the Yee algorithm for non-uniform,
anisotropic dielectric media, with particular attention to sharp
transitions between different dielectrics. (This generalization
is also suitable for the intermediate cases, such as
continuously-varying dielectrics, whether isotropic or not.)

When different dielectric materials meet at a sharp interface,
the discontinuity in the fields introduces greater error into the
simulation.  If the discontinuity is disregarded, 
operations such as field-interpolation typically have local
$O(1)$ error at the interface (the error remains constant as
$\Delta x \rightarrow 0$ because the field discontinuity remains
constant).  However, because the ratio of the cells cut by the
interface to the total number of cells is $O(\Delta x)$, the
local error $O(1)$ is watered down by a factor of $O(\Delta x)$,
leading to a ``global error'' of $O(\Delta x)$. (By global error,
we refer to the error in a mode frequency or the average field error 
over an entire
eigenmode, whereas local error is field error in a single
cell.  This relationship between global error and relatively high
local error on a subset of cells has been proven rigorously for
1D waves \cite{Gustafsson:1975}, and demonstrated empirically for
2D and 3D electromagnetic problems, e.g., with curved metal
boundaries \cite{Nieter:2004} and curved dielectric interfaces
\cite{Werner:2007,Bauer:2011}.)

When the dielectric constant varies continuously,
the variation across a cell vanishes
as $\Delta x \rightarrow 0$, so it is not difficult to obtain 
local $O(\Delta x)$ error, hence
global $O(\Delta x^2)$ error \cite{Werner:2007}.

Of course we would like the same $O(\Delta x^2)$ error even
with sharp dielectric boundaries; to the best of our knowledge,
the first finite-difference
approach to accomplish this was \cite{Bauer:2011}, 
which obtained global second-order
error with local first-order error at the
interface.  Instead of considering the error of the discretized
system for fixed frequency or wavelength $\lambda$ and vanishing
$\Delta x$, accuracy was demanded for fixed $\Delta x$ and
infinite $\lambda$.  Ref.~\cite{Bauer:2011} showed how to convert
$\mathbf{D}$ to $\mathbf{E}$ \emph{exactly} in the limit of
$\lambda \rightarrow \infty$.  This led to global $O(\Delta x^2)$
error.

Unfortunately, the accurate method of \cite{Bauer:2011} is
unstable in the time-domain, because it uses an asymmetric {\em
inverse dielectric matrix} (the linear operator that transforms
the $\mathbf{D}$ field to the $\mathbf{E}$ field on the Yee
mesh), which has complex eigenvalues, hence complex mode
frequencies. While the imaginary parts of the frequencies are
within the error [i.e., $\textrm{Im}(\omega) \lesssim O(\Delta
x^2/\lambda^2)$], they lead to unphysical growth that becomes
significant after sufficient time.   Moreover, while
well-resolved modes ($\Delta x \ll \lambda$) may have slow
growth, there are always modes with $\lambda \sim \Delta x$ which
may grow quickly; thus machine-precision-level noise eventually
grows to overwhelm the desired signal.

A \emph{symmetric} inverse dielectric matrix was given by
\cite{Werner:2007}, yielding an algorithm with $O(\Delta x)$
global error (we will refer to this algorithm as ``wc07'').
This algorithm is stable at the low dielectric contrasts studied
in \cite{Werner:2007} (where ``dielectric contrast'' is the ratio
of dielectric constants between neighboring media in a
simulation). However, we have recently found 
that wc07 is unstable for high dielectric contrast (see 
Sec.~\ref{sec:oldAlg}), because the dielectric matrix is not
always positive definite.

In this paper we use the ``triplets'' concept of \cite{Bauer:2011} to
obtain stable algorithms in the time domain.
If one can find symmetric and positive-definite (SPD) effective
dielectric tensors acting on triplets of field components, 
then the tensors can be combined into an inverse dielectric matrix
that is also SPD, which ensures stability
in the time domain.  We show that a small change to the wc07 
algorithm \cite{Werner:2007} yields a new 
algorithm (``wc07mod'') with an SPD
inverse dielectric matrix; wc07mod is therefore stable for
arbitrary dielectric contrast.  Using the same framework for
stability, we provide yet another algorithm
that is stable for arbitrary dielectric
contrast; this algorithm simply uses the symmetrized
matrix of \cite{Bauer:2011}.  The act of symmetrizing increases
the error from $O(\Delta x^2)$ to $O(\Delta x)$, but we find that
this algorithm still has smaller error than the other
effective-dielectric methods \cite{Werner:2007,Kottke:2008,Oskooi:2009}.

For example, for a dielectric contrast around 10, the new method
reduces the frequency error by a factor of 2--3 at high
resolutions, and it reduces the error of fields by a factor of
2--10. Reducing the error by a factor of 2 can reduce computation
by a factor of $16$, because the error scales as $O(\Delta x)$
while the computation time typically scales as $O(1/\Delta x^4)$
(for a 3D problem).

While examining the error, we show that \emph{for low dielectric
contrast} (less than 10 or so), the new algorithm yields very
similar results to other effective dielectrics
(\cite{Werner:2007,Kottke:2008,Oskooi:2009}), and the error is
very nearly second-order up to high resolution. In other words,
for low dielectric contrast, the error is dominated not by the
field discontinuities, but by the bulk Yee algorithm. At
sufficiently high resolution, however, we believe all these
algorithms transition to first-order.  As the dielectric contrast
increases, the transition point moves to lower resolutions.

We also examine the error in fields: the error at (or within a
fixed number of cells of) the dielectric surface is $O(1)$.
However, the field error is $O(\Delta x)$ at a fixed physical distance
from the surface (N.B., that distance spans more cells as 
$\Delta x$ diminishes).  This supports the application of
\cite{Gustafsson:1975} to this problem; in other words, the local
surface error is $O(1)$, but only $O(\Delta x)$ cells are cut by
the surface, so the global error is $O(\Delta x)$.

The latter point may be very important for applications
attempting to characterize surface fields with effective
dielectric algorithms. Such attempts should be wary of errors in
the surface fields, because the error does not decrease with cell
size.  However, the error is probably small enough in many cases
that it will not be a problem.  And, the fields a fixed distance
from the surface do become more accurate as the cell size is
reduced.

After a brief outline of algorithms discussed in this paper, we
will present the discretization of Maxwell's equations, reducing
the problem of introducing dielectric to the problem of finding
an inverse dielectric matrix $\Xi$. Section~\ref{sec:stability}
then describes how to create $\Xi$ to guarantee stability,
assuming the ability to create local $3\times 3$ effective
dielectric tensors that are SPD (but
otherwise unrestricted), thus reducing the problem to finding the
local effective dielectric.

Sections~\ref{sec:oldAlg}, \ref{sec:simpleStable}, and
\ref{sec:symAcc} present three different methods for calculating local
effective dielectric tensors. First we describe the effective
dielectric that reproduces the wc07 algorithm---this effective
dielectric is not (always) positive definite, so it doesn't
guarantee stability. Second, we modify wc07 slightly to guarantee
stability. Third, we present an effective dielectric that
guarantees stability, and has similar or better accuracy than the
second option.  All these methods require the same amount of
computation for every time step.

Subsequent sections present the error, in both frequency and
field, of the different algorithms for different problems: 2D and
3D, isotropic and anisotropic, over a range of dielectric
contrast from 5--100.

\section{Algorithms in this paper}

The following algorithms will be discussed in this paper:
\begin{itemize}
  \item wc07: the algorithm recommended by \cite{Werner:2007}
    [therewithin called variant (c)$+$(e)]---it is
    unstable for high dielectric contrast;
  \subitem (Ref.~\cite{Oskooi:2009} used wc07, with the improved
    dielectric averaging of \cite{Kottke:2008} instead of
    \cite{Johnson:2001}.  These two averaging methods are
    identical for isotropic
    dielectrics; accuracy for anisotropic dielectrics may improve,
    but the error still converges as $O(\Delta x)$.)
  \item wc07mod: a stable algorithm, almost the same as wc07;
  \item the ``new'' method: a stable and more accurate algorithm, but
    still with $O(\Delta x)$ error
    (sections \ref{sec:stability} and \ref{sec:symAcc});
  \item the second-order method of \cite{Bauer:2011}: the only
    finite-difference algorithm with second-order error is
    unfortunately asymmetric, rendering it unstable for time-domain use
    (but still good for frequency-domain eigensolvers).
\end{itemize}

\section{The basic algorithm}
\label{sec:basicAlg}

\begin{figure}[tp]
\centering
\begin{tabular}{rr}
\includegraphics*[trim = 0mm 0mm 0mm 0mm,width=80mm]{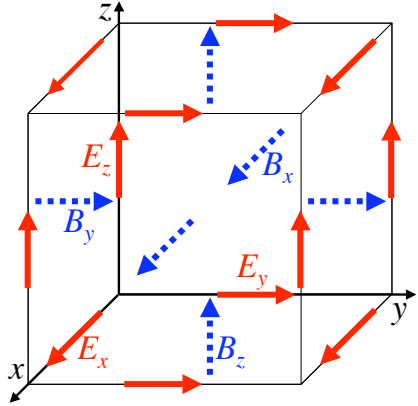}
\end{tabular}
\caption{(Color online.) Field components in one grid cell of the Yee mesh.  If
the above cell has the 3-integer index $(i,j,k)$, then the labeled
field components have the same index, e.g., $E_{ijkx}$.  The
$\mathbf{D}$ components are collocated with the corresponding
$\mathbf{E}$ components.
\label{fig:yeeMesh}}
\end{figure}

We want to simulate the dynamic Maxwell equations with dielectric:
\begin{eqnarray}
	\frac{\partial \bf B}{\partial t} &=& -\nabla \times {\bf E}
	\label{eq:yeeFaraday}\\
	\frac{\partial \bf D}{\partial t} &=& \nabla \times {\bf B}
	\\
	\mathbf{E} &=& \underline{\underline{\xi}}(x,y,z) \mathbf{D}
\end{eqnarray}
where $ \underline{\underline{\xi}} = \underline{\underline{\varepsilon}}^{-1}$
is the inverse dielectric tensor (which can vary in space).

To discretize Maxwell's equations for computational work, we
follow Yee \cite{Yee:1966}. To analyze the discretization, we
treat the fields as large vectors, with each component
representing a field at one point of the Yee mesh, shown in
Fig.~\ref{fig:yeeMesh}.  (Discretization in time is irrelevant
to this paper, so we retain continuous time
derivatives.)

For convenience, we label each component of a field vector with 4
sub-indices: e.g., the component $E_{ijkx}$ represents the
electric field in the $x$ direction at its Yee location in cell
$(i,j,k)$. The differential operators become matrices, with rows
and columns each indexed by 4 sub-indices: e.g., one element of a
matrix $M$ is $M_{(ijk\mu),(i'j'k'\nu)}$.

We will not review the Yee discretization, since
\cite{Werner:2007} details the relevant aspects, instead taking
it for granted that the matrices $C$ and $C^T$ represent the curl
operators (the matrix representation of the curl of $E$ is the
transpose of the curl of $B$). We depart from Yee when we
introduce the matrix $\Xi$ to discretize the linear relationship
between $E$ and $D$, yielding:
\begin{eqnarray}
	\frac{\partial }{\partial t} B_{ijk\mu} &=& - [ C E ]_{ijk\mu}
	\\
	\frac{\partial }{\partial t} D_{ijk\mu} &=& [C^T B]_{ijk\mu}
	\\
	E_{ijk\mu} &=& [\Xi D]_{ijk\mu} =
	  \sum_{i'j'k'\nu} \Xi_{(ijk\mu),(i'j'k'\nu)} D_{i'j'k'\nu}
.\end{eqnarray}
These equations are a prescription for advancing the fields in time:
$B$ is advanced by a short time (using $E$), then $D$ is advanced 
(using $B$), then $E = \Xi D$, and the cycle repeats (in practice, this
can be implemented with only two fields, $E$ and $B$).
Combining these into one equation
\begin{eqnarray}
	\frac{\partial^2 }{\partial t^2} B_{ijk\mu} &=&
	- [ C \Xi C^T B]_{ijk\mu}
\end{eqnarray}
shows that the eigenvalues $\omega^2$ (frequencies squared) of the
$-\partial^2/\partial t^2$ operator are the eigenvalues of the
$C \Xi C^T$ matrix.

Simulating dielectrics therefore reduces to the determination of
$\Xi$ such that
\begin{enumerate}
  \item $\Xi$ accurately represents the inverse
    dielectric tensor, $\xi(x,y,z)$, and
  \item $C \Xi C^T$ is diagonalizable with
    only real, non-negative eigenvalues.
\end{enumerate}
The first point addresses accuracy.  The second addresses
robustness/stability: if $C \Xi C^T$ has negative or complex
eigenvalues, some frequencies $\omega$ will be complex, and some
fields will grow exponentially (unphysically), ultimately
overwhelming the simulation with noise.

If $\Xi$ is symmetric and positive definite (SPD), then the second
point above will be guaranteed: all modes will oscillate (with
real frequency) without growing (cf., \cite{Werner:2007}, or a
standard linear algebra text such as \cite{Artin:Algebra})---at
least for a sufficiently small time step. (As stated, temporal
discretization is outside the focus of this paper).

This paper is devoted to finding an SPD
matrix $\Xi$ that accurately represents dielectric media.

\section{Creating a stable \texorpdfstring{$\Xi$}{inverse
dielectric} matrix from local \texorpdfstring{$\xi$}{inverse
dielectric} tensors}
\label{sec:stability}

Our approach to dielectric simulation falls under the ``effective
dielectric'' category, specifying the matrix $\Xi$ to find
$E = \Xi D$, while advancing $D$ (and $B$) in time according to
the unaltered Yee algorithm \cite{Yee:1966}.

As in Ref.~\cite{Bauer:2011}, we demand  that
$\Xi$ have the following properties.
\begin{enumerate}
  \item In case of uniform (anisotropic) dielectric, $\Xi$
    involves only centered interpolations of fields
    (i.e., interpolations with $O(\Delta x^2)$ error for continuous
    fields).
  \item Within uniform, \emph{isotropic} dielectric, $\Xi$
    reduces to a multiple of the identity.
  \item $\Xi$ must be symmetric.
  \item $\Xi$ must be positive definite.
\end{enumerate}
The first property yields a local $O(\Delta x^2)$ error within
uniform dielectric (or even continuously-varying dielectric 
\cite{Werner:2007}).  The second is for convenience and
minimalism---we can still use the plain Yee algorithm for fields
within any bulk isotropic region. Together, the third and fourth
properties guarantee stability. (For comparison: wc07 satisfies
1, 2, and 3.)

Before defining the $\Xi$ matrix, we need to introduce notation
for an effective dielectric tensor involving field components of a 
single Yee cell.  For example, we consider
the ``triplet'' of three $E_\mu$ values labeled in Fig.~\ref{fig:yeeMesh} 
and the three $D_\mu$ values at the same locations,
along edges that touch a common node (corner) of the 
cell.  If the node index is $(ijk)$, then
this triplet comprises $(E_{ijkx},E_{ijky},E_{ijkz})$, and similarly
$(D_{ijkx}, D_{ijky}, D_{ijkz})$.  
(We call these components a triplet, instead of a 3-vector,
because they are not collocated.)
The triplets for $E$ and $D$ can be related by a $3\times 3$ tensor, 
$\xi_{ijk}^{+++}$:
\begin{equation} \label{eq:onecellrelation}
  \mathbf{E}_{ijk} = \xi_{ijk}^{+++} \mathbf{D}_{ijk}
  \quad \textrm{or} \quad
  E_{ijk\mu} = \sum_{\nu=x,y,z} \xi_{ijk,\mu\nu}^{+++} D_{ijk\nu}
\end{equation}
(the meaning of the $+\!+\!+$ superscript will be explained shortly).
Defined thus, $\xi_{ijk}^{+++}$ is an 
effective (inverse) dielectric tensor for
this particular triplet of $E$ and $D$ values.

Touching the same cell node are seven other (eight, in all)
geometrically-identical triplets.
Above, we chose a triplet with edges extending
positively in each direction from their common node, but 
(due to the symmetries of the Yee mesh)
we could equally well have chosen, e.g.,
the $x$-edge extending in the \emph{negative} $x$ direction,
namely $(E_{(i-1)jkx},E_{ijky},E_{ijkz})$, 
and $(D_{(i-1)jkx},D_{ijky},D_{ijkz})$.  With this triplet, 
we associate a different effective
dielectric, $\xi_{ijk}^{-++}$, with the minus sign indicating
that the $E_x$ and $D_x$ edges extend in the negative $x$ direction
from node $(ijk)$.

We will construct (the large matrix) $\Xi$ from the 
(small) $\xi_{ijk}^{\pm\pm\pm}$ tensors, through an intermediate stage
involving (large, block-diagonal matrices) $\Xi^{\pm\pm\pm}$.  
Symmetry and positive definiteness transfer 
easily from one stage to the next:
we will show that $\Xi$ is SPD
if all the $\xi_{ijk}^{\pm\pm\pm}$ are SPD; since the latter
are mere $3\times 3$ matrices, evaluating their positive
definiteness is easy, numerically if not analytically.

We start by creating block-diagonal matrices
$\Xi^{+++}$, $\Xi^{-++}$, \ldots, $\Xi^{---}$, where the blocks are
the $\xi_{ijk}^{\pm\pm\pm}$; for example, each block of $\Xi^{-++}$
is $\xi_{ijk}^{-++}$ for some node $(ijk)$.
Precisely we define
\begin{equation} \label{eq:offcenterScheme}
      \begin{array}{r@{\;}c@{\;}l}
	\Xi^{+++}_{(ijk\mu)(i'j'k'\nu)}
	& \equiv &
   \delta_{(ijk)(i'j'k')} \xi_{ijk,\mu\nu}^{+++}
   \\
	\Xi^{++-}_{(ijk\mu)(i'j'k'\nu)}
	& \equiv &
	\xi^{++-}_{ij,k+\delta_{\mu z},\mu\nu}
	\delta_{(ij,k+\delta_{\mu z}),(i'j',k'+\delta_{\nu z})}
	\\
	\Xi^{+-+}_{(ijk\mu)(i'j'k'\nu)}
	& \equiv &
	\xi^{+-+}_{i,j+\delta_{\mu y},k,\mu\nu}
	\delta_{(ij+\delta_{\mu y},k),(i'j'+\delta_{\nu y},k')}
	\\
	\Xi^{+--}_{(ijk\mu)(i'j'k'\nu)}
	& \equiv &
	\xi^{+--}_{i,j+\delta_{\mu y},k+\delta_{\mu z},\mu\nu}
	\delta_{(ij+\delta_{\mu y},k+\delta_{\mu z}),
	        (i'j'+\delta_{\nu y},k'+\delta_{\nu z})}
	\\
	&\vdots &
	\\
	\Xi^{---}_{(ijk\mu)(i'j'k'\nu)}
	& \equiv &
	\xi^{---}_{i+\delta_{\mu x},j+\delta_{\mu y},k+\delta_{\mu z},\mu\nu}
	\delta_{(i+\delta_{\mu x},j+\delta_{\mu y},k+\delta_{\mu z}),
	        (i'+\delta_{\nu x},j'+\delta_{\nu y},k'+\delta_{\mu z})}
	\\
	\end{array}
\end{equation}
where $\delta_{(ijk)(i'j'k')}$ is the Kronecker delta, equal to
one when $i=i'$, $j=j'$, and $k=k'$, and otherwise equal to zero.
A block-diagonal matrix is SPD if and
only if each block is SPD; therefore,
if each $\xi_{ijk}^{+++}$ is SPD, 
then $\Xi^{+++}$ is SPD, and likewise for all $\Xi^{\pm\pm\pm}$.

\begin{figure}[tp]
\centering
\includegraphics*[trim = 40mm 60mm 15mm 0mm,width=100mm]{%
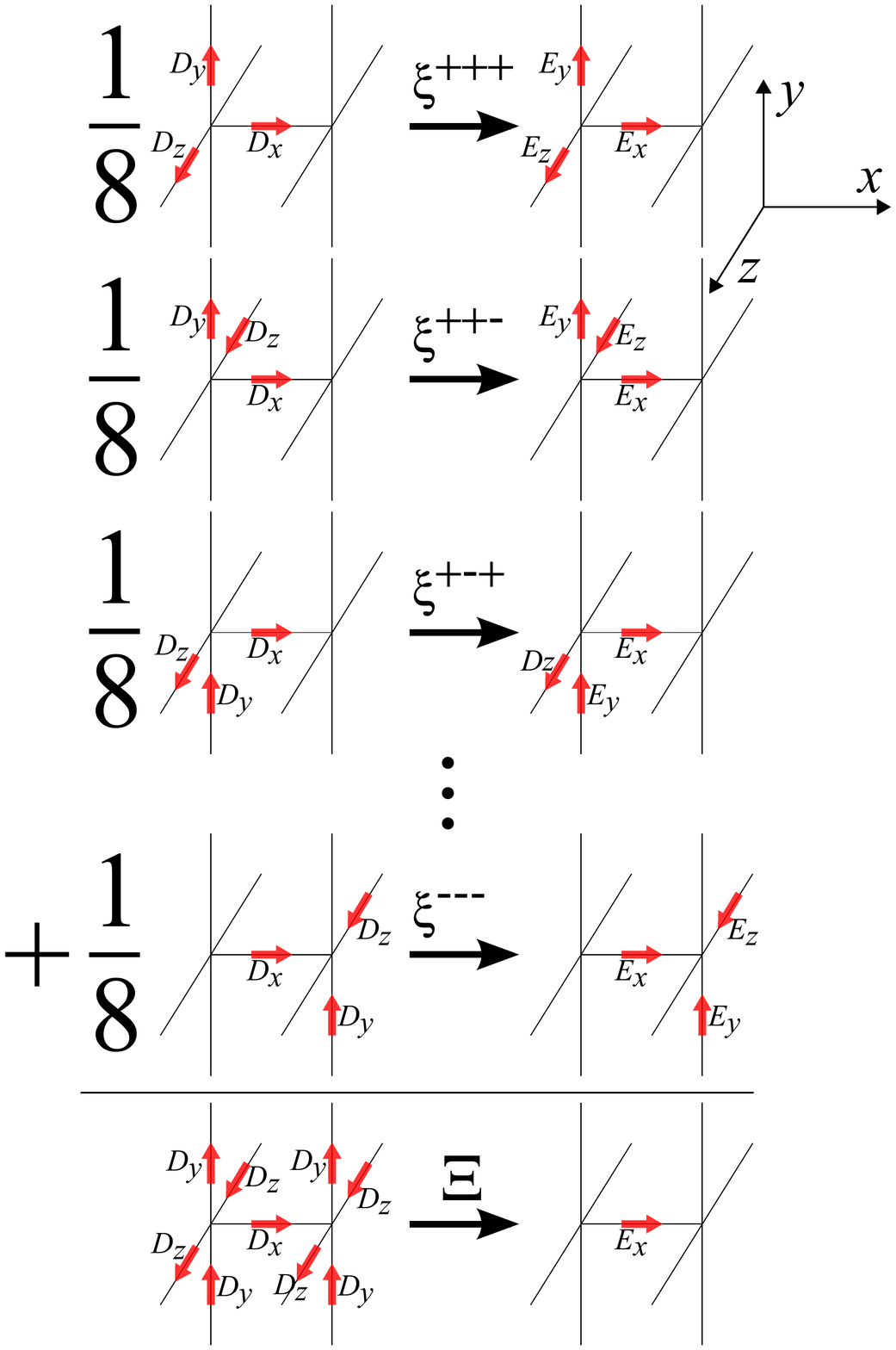}
\caption{(Color online.) A pictorial representation $E=\Xi D$, showing how
$E_{x}$ in cell $(i,j,k)$ is found from neighboring components of
$D$.  For each of 8 ``triplets'' a local effective inverse
dielectric $\xi$ (a $3\times 3$ matrix) converts $D$ to $E$.
Ultimately $E_x$ is found using 8 different $\xi$ matrices, one
for each triplet involving $E_x$. By averaging over all 8
triplets, $E_x$ depends symmetrically on its neighboring $D_y$
and $D_z$, which yields a centered algorithm that, in uniform (or
continuously-varying) dielectric, has second-order error
\cite{Werner:2007}.  We note that 4 of the triplets
use $\xi^{+\pm\pm}_{ijk}$ and 4 use $\xi^{-\pm\pm}_{(i+1)jk}$.
\label{fig:algorithm}}
\end{figure}

We finish by taking the $\Xi$ matrix to be the average,
\begin{equation} \label{eq:stableScheme}
	\Xi \equiv \frac{1}{8} \left(
	\Xi^{+++} +\Xi^{++-} +\Xi^{+-+} +\Xi^{-++}
	+\Xi^{+--} +\Xi^{-+-} +\Xi^{--+} +\Xi^{---} \right)
.\end{equation}
Since the sum of SPD matrices is again SPD, $\Xi$ is SPD.
Figure~\ref{fig:algorithm} illustrates
how this scheme determines $E_{ijkx}$ from neighboring $D$ values. 

Thus we can create a stable algorithm
independent of the details 
of the $\xi_{ijk}^{\pm\pm\pm}$,
which can be chosen to improve accuracy at the
dielectric interface, within broad constraints: the $\xi_{ijk}^{\pm\pm\pm}$ 
must be SPD.  In addition, 
when node $(ijk)$ is within uniform dielectric and 
far from a dielectric interface,
$\xi_{ijk}^{\pm\pm\pm}$ must equal $\xi(x,y,z)$, 
where $(x,y,z)$ is the position of node $(i,j,k)$.
This guarantees that, within isotropic dielectric,
each $\Xi_{\pm\pm\pm}$ is a multiple of the identity, and so $\Xi$ 
satisfies requirement 2, above.  
Furthermore, within uniform dielectric, the algorithm becomes
identical to wc07 (see Fig.~\ref{fig:oldAlgorithm}), which uses
centered interpolation to yield $O(\Delta x^2)$ error (within uniform
dielectric \cite{Werner:2007}).  
Therefore, $\Xi$ satisfies requirements 1 and 2 (as
well as 3 and 4).

It is easy to forget, when viewing the algorithm as a way to find
a single $E_x$ (as in Fig.~\ref{fig:algorithm}), that
$\xi^{+-+}_{ijk,xx}$, $\xi^{+-+}_{ijk,xy}$, and $\xi^{+-+}_{ijk,xz}$ must come from the
same (SPD) tensor $\xi^{+-+}_{ijk}$. 
Indeed, it was forgetting which
$\xi_{ijk,\mu\nu}$ had to be mathematically related to each other
that led to the instability of wc07.

To reiterate: each $\Xi^{\pm\pm\pm}$ matrix is block-diagonal,
with $3\times 3$ blocks ($\xi^{\pm\pm\pm}_{ijk}$), each of which
represents the effective dielectric tensor around node $(ijk)$. As
long as the $\xi_{ijk}^{\pm\pm\pm}$ are SPD,
the $\Xi^{\pm\pm\pm}$ are SPD.
This further implies that the average,
Eq.~\ref{eq:stableScheme}, is SPD.

We have thus shown how to find a stable matrix $\Xi$, given the
ability
to find $3\times 3$ effective inverse dielectric tensors that map
a triplet of neighboring components $(D_x, D_y, D_z)$ to the
$(E_x, E_y, E_z)$ at the same locations. Moreover, within uniform
(isotropic or anisotropic) dielectric, 
this algorithm is identical to wc07 (hence it
satisfies requirements 1 and 2).

It remains to find the effective dielectric tensors
$\xi^{\pm\pm\pm}_{ijk}$ that will accurately represent the real
dielectric.  Reference~\cite{Bauer:2011} showed how to find
$\xi^{\pm\pm\pm}_{ijk}$ with local $O(\Delta x)$ error, yielding
global $O(\Delta x^2)$ error; unfortunately, those
$\xi^{\pm\pm\pm}_{ijk}$ are asymmetric.

In the next section, we will describe the effective dielectric
tensors that yield the wc07 algorithm; some of those tensors may
not be positive definite, so that algorithm can be unstable.  We
then describe a modification to make it stable.  However, another
effective dielectric turns out to yield lower error
(Sec.~\ref{sec:symAcc}).  We have tried several other effective
dielectrics, and found them less accurate, but always yielding
ultimate $O(\Delta x)$ global error (even stairstepped
dielectrics have $O(\Delta x)$ error).

\section{The wc07 method, unstable at high contrast}
\label{sec:oldAlg}

Experiment tells us that wc07 (the algorithm of \cite{Werner:2007})
can be unstable;
therefore, we need
concern ourselves no more with that algorithm.  However, it is
interesting to see exactly how these algorithms differ,
so in this section we describe wc07 within the framework of this paper,
and show why
it does not meet the previously-described (sufficient, but not necessary)
conditions for a stable algorithm.

\begin{figure}[tbp]
\centering
\includegraphics*[trim = 0mm 250mm 0mm 0mm,width=130mm]{%
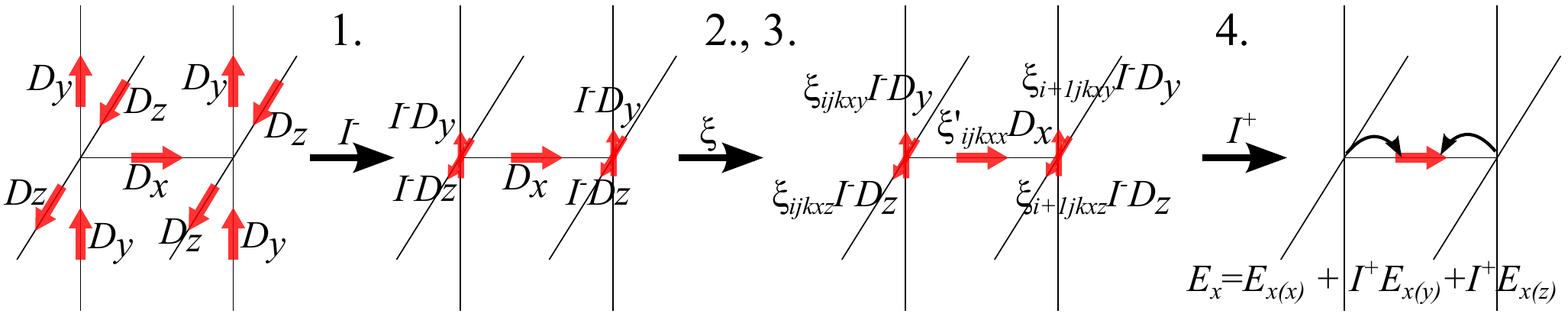}
\caption{(Color online.) A pictorial representation of $E=\Xi D$ as
suggested in \cite{Werner:2007}.  $E_x$ in cell $(i,j,k)$
is found from $D_x$
at the same location and the nearest $D_y$ and $D_z$ values.
The $I^\pm$ are interpolation operators (see \cite{Werner:2007}).
\label{fig:oldAlgorithm}}
\end{figure}

Figure~\ref{fig:oldAlgorithm} depicts the wc07 method recommended in
\cite{Werner:2007} [where it is called (c)$+$(e)].
To find $E_x$ from its nearest-neighbor
$D_\mu$ values:
\begin{enumerate}
  \item perform a centered interpolation of $D_y$ and $D_z$
    to the nearest cell node;
  \item find $E_{x(x)} = \xi^{(e)}_{xx} D_x$, using the effective
    dielectric $\xi^{(xe)}$ at the $x$-edge center;
  \item find $E_{x(y)} = \xi^{(n)}_{xy} D_y$ at each node, using
    the effective dielectric $\xi^{(n)}$ at that node;
    similarly, find $E_{x(z)} = \xi^{(n)}_{xz} D_z$ at each node;
  \item interpolate the two $E_{x(y)}$ from each node to the center
    of the $x$-edge where $E_x$ and $D_x$ are located;
    interpolate $E_{x(z)}$ similarly.
\end{enumerate}
Finally, $E_x$ is the sum of parts coming from $E_{x(x)}$,
$E_{x(y)}$, and $E_{x(z)}$.

In the framework of the previous section, wc07 is a
choice of effective $\xi^{\pm\pm\pm}$:
\begin{eqnarray}
  \label{eq:oldEffDiel}
  \xi^{\pm\pm\pm}_{ijk,\mu\nu} = \xi^{\pm\pm\pm}_{ijk,\nu\mu}
  &=& \xi^{(n)}_{ijk,\mu\nu} \quad (\textrm{for\ } \mu \neq \nu)
    \nonumber \\
  \xi^{+\pm\pm}_{ijk,xx} = \xi^{-\pm\pm}_{(i+1)jk,xx} &=& \xi^{(xe)}_{ijk,xx} \\
  \xi^{\pm+\pm}_{ijk,yy} = \xi^{\pm-\pm}_{i(j+1)k,yy} &=& \xi^{(ye)}_{ijk,yy} \nonumber \\
  \xi^{\pm+\pm}_{ijk,zz} = \xi^{\pm\pm-}_{ij(k+1),yy} &=& \xi^{(ze)}_{ijk,zz} \nonumber
\end{eqnarray}
The tensors $\xi^{(n)}_{ijk}$, $\xi^{(xe)}_{ijk}$, $\xi^{(ye)}_{ijk}$,
and $\xi^{(ze)}_{ijk}$ are all to be found from the averaging method
of \cite{Kottke:2008}, where $\xi^{(n)}_{ijk}$ is the ``average'' of
$\xi(x,y,z)$ over a cell volume centered at the node (lowest corner)
of cell $(i,j,k)$, and $\xi^{(xe)}_{ijk}$ is the ``average'' over
a cell volume centered at the location of $E_{ijkx}$ (the $x$-edge-center),
etc.

These $\xi_{ijk}^{\pm\pm\pm}$ do not satisfy the conditions
for the effective dielectric required by the previous section.
E.g., $\xi^{+++}_{ijk}$ is symmetric, but it is not necessarily
positive definite.  The reason is that the diagonal and off-diagonal
elements of $\xi^{+++}_{ijk}$ come from different
tensors: $\xi^{(n)}_{ijk}$, $\xi^{(xe)}_{ijk}$, $\xi^{(ye)}_{ijk}$, and
$\xi^{(ze)}_{ijk}$.  Each of these four tensors is SPD
(using the averaging method of \cite{Kottke:2008},
cf. \ref{sec:posDefEffDiel}),
but there's no guarantee that a tensor with a mixture of elements from
those tensors is positive definite.

Indeed, we have found experimentally that the wc07 algorithm
can yield an instability when the dielectric contrast
is high enough.  We hasten to point out that
we have used that algorithm successfully on a wide range of
problems without noticing any instability.
An instability seems to be more likely for higher contrast,
and for larger and more complicated dielectric shapes.

By using the Gershgorin circle theorem to place a lower bound on the
eigenvalues of the $\Xi$ matrix (if the lower eigenvalue bound is positive,
then $\Xi$ is positive definite, assuming $\Xi$ is symmetric), we can prove
for many particular simulations that the algorithm is in fact stable.  For
example, we usually find that simulations with $\epsilon \leq 10$ are
provably stable (on an individual basis, by examining $\Xi$ with
the Gershgorin circle theorem).

Many simulations appear stable for long times even when
not provably stable.  Of course, it's hard to know whether there might be
unstably-growing modes that would dominate the simulation if run 100 times
longer; and such unstable modes might interfere with precision
measurements well before they become obviously apparent.

For a 2D simulation of photonic crystal modes of
a square lattice of isotropic dielectric discs in vacuum
(radius $r=0.37a$, where $a$ is the lattice constant), we
have seen that at $\epsilon=60$ we can run simulations with
$48^2$ cells for a time 3000$a/c$ (where $c$ is the speed of light) 
\emph{without}
seeing an instability (which rules out instabilities growing faster
than $\gamma \sim 0.01c/a$, which corresponds to growth of
16 orders of magnitude in 3000$a/c$).  For $64^2$ cells, however, an
instability grows as $\exp(\gamma t)$, where $\gamma \approx 0.5c/a$.

For the same problem, with contrast $\epsilon = 100$, it's harder to
find stability for any resolution.  
For $32^2$ cells, an instability grows with
$\gamma \approx 3c/a$, and for $64^2$ cells, $\gamma \approx 6c/a$.

\section{wc07mod: a small change yields stability}
\label{sec:simpleStable}

In this section we make a small change to the wc07 algorithm
that renders it stable.  Although this algorithm, which we will call
``wc07mod,'' is not the most accurate,
we present it because it is a relatively trivial
modification of wc07, and the resulting degradation
of accuracy is interesting, in light of the small modification,
which still uses the same averaging method to find the effective
dielectric within a given cell-sized volume.

In the language of this paper, this algorithm is described simply as
\begin{eqnarray}
  \xi^{\pm\pm\pm}_{ijk\mu\nu} = \xi^{\pm\pm\pm}_{ijk\nu\mu}
  &=& \xi^{(n)}_{ijk,\mu\nu}
\end{eqnarray}
where $\xi^{(n)}_{ijk}$ is the ``average''
 inverse dielectric tensor for
a cell volume centered at the node of cell $(i,j,k)$---where
averaging is done according to \cite{Kottke:2008}.

Only the diagonal elements of the effective dielectric change,
compared to Eq.~(\ref{eq:oldEffDiel}).

In the
language of \cite{Werner:2007}, we need simply replace,
e.g., in 
Eq.~(27e) of \cite{Werner:2007} or in step 2 of the wc07 algorithm:
\begin{eqnarray}
  \xi^{(xe)}_{ijkxx} \quad &\rightarrow& \quad
  \frac{1}{2} \left[ \xi^{(n)}_{ijkxx} + \xi^{(n)}_{(i+1)jkxx} \right]
\end{eqnarray}
and similarly for the $yy$ and $zz$ elements.

\section{The new method}
\label{sec:symAcc}

The most accurate local effective dielectric, from \cite{Bauer:2011},
yields local $O(\Delta x)$ error, but is unfortunately asymmetric
(except for a few surface cuts: e.g., when a planar surface is parallel
to a grid plane).
Simply symmetrizing it increases the error to $O(1)$, but turns out
to be more accurate than other (symmetric) effective dielectrics.

Section~\ref{sec:stability} reduced the problem of finding a stable
$\Xi$ matrix to the problem of finding SPD
$3\times 3$ matrices, e.g., $\xi^{+-+}_{ijk}$, that
map a triplet of neighboring components $(D_x, D_y, D_z)$ to
the $(E_x, E_y, E_z)$ at the same locations---in a way that accurately
represents the real dielectric.  In this section, we describe the
best such recipe that we have found.

While we focus on finding the effective dielectric for a single triplet,
we will omit the $(+\!-\!+)$
and $(ijk)$ super- and sub-scripts, which identify the triplet.

This effective dielectric, which is not a volume-averaged dielectric as
in \cite{Johnson:2001,Kottke:2008}, derives
from Ref.~\cite{Bauer:2011}, which finds the unique
$3\times 3$ tensor $\xi_{\rm acc}$ that
guarantees that $(E_x, E_y, E_z)^T = \xi_{\rm acc} (D_x, D_y, D_z)^T$ 
will be exactly
accurate in the
limit of infinite wavelength and planar interface.
In other words,
$\xi_{\rm acc}$ will convert $(D_x,D_y,D_z)$ to $(E_x,E_y,E_z)$
with no error, given that the triplets are from the finite-difference
(or rather, finite integration) representation of an infinite-wavelength
solution of Maxwell's equations.
Unfortunately, as we have mentioned, $\xi_{\rm acc}$ is not symmetric.

To achieve stability in the time-domain, we will use
\begin{eqnarray}
  \xi_{\rm eff} = \frac{1}{2} \left( \xi_{\rm acc} + \xi_{\rm acc}^T \right)
.\end{eqnarray}
This will be stable; its accuracy will be evaluated empirically.

Finding $\xi_{\rm acc}$ is a lengthy process
fully described in \cite{Bauer:2011}, so we present only a terse recipe
for converting a triplet $(D_x,D_y,D_z)$ to $(E_x,E_y,E_z)$ in
the presence of
two dielectric regions, $\epsilon_1$ and $\epsilon_2$ (both
symmetric tensors).
\begin{enumerate}
  \item Within a small region around the triplet (we use the cell volume
    centered at the nearest node),
    the dielectric interface is nearly
    planar; find the unit surface normal $\hat{\bf n}$.
  \item Each electric field component $E_\mu$ is associated with a
    cell edge $L_\mu$; for each edge, determine the fraction
    $\ell_{\mu}$ of its length in dielectric $\epsilon_1$.
    See \cite{Bauer:2011} for explicit definition
    of $L_\mu$ (and $A_\mu$ in the following).
  \item Each component $D_\mu$ is associated with a dual-face area
   $A_{\mu}$ (centered at the Yee location of $D_\mu$,
   perpendicular to $\mu$);
   for each area, determine the fraction $a_{\mu}$ of the area
   in $\epsilon_1$.
  \item Form the $3\times 3$ matrices (for $p=1,2$)
    \begin{eqnarray}
       \Gamma_p &\equiv & I + \frac{1}{\hat{\bf n}^T \epsilon_p \hat{\bf n}}
	   [\hat{\bf n} \hat{\bf n}^T] (I - \epsilon_p) \\
	 \Pi_p & \equiv & \epsilon_p \Gamma_p \\
	 \Gamma & \equiv &
	  \left( \begin{array}{ccc}
	    \ell_{x} & 0 & 0 \\
	    0 & \ell_{y} & 0 \\
	    0 & 0 & \ell_{z} \\
	  \end{array} \right) \Gamma_1 +
	  \left( \begin{array}{ccc}
	    1-\ell_{x} & 0 & 0 \\
	    0 & 1-\ell_{y} & 0 \\
	    0 & 0 & 1-\ell_{z} \\
	  \end{array} \right) \Gamma_2
	  \\
	 \Pi & \equiv &
	  \left( \begin{array}{ccc}
	    a_{x} & 0 & 0 \\
	    0 & a_{y} & 0 \\
	    0 & 0 & a_{z} \\
	  \end{array} \right) \Pi_1 +
	  \left( \begin{array}{ccc}
	    1-a_{x} & 0 & 0 \\
	    0 & 1-a_{y} & 0 \\
	    0 & 0 & 1-a_{z} \\
	  \end{array} \right) \Pi_2
	\end{eqnarray}
    where $[\hat{\bf n}\hat{\bf n}^T]$ is the dyadic matrix with elements
    $[\hat{\bf n}\hat{\bf n}^T]_{\mu\nu} = \hat{n}_\mu \hat{n}_\nu$, 
    and $I$ is the identity.
  \item The accurate effective (inverse) dielectric tensor is:
    \begin{eqnarray}
      \xi_{\rm acc} &=& \Gamma \Pi^{-1}
    \end{eqnarray}
    We then symmetrize that to find
    \begin{eqnarray}
      \xi_{\rm eff} &=& \frac{1}{2} ( \xi_{\rm acc} + \xi_{\rm acc}^T )
    \end{eqnarray}
  \item The above can fail if $\Pi$ is
    not invertible, and if the resulting $\xi_{\rm eff}$ is not
    positive definite.  Failure is ruled out for
    isotropic dielectrics \cite{Bauer:2011},
    and for anisotropic dielectrics, it has
    not yet happened in our experience.  Nevertheless, it's important to
    guard against pathological cases.  We suggest checking
    every $\xi_{\rm eff}$ for these two problems; if one should occur,
    then substitute the effective dielectric from \cite{Kottke:2008}
    (which is proven suitable in \ref{sec:posDefEffDiel}).
    This will happen so rarely, if ever, that the global error will not
    be significantly affected.
\end{enumerate}

\section{Simulation Results}

We tested various FDTD algorithms on different dielectric problems:
a square lattice of 2D isotropic and anisotropic dielectric discs in
vacuum;
a square lattice of 2D vacuum discs (holes) in isotropic dielectric;
and a cubic lattice of 3D dielectric spheres in vacuum,
for both isotropic and anisotropic dielectric.
For many of these cases, we also tested different dielectric contrasts.
We define $a$ to be the lattice constant, and $N_a$ the number of
(square or cubic) cells per lattice constant, hence $\Delta x = a/N_a$.

Ultimately, the FDTD algorithms all show
first-order error in frequency; the error in a mode frequency falls as
$O(\Delta x)$, or $O(\Delta x/\lambda)$, where $\lambda$ is a
characteristic wavelength of the mode,
with decreasing cell size $\Delta x$.
However, at coarse resolutions (large $\Delta x$), the error often
falls as $O(\Delta x^2)$ for low dielectric contrast.  This may explain
why previous studies have concluded that methods such as wc07 have
second-order error---they did not explore high-enough resolution
or contrast (of course, in
practice, one may often reach a tolerable error level within the
second-order regime, in which case
the ultimate order of error may be irrelevant).

The error convergence in surface fields was the same as in frequencies 
when
we considered the surface fields a fixed distance (e.g., $a/8$)
away from the dielectric boundary.
However, the error in fields a fixed number of cells (e.g., $3\Delta x$) away
from the boundary, is $O(1)$, not $O(\Delta x)$.  This supports
our assertion that the local error at the boundary,
due to discontinuous fields, is $O(1)$, but the global error is
$O(\Delta x)$ because the ratio of boundary cells to total cells is
$O(\Delta x)$.

We performed the FDTD simulations with \textsc{Vorpal} \cite{Nieter:2004}
using the FDM method \cite{Werner:2008} to extract accurate mode fields and
frequencies.  We
compared these results with the frequency-domain
algorithm of \cite{Bauer:2011}, which was shown to have second-order
global error.  For 2D simulation frequencies, we extrapolated results from
$N_a=512$ and $N_a=1024$ assuming second-order convergence to get
a normative value with approximately $O(\Delta x^3)$ error.

We will show the most detailed convergence results for the ``new''
algorithm (Sec.~\ref{sec:symAcc}) for 2D anisotropic
discs.  Isotropic and 3D dielectrics show similar convergence,
so we present only a few examples.

We will show that the other FDTD algorithms, wc07 and wc07mod
generally have similar or greater
error compared to the new algorithm;
and the other examples (3D and isotropic)
show qualitatively similar convergence.

For comparison, we also show frequency convergence for the second-order
(but unstable in the time-domain) method of \cite{Bauer:2011} in
\ref{sec:secondOrderConvergence}.

\subsection{Convergence: 2D anisotropic discs}
\label{sec:2dAnisoDiscs}

We simulated TE modes (with $E_z=0$, $B_x=0=B_y$, and no variation
in $z$) in
a square lattice (lattice constant $a$) of dielectric
discs of radius $r = 0.37a$ in vacuum; the discs were of dielectric
\begin{eqnarray}
  \epsilon &=& \epsilon_b
    \left( \begin{array}{ccc}
      1.025 & -\sqrt{3}/40 & 0 \\
	-\sqrt{3}/40 & 1.075 & 0 \\
	0 & 0 & 1. \\
    \end{array} \right)
\end{eqnarray}
where we vary the scalar $\epsilon_b$ to vary the dielectric contrast.
For $\epsilon_b=10$, the above is
the diagonal matrix $(10,11,10)$ rotated by 30 degrees around the
$z$-axis.

\begin{figure}[tp]
\centering
\includegraphics*[trim=0mm 0mm 0mm 0mm,width = 65mm]{%
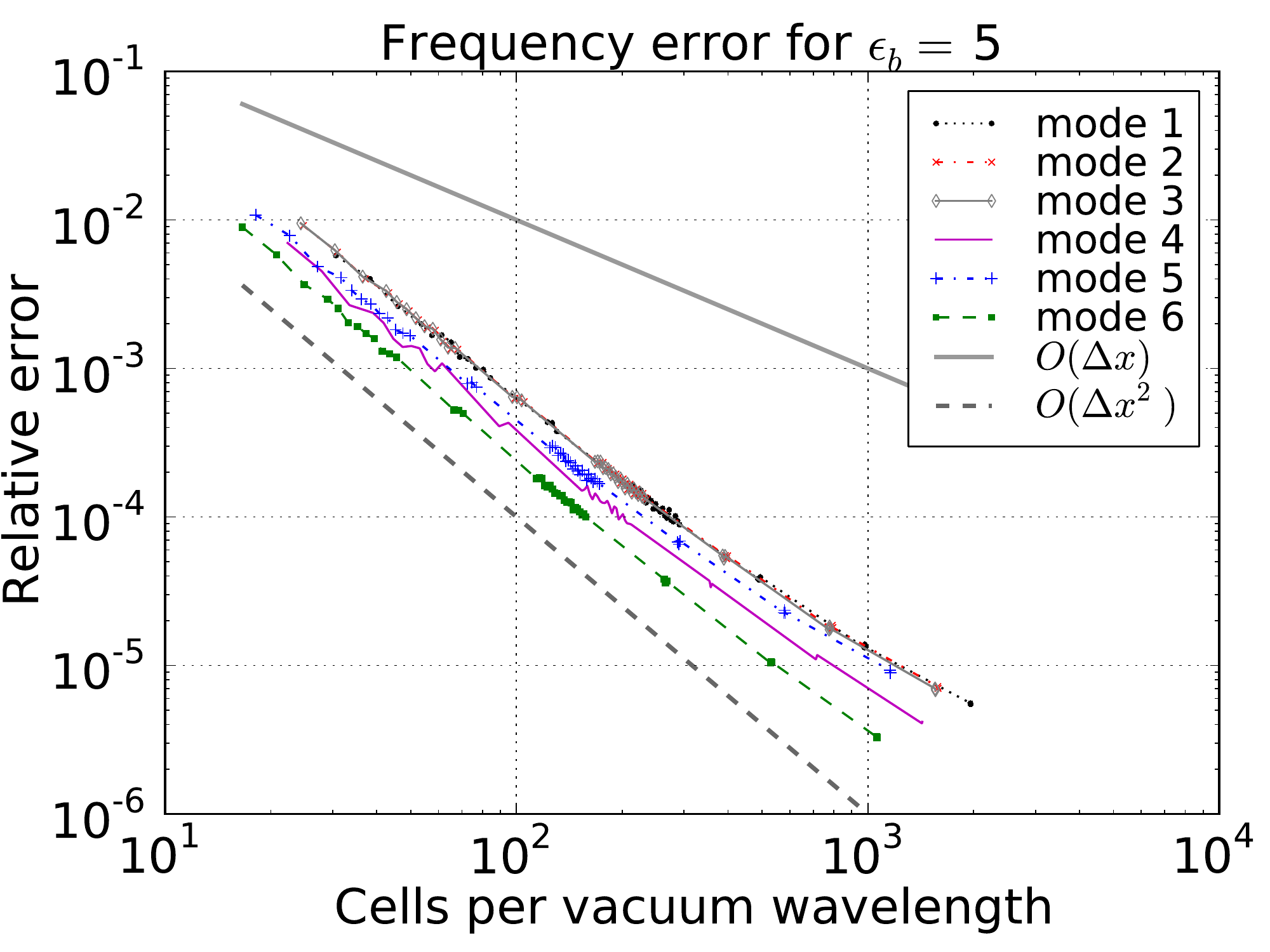}%
\includegraphics*[trim=0mm 0mm 0mm 0mm,width = 65mm]{%
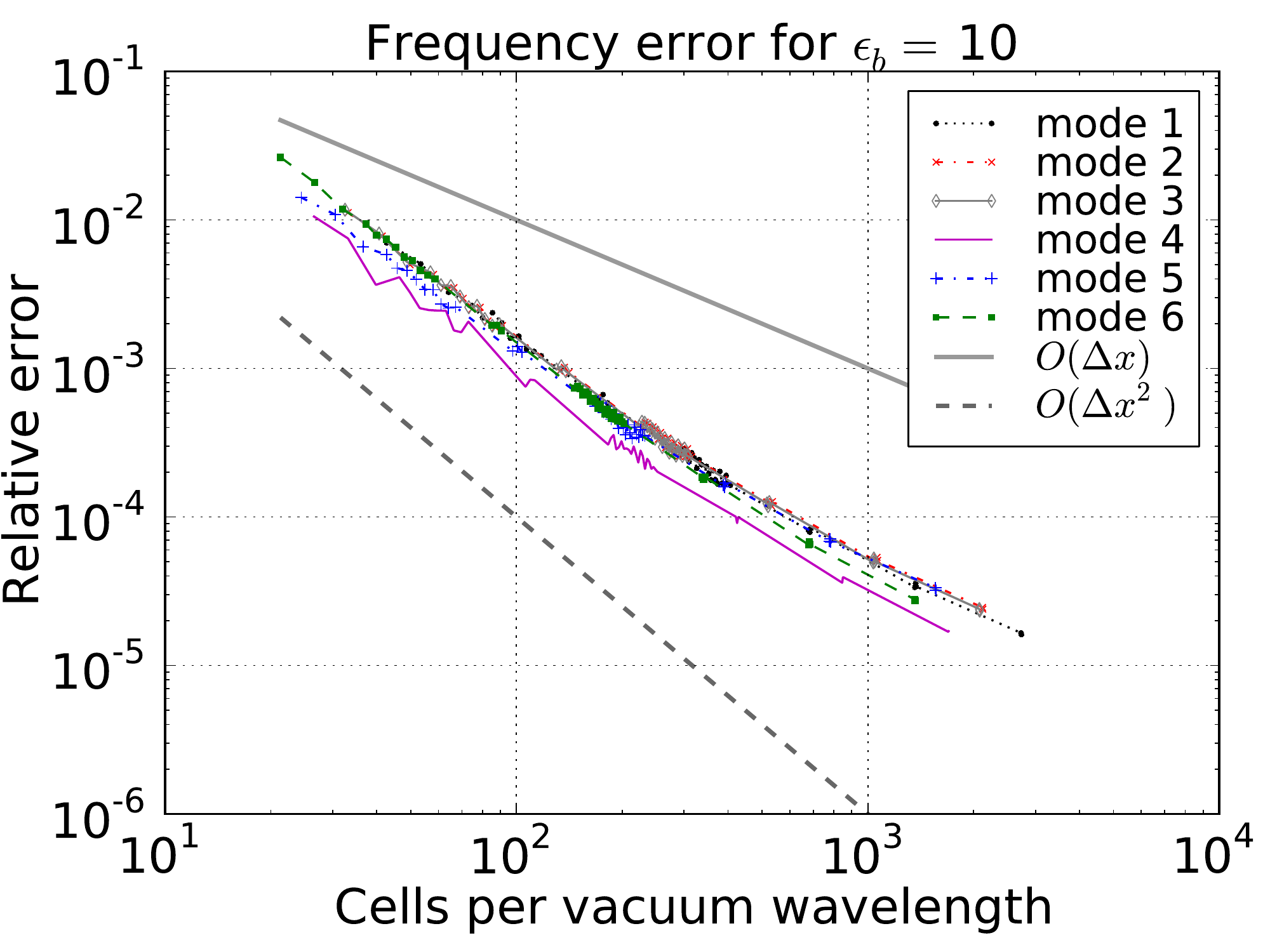}
\includegraphics*[trim=0mm 0mm 0mm 0mm,width = 65mm]{%
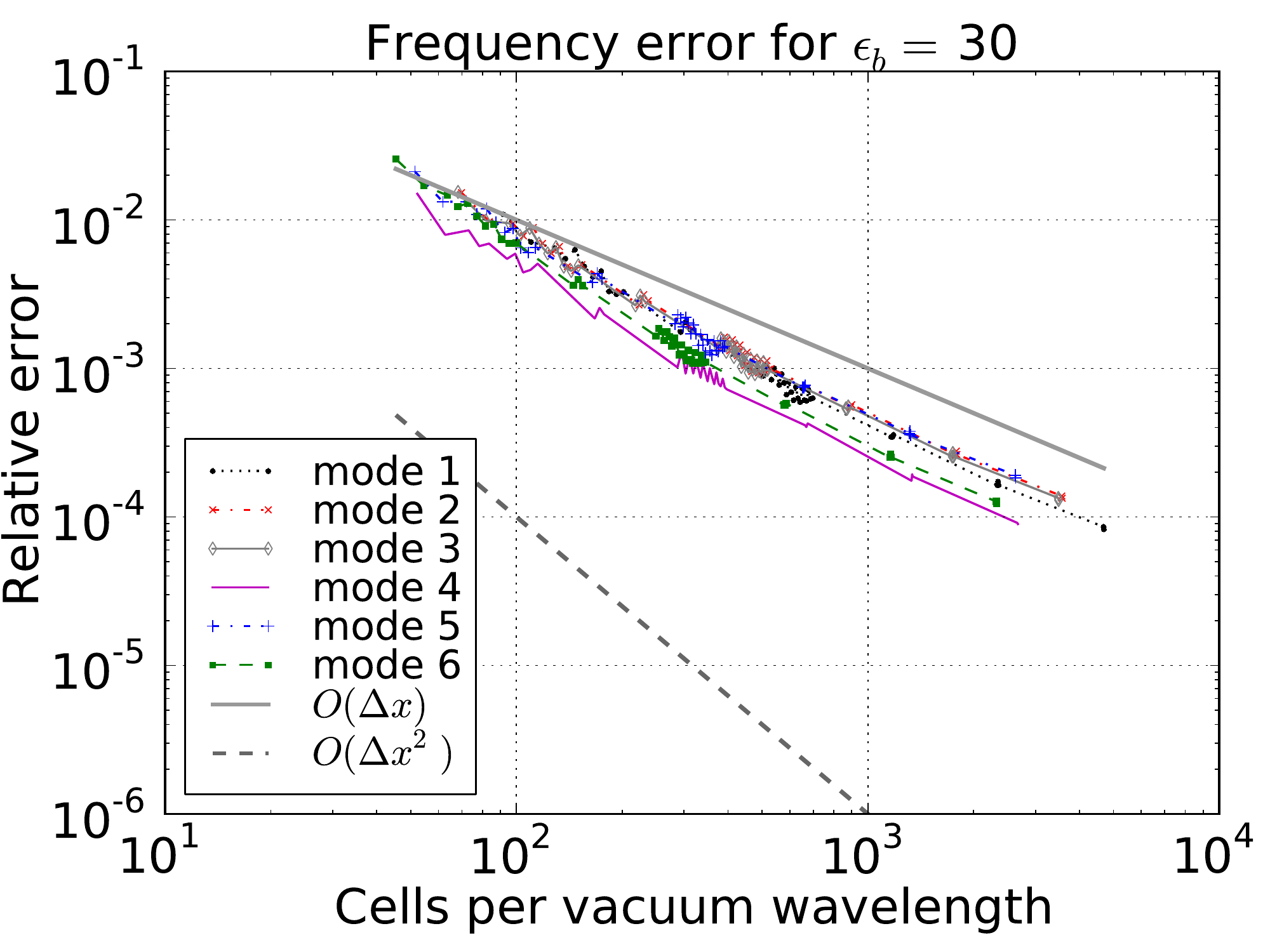}%
\includegraphics*[trim=0mm 0mm 0mm 0mm,width = 65mm]{%
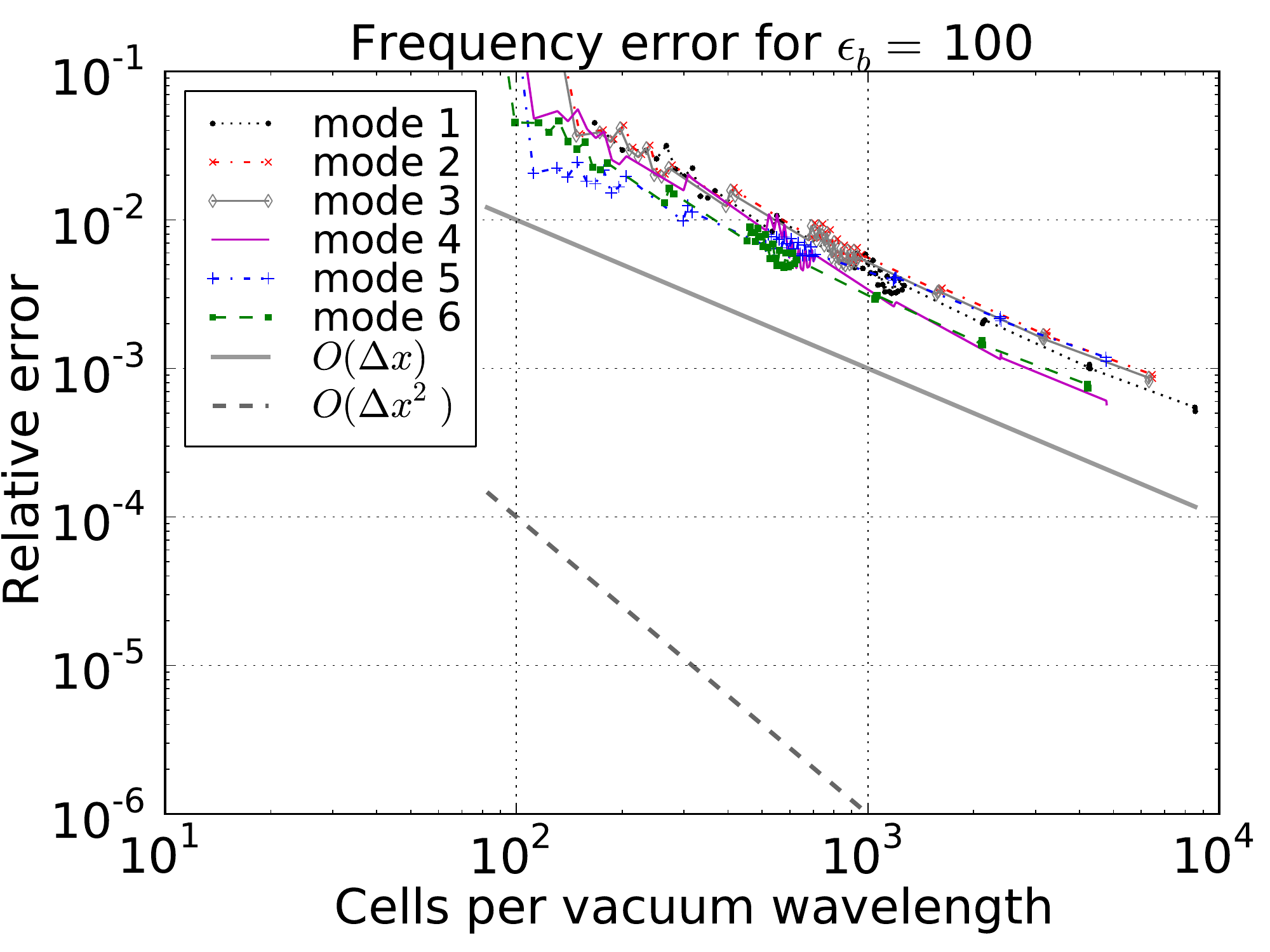}
\caption{(Color online.) For the new algorithm presented in this paper,
relative errors vs. resolution
for mode frequencies for a 2D photonic crystal of
$r/a=0.37$ anisotropic discs with varying dielectric contrast $\epsilon_b$.
The error transitions from
second-order to first-order at higher resolutions.  The transition
point occurs at coarser resolution for higher dielectric contrast.
\label{fig:anisoDiscFreqConv}}
\end{figure}

Figure~\ref{fig:anisoDiscFreqConv} shows the relative
error in the frequency of the lowest several modes vs. the number of
cells per vacuum wavelength (or $c$ divided by the mode frequency),
for dielectric contrast of $\epsilon_b = 5, 10, 30, 100$.
Low contrast simulations, $\epsilon_b \lesssim 10$, yield second-order
error to rather high resolutions.  At sufficiently high resolution,
the error becomes first-order; this is more clearly seen in the
mid-contrasts.  For high contrast, $\epsilon_b \gtrsim 30$, the second-order
region is too small to notice, and first-order convergence is clear.

There is a problem in examining the convergence of the fields.  The fields
are discontinuous at the dielectric interface, and it is not obvious
how best to interpolate the fields near the interface.  There is a danger,
when choosing an interpolation method, that it might not be the best
interpolation method.  Therefore, we avoid the interface.  Staying at
least $3\Delta x$ (where $\Delta x$ is the cell size) away from the surface,
there is no serious ambiguity in interpolation: a simple bilinear
(or, in 3D, trilinear)
interpolation should be sufficient for errors of at least $O(\Delta x^2)$.

We examine field-convergence in two ways.

\begin{figure}[tp]
\centering
\includegraphics*[trim=0mm 0mm 0mm 0mm,width = 65mm]{%
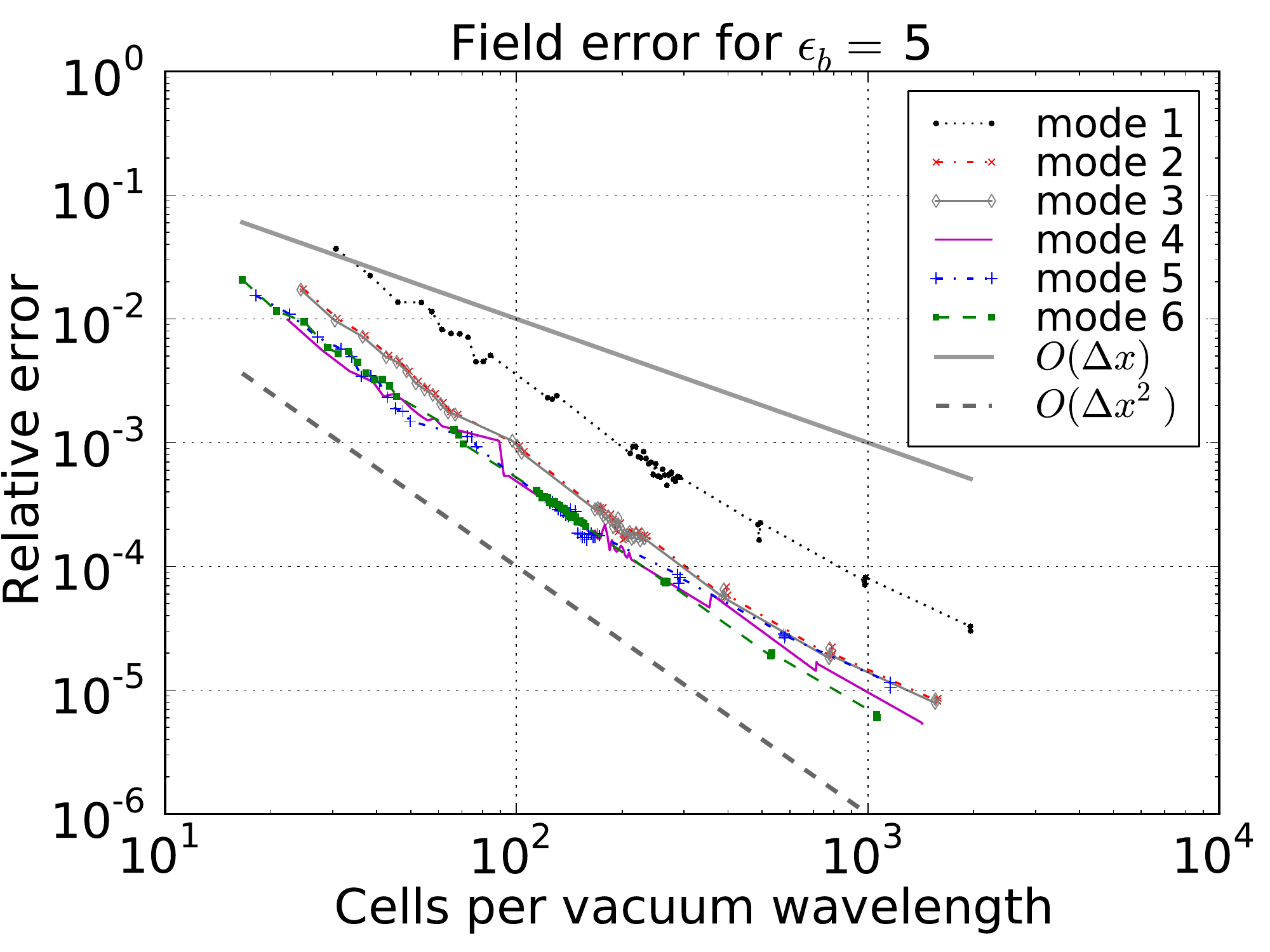}%
\includegraphics*[trim=0mm 0mm 0mm 0mm,width = 65mm]{%
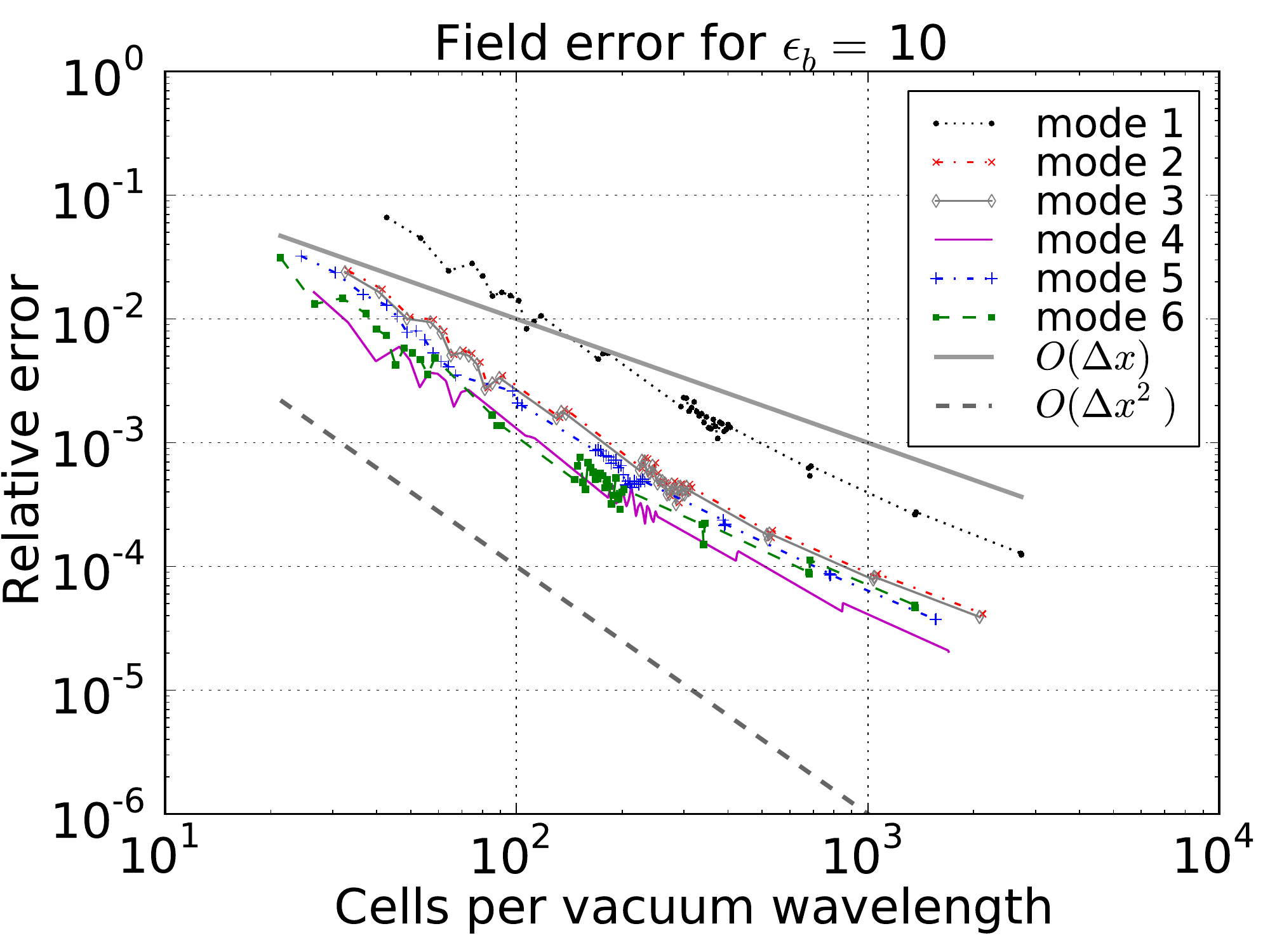}
\includegraphics*[trim=0mm 0mm 0mm 0mm,width = 65mm]{%
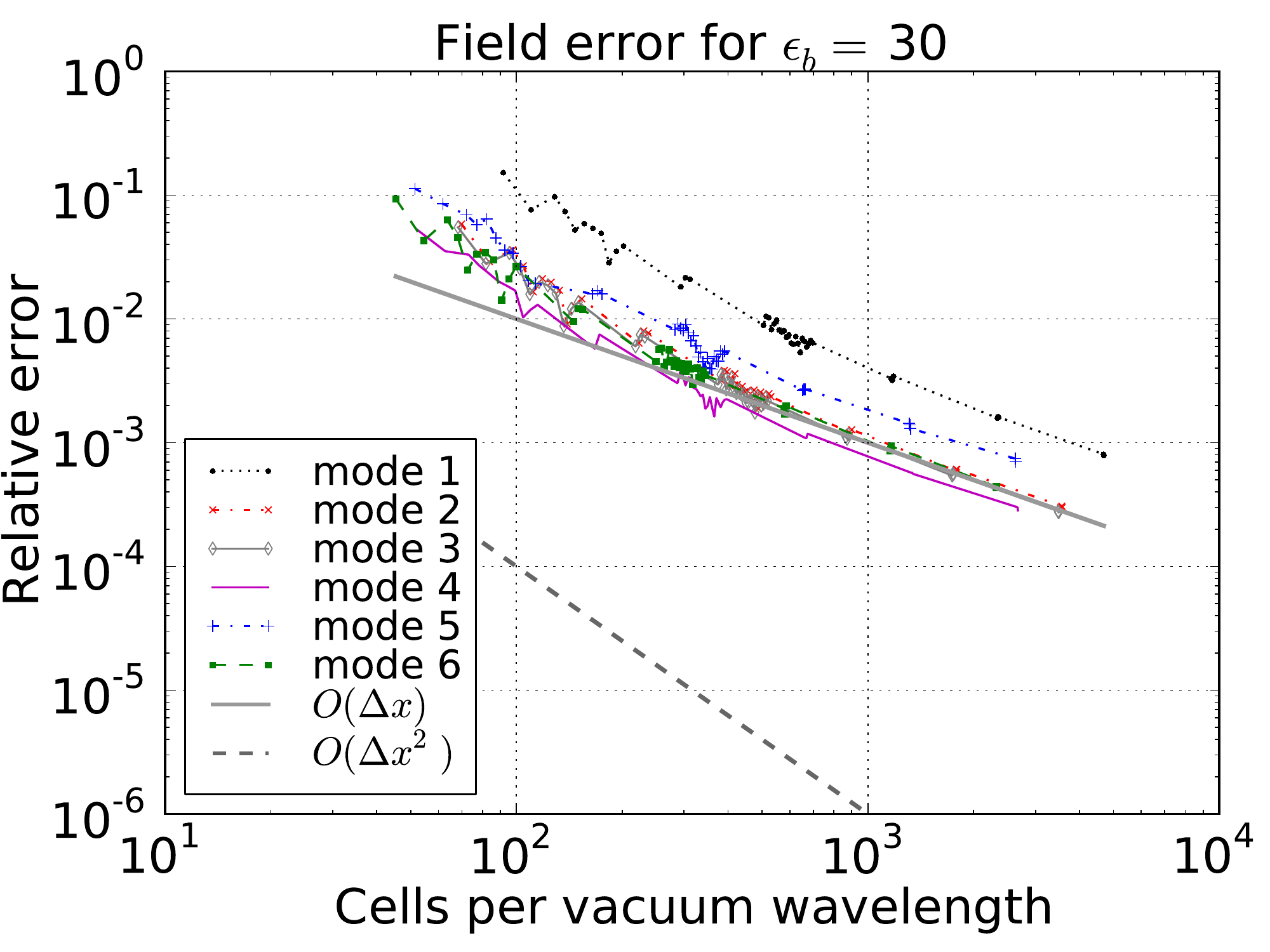}%
\includegraphics*[trim=0mm 0mm 0mm 0mm,width = 65mm]{%
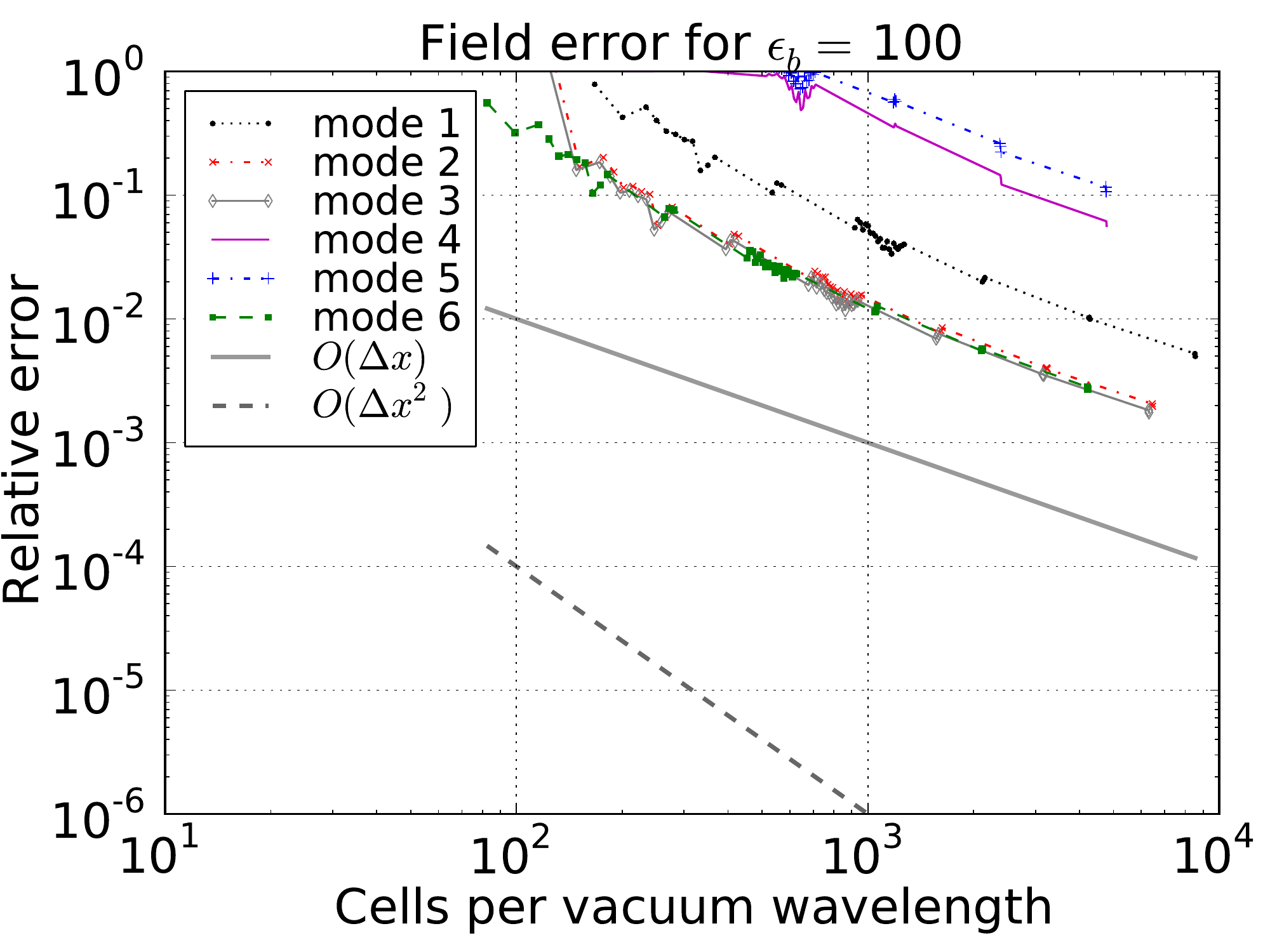}
\caption{(Color online.) For the new method, 2D, anisotropic:
relative error (in an $\ell_2$-norm) vs. resolution
in $\mathbf{E}$ at points on a circle of
radius $a/8$ outside the dielectric interface,
for a 2D photonic crystal of
anisotropic discs with varying dielectric contrast $\epsilon_b$.
At high resolutions, the error is first-order.
\label{fig:symAccSurfE}}
\end{figure}

First, we generate a set of (evenly-distributed) points on a circle at
a radius $a/8$ larger than the interface, and, at each resolution,
interpolate the mode fields to those points, comparing against
our normative values (from the algorithm of \cite{Bauer:2011}) using
the $\ell_2$-norm over the set of points 
(after normalizing the entire eigenmodes).
We then graph the relative error vs. resolution.
We find (not surprisingly) that
it converges at the same rate (ultimately $O(\Delta x)$)
as the mode frequencies (Fig. \ref{fig:symAccSurfE}).

For $\epsilon_b = 100$, modes 4 and 5 have relatively large errors
in the surface field; qualitatively, however, the modes look
more accurate than the $\ell_2$ norm suggests.  
These modes place most of the field
energy inside the dielectric; they resemble a pair of quadrupole
(e.g., $\textrm{TE}_{21}$) modes in a circular waveguide: that is,
the field patterns have nearly
(but not exactly, due to the square lattice)
an azimuthal dependence
$\cos(2\theta + \theta_0)$ and $\cos(2\theta + \theta_0 + \pi/2)$ for
some angle $\theta_0$.  If the dielectric were isotropic, these modes
would be degenerate, and $\theta_0$ could be chosen arbitrarily
(since there is a linear combination of the above two terms that yields
$\cos(2\theta + \theta_0')$ for any $\theta_0'$).
With the anisotropic dielectric, the mode frequencies differ by
about 0.3\%, and so $\theta_0$ is determined.  It appears that the
error is so high because the eigenmodes have a large error in $\theta_0$.
In other words, the field patterns look very similar to the correct
fields, except they are rotated slightly.  This is a consequence
of the difficulty of eigensolving for nearly-degenerate modes;
as two modes approach degeneracy it becomes impossible to separate
them correctly (without recourse to some other operator).  In this
light, it is not surprising that when the error in frequency is larger
than the actual separation between the two modes, the resulting
eigenmodes may be the wrong linear combinations of the exact
eigenmodes.
Indeed the surface error starts diminishing for modes 4 and 5
approximately when the frequency errors
approach 0.3\% (see Fig.~\ref{fig:anisoDiscFreqConv}).

\begin{figure}[tp]
\centering
\includegraphics*[trim=0mm 0mm 0mm 0mm,width = 65mm]{%
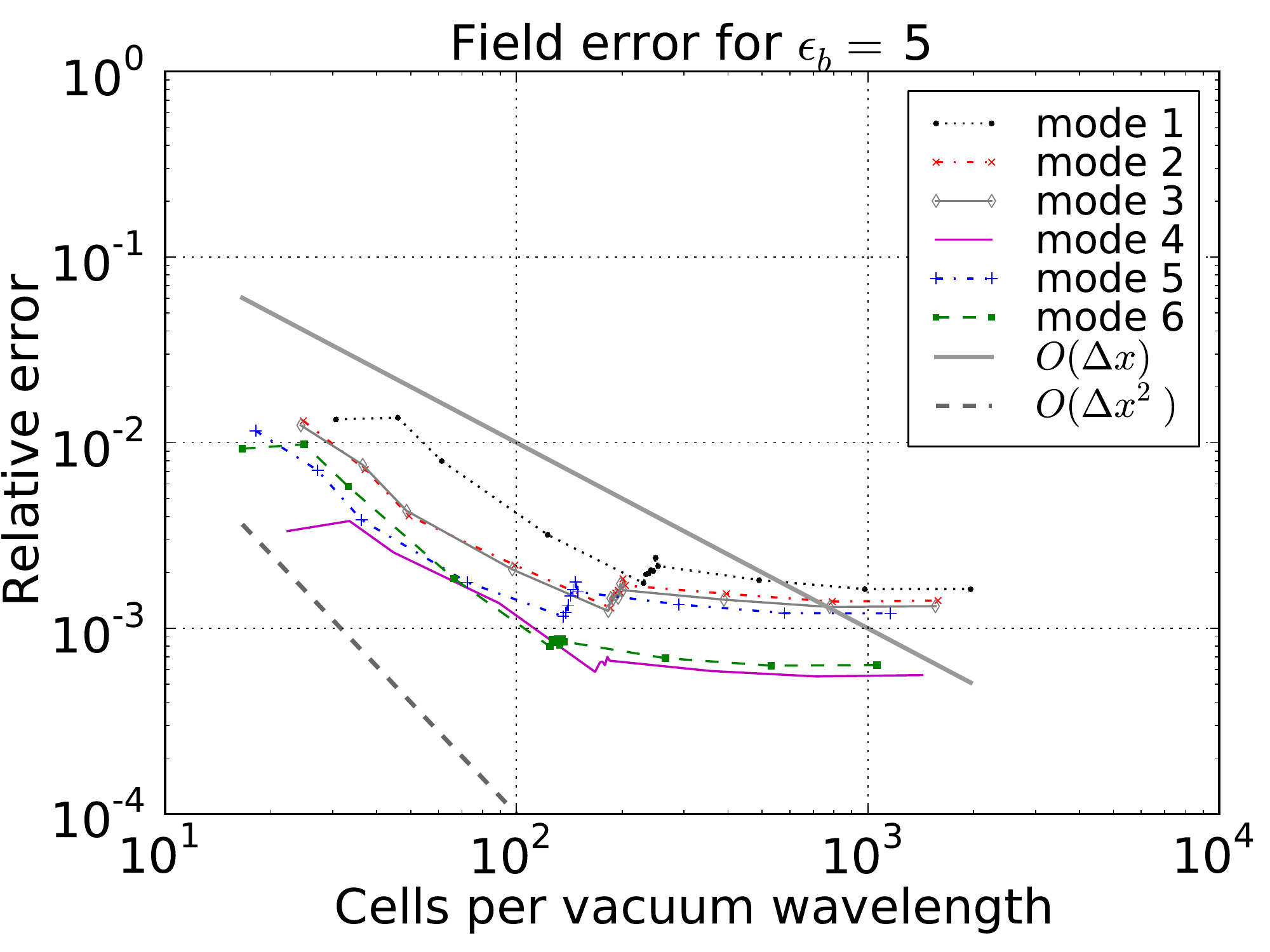}%
\includegraphics*[trim=0mm 0mm 0mm 0mm,width = 65mm]{%
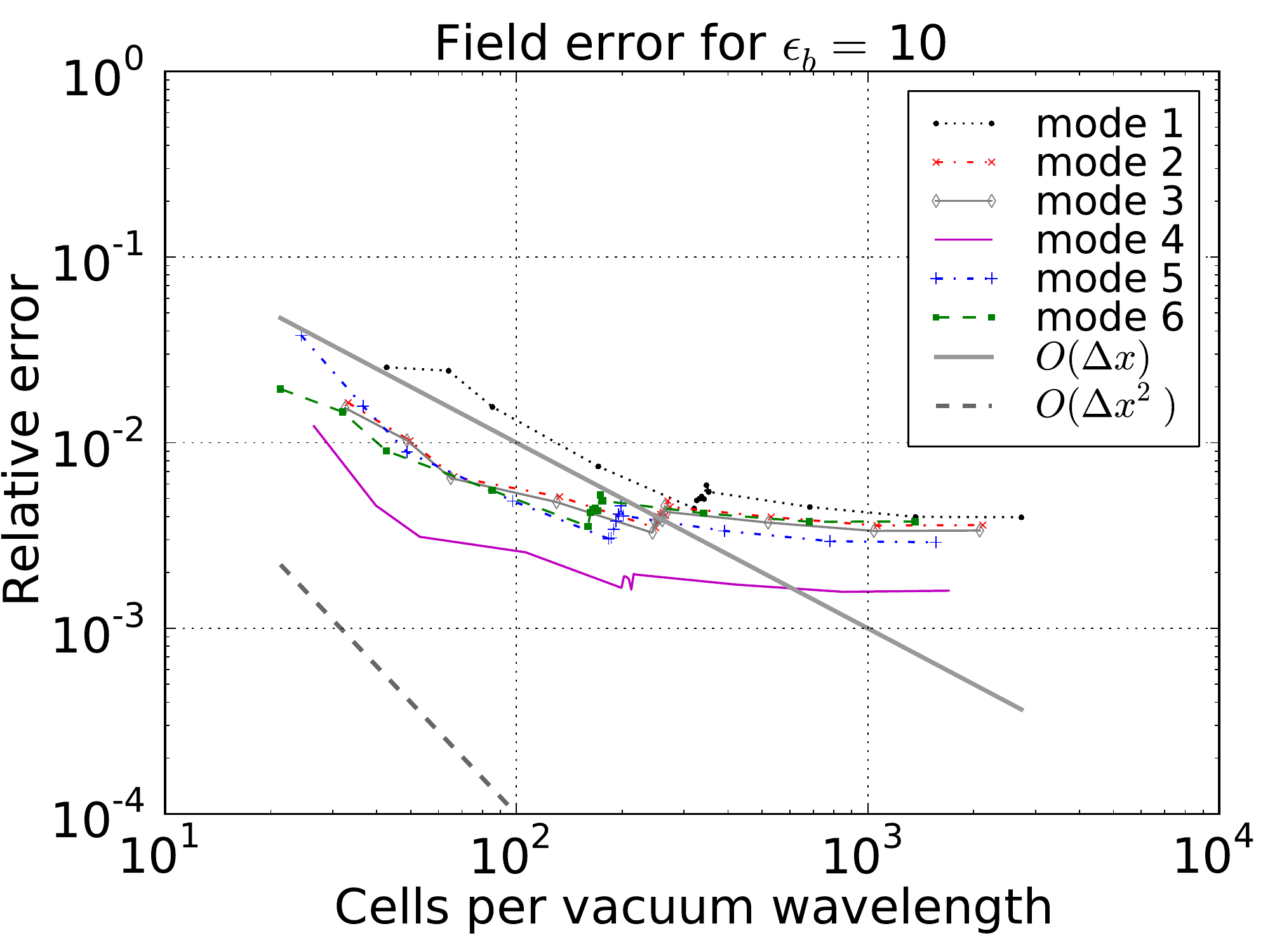}
\includegraphics*[trim=0mm 0mm 0mm 0mm,width = 65mm]{%
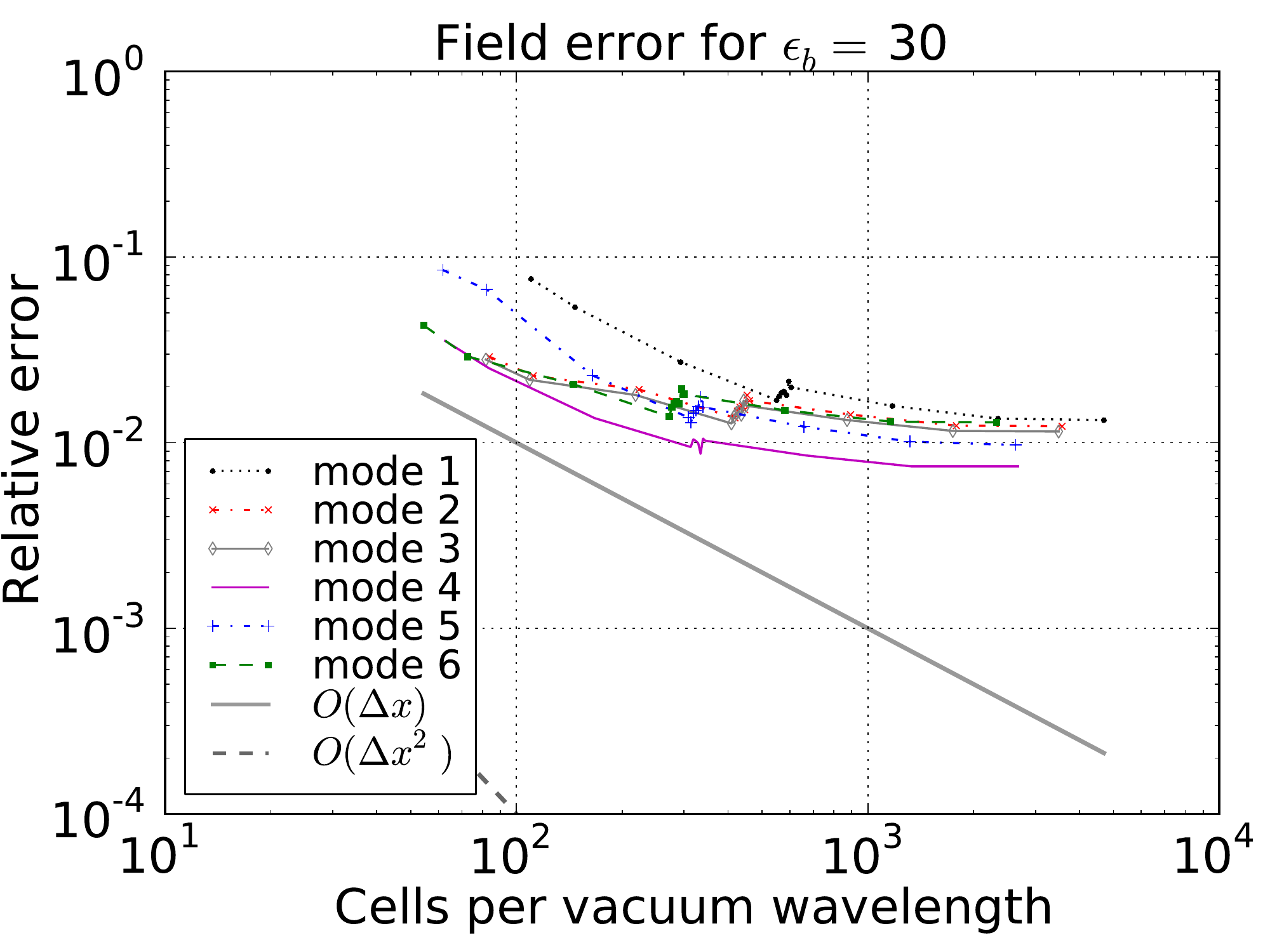}%
\includegraphics*[trim=0mm 0mm 0mm 0mm,width = 65mm]{%
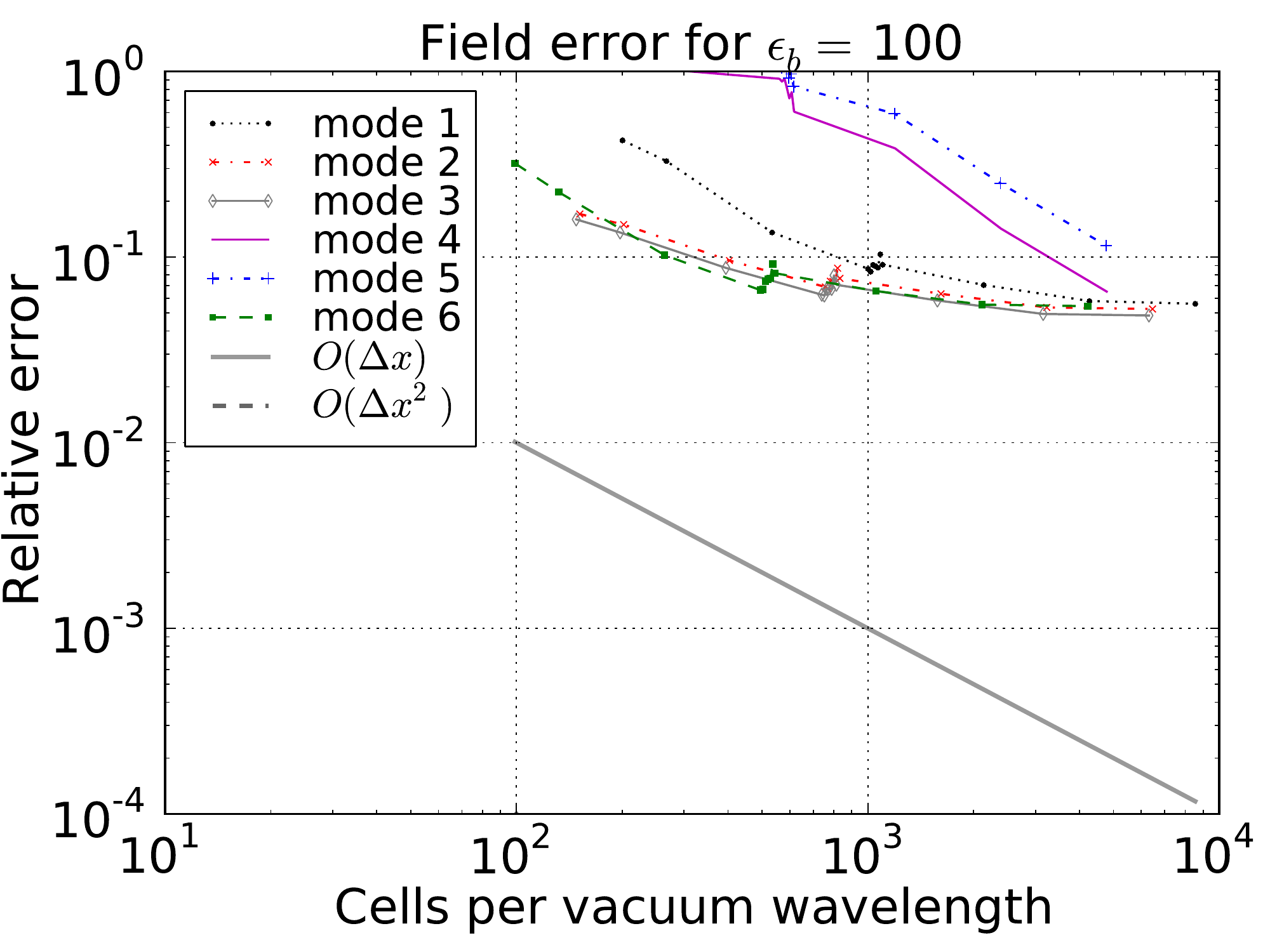}
\caption{(Color online.) For the new method, 2D, anisotropic:
relative error (in an $\ell_2$-norm) vs. resolution
in $\mathbf{E}$ at points on a sphere of
radius $(0.37 + 3\Delta x)a$ for a 2D photonic crystal of
anisotropic discs with varying dielectric contrast $\epsilon_b$.
This error does not go to zero as $\Delta x \rightarrow 0$.
\label{fig:symAccCSurfE}}
\end{figure}

Second, we generate a set of points on a circle that is a radius
$3\Delta x$ outside the interface; thus each different resolution has
a different set of points; as simulation resolution increases, the points
move closer to the actual interface.  For each resolution, the fields
are compared to our normative high-resolution simulation.
In this case, the error does not vanish with $\Delta x$;
in other words, it is $O(1)$
(Fig. \ref{fig:symAccCSurfE}).

This latter conclusion is disquieting for those who want particularly
to measure fields at the interface.  However, we point at that the error,
although $O(1)$, can be quite low, especially for low dielectric contrast.
For dielectric contrast $\leq 15$, the relative
error in field is less than one percent.
The error drops significantly as one moves away from the
interface, and we believe it might be possible to extrapolate
the fields from several cells away to find surface fields with
vanishing error as $\Delta x \rightarrow 0$.

The wc07 method \cite{Werner:2007,Oskooi:2009}, and
indeed all the effective FDTD dielectric methods we've tried, 
have very similar
error convergence, transitioning from second- to first-order at a
resolution that decreases as the dielectric contrast increases.
We believe this
may explain a discrepancy that has puzzled us: this work
and \cite{Werner:2007} see first-order error (in mode frequencies) 
for this effective
dielectric, while the results of \cite{Oskooi:2009} show
second-order error.  The latter uses two anisotropic dielectrics,
one with eigenvalues (1.45, 2.81, 4.98), and another with
(8.49, 8.78, 11.52), both rotated by random orthogonal matrices.
At contrast $11.52/1.45 \approx 8$, we do in fact see
second-order behavior up to hundreds of cells per wavelength, and for
contrasts between other pairs (e.g., $8.49/4.98$), second-order
behavior persists up to higher resolutions than we have explored.

In fact, practically, it must be said that the effective dielectric in
\cite{Kottke:2008,Oskooi:2009} does yield second-order behavior for
low-contrast dielectric up to nearly the highest resolutions that one
might practically use.  Ultimately, however, it has
first-order error.

For low contrasts, the local error associated with the
dielectric interface doesn't see to have much effect, and so the results
are similar for different algorithms.

\subsection{Convergence: 2D isotropic discs}

\begin{figure}[tp]
\centering
\includegraphics*[trim=0mm 0mm 0mm 0mm,width = 65mm]{%
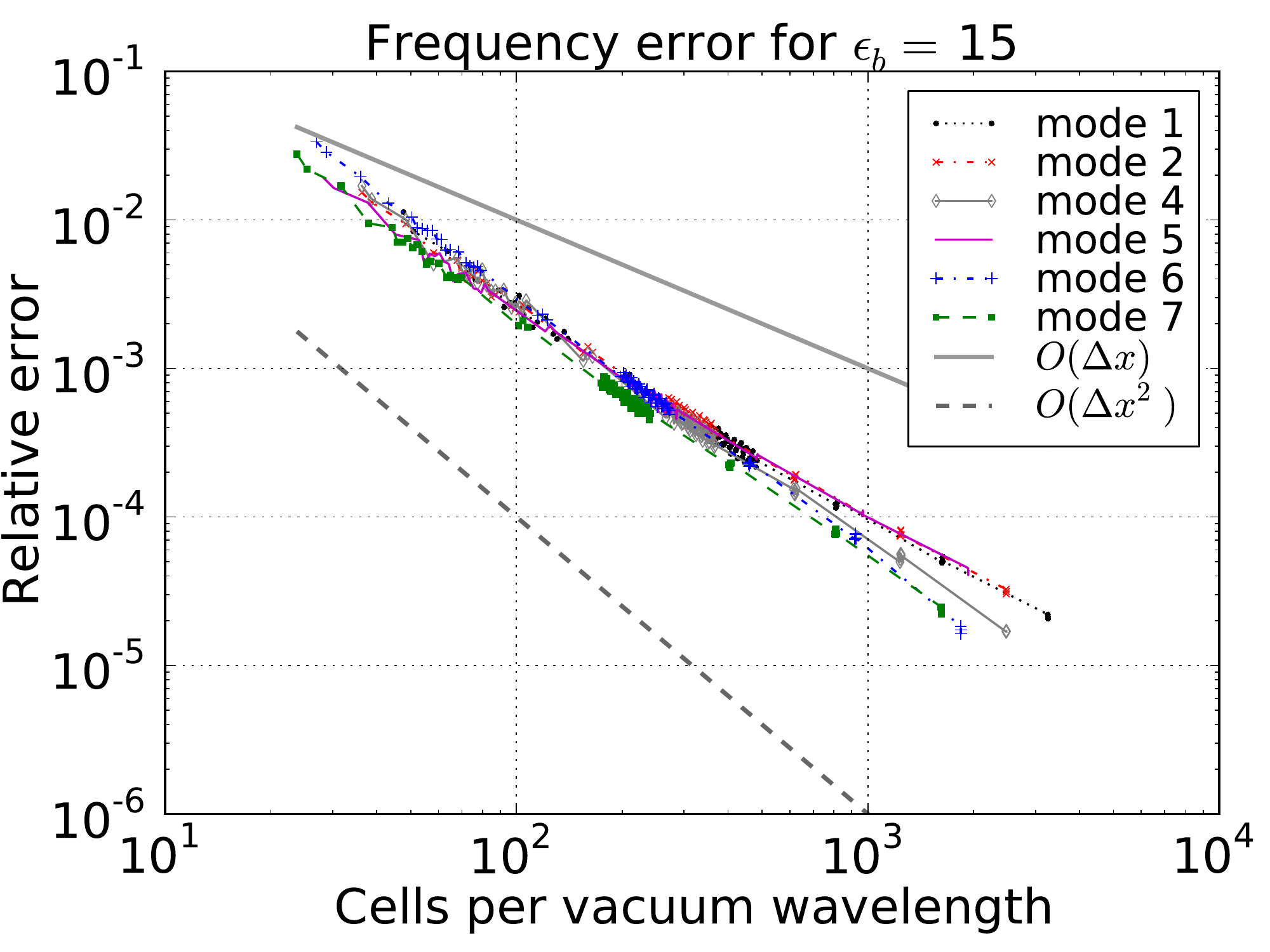}%
\includegraphics*[trim=0mm 0mm 0mm 0mm,width = 65mm]{%
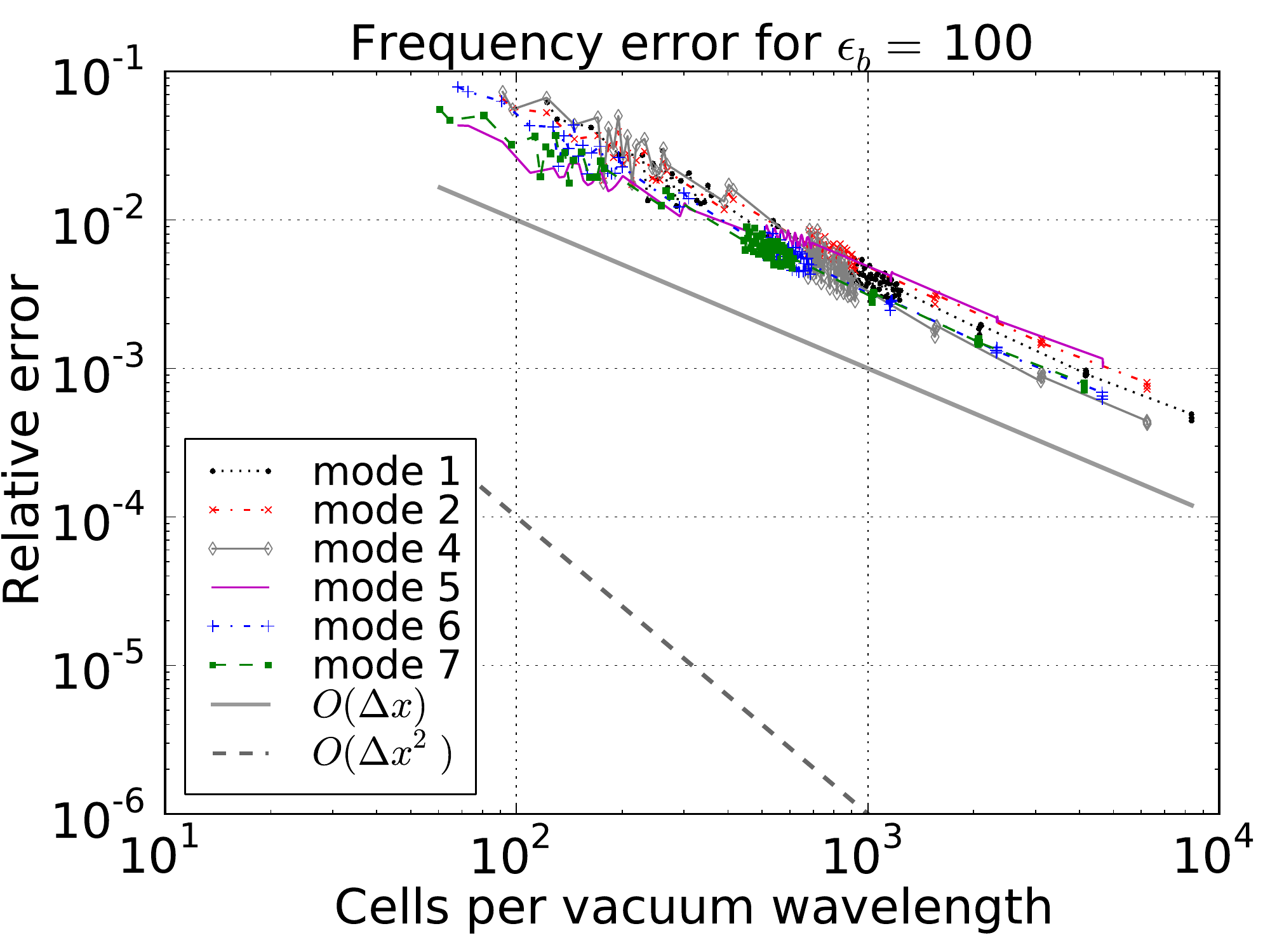}
\caption{(Color online.) For the new method, 2D, isotropic: relative errors vs. resolution
for mode frequencies for a 2D photonic crystal of
$r/a=0.37$ isotropic discs with dielectric contrast $\epsilon_b = 15$
(left) and $\epsilon_b=100$ (right).  Again, the error transitions from
second-order to first-order.  The transition
point occurs at coarser resolution for higher dielectric contrast.
\label{fig:symAccIsoDiscFreqs}}
\end{figure}

To show that using isotropic dielectric (instead of anisotropic) does
not change the order of error, we present frequency convergence results
for the same problem in the previous section, except that the dielectric
is replaced with an isotropic dielectric $\epsilon = \epsilon_b$.
Figure~\ref{fig:symAccIsoDiscFreqs} shows frequency error convergence
for $\epsilon_b=15$ and $\epsilon_b=100$.  The former shows a gradual 
transition
from second-order to first-order error, while the the higher contrast
shows just first-order error.

\subsection{Convergence: 3D isotropic spheres}

\begin{figure}[tp]
\centering
\includegraphics*[trim=0mm 0mm 0mm 0mm,width = 65mm]{%
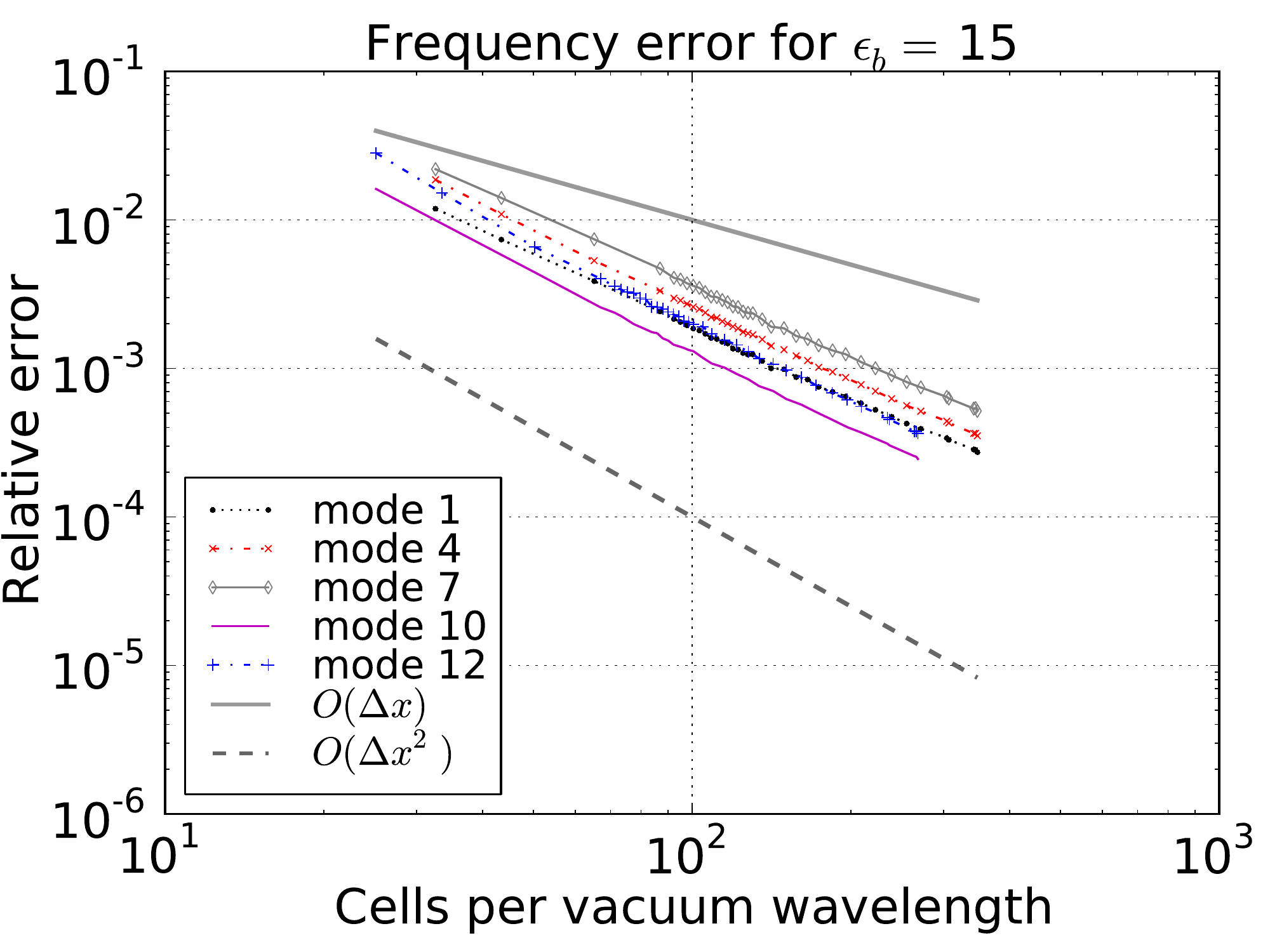}%
\includegraphics*[trim=0mm 0mm 0mm 0mm,width = 65mm]{%
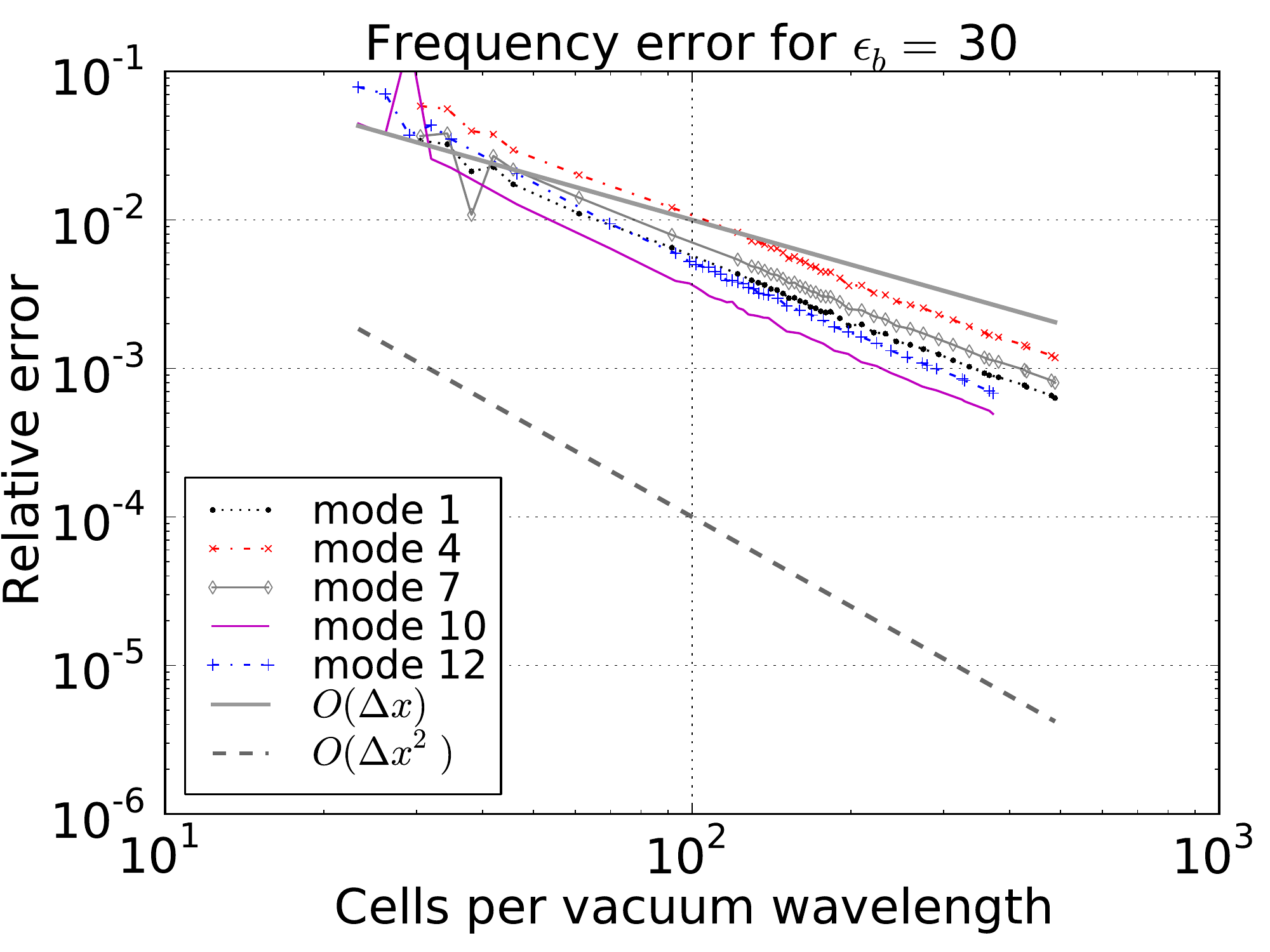}
\caption{(Color online.) For the new method, 3D, isotropic:
relative errors vs. resolution for mode frequencies for
a 3D cubic lattice of spheres with isotropic $\epsilon=15$ (left) and
$\epsilon=30$ (right), using the algorithm of Sec.~\ref{sec:symAcc}.
For $\epsilon=15$ convergence is second-order for low resolution, but
gradually moves toward first-order; for $\epsilon=30$, convergence
is nearly first-order.
\label{fig:isoSpheresConv}}
\end{figure}

Figure \ref{fig:isoSpheresConv} shows frequency convergence for
a 3D cubic lattice of isotropic spheres ($r/a=0.37$)
with $\epsilon=15$ and
$\epsilon=30$.
Although the computational requirements prevent
exploration over the wide range of resolutions of the 2D simulations, the
$\epsilon=15$ case shows mostly second-order behavior starting to
transition to first-order, while $\epsilon=30$ shows
nearly first-order behavior.
This is very similar to Fig.~\ref{fig:anisoDiscFreqConv}, considering
the resolutions where they overlap.  There is no reason to suspect that
error convergence in 3D is any different from 2D.

\subsection{Comparison to wc07: 2D and 3D, aniso- and iso-tropic}

In this section we compare the ``new'' method recommended in this paper
(Sec.~\ref{sec:symAcc})
to wc07 (which uses the effective dielectric
of \cite{Johnson:2001}), except we improved wc07 for anisotropic
dielectrics by using the effective dielectric of
\cite{Kottke:2008,Oskooi:2009}.

The main advantage of the new method is that it is always stable;
since wc07 is usually stable for contrasts less than $\epsilon=30$,
and can be used practically, we wanted to compare their accuracies.
The new method is generally better than wc07, by a small factor
(2--10), for higher contrasts and resolutions.
We cannot compare them for contrasts at $\epsilon=100$ because
the old method becomes unstable.

For low contrast, and low resolution, there is less reason
to choose one algorithm over the other---for some modes one is better,
for other modes the other is better.  However, even for $\epsilon=5$,
the new method can yield more accurate fields, even while the frequencies
are more or less equally accurate.  Of course, there may be
geometries and contrasts for which wc07 is better.

\begin{figure}[tp]
\centering
\includegraphics*[trim=0mm 0mm 0mm 0mm,width = 65mm]{%
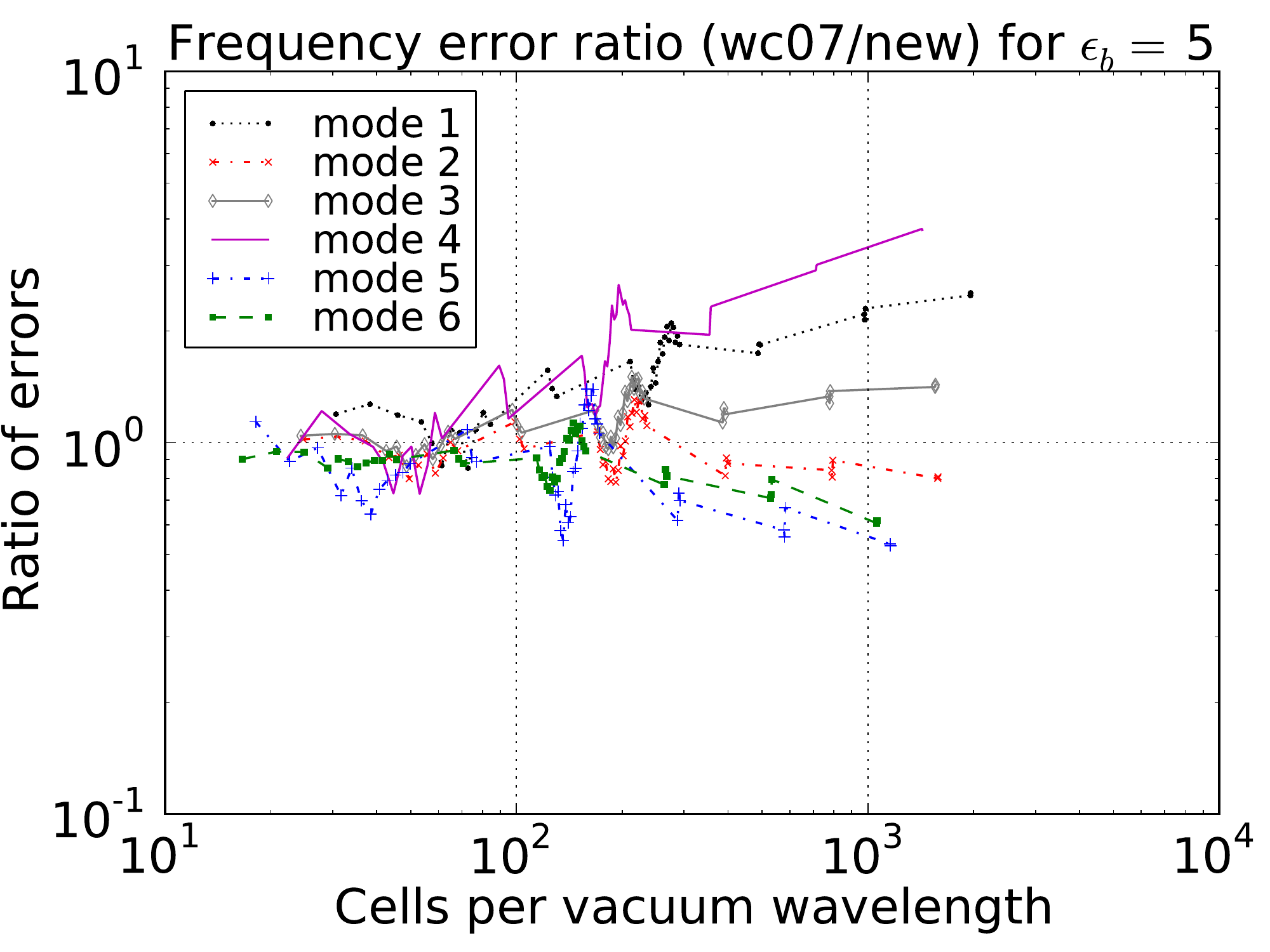}

\includegraphics*[trim=0mm 0mm 0mm 0mm,width = 65mm]{%
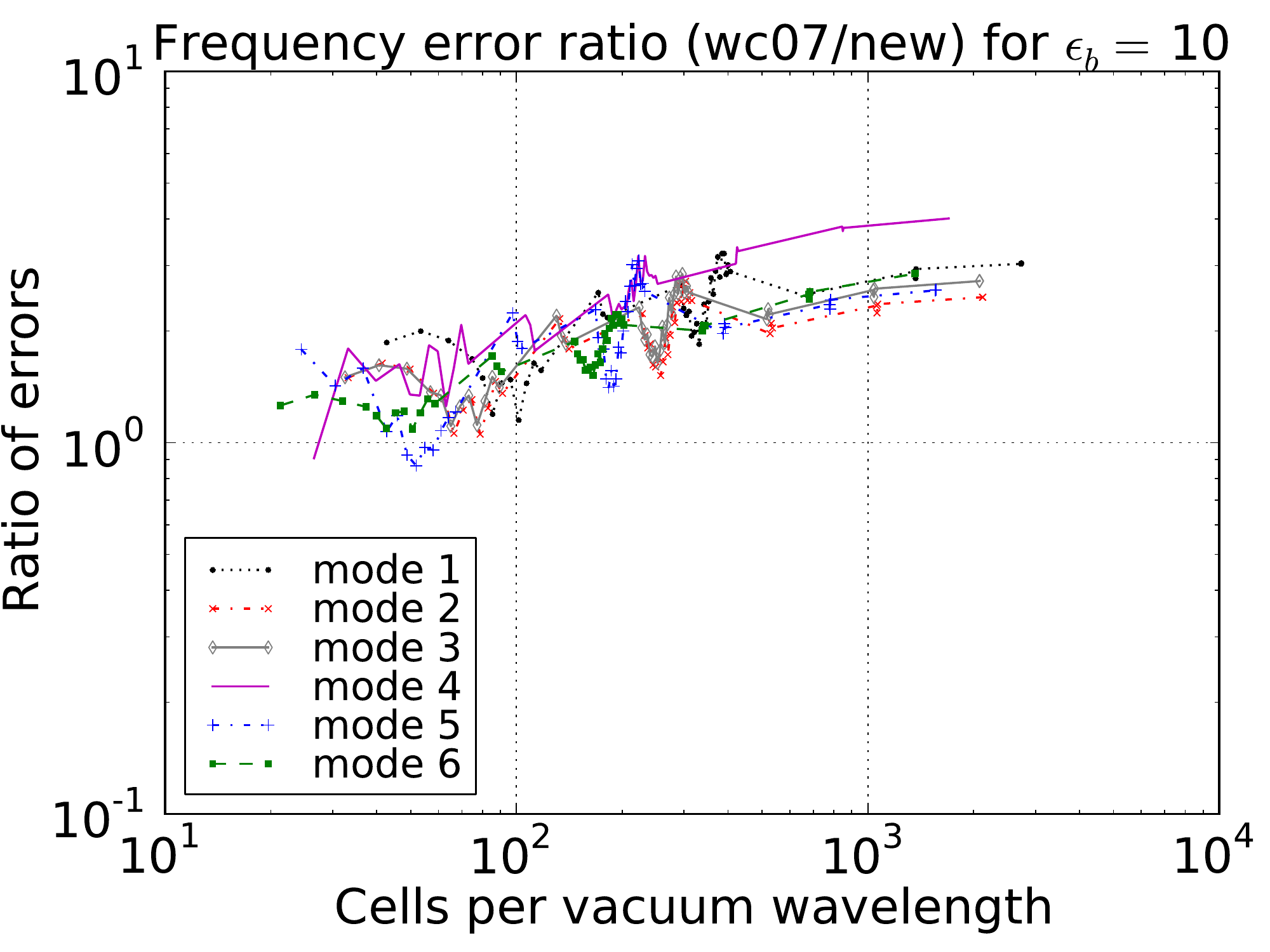}

\includegraphics*[trim=0mm 0mm 0mm 0mm,width = 65mm]{%
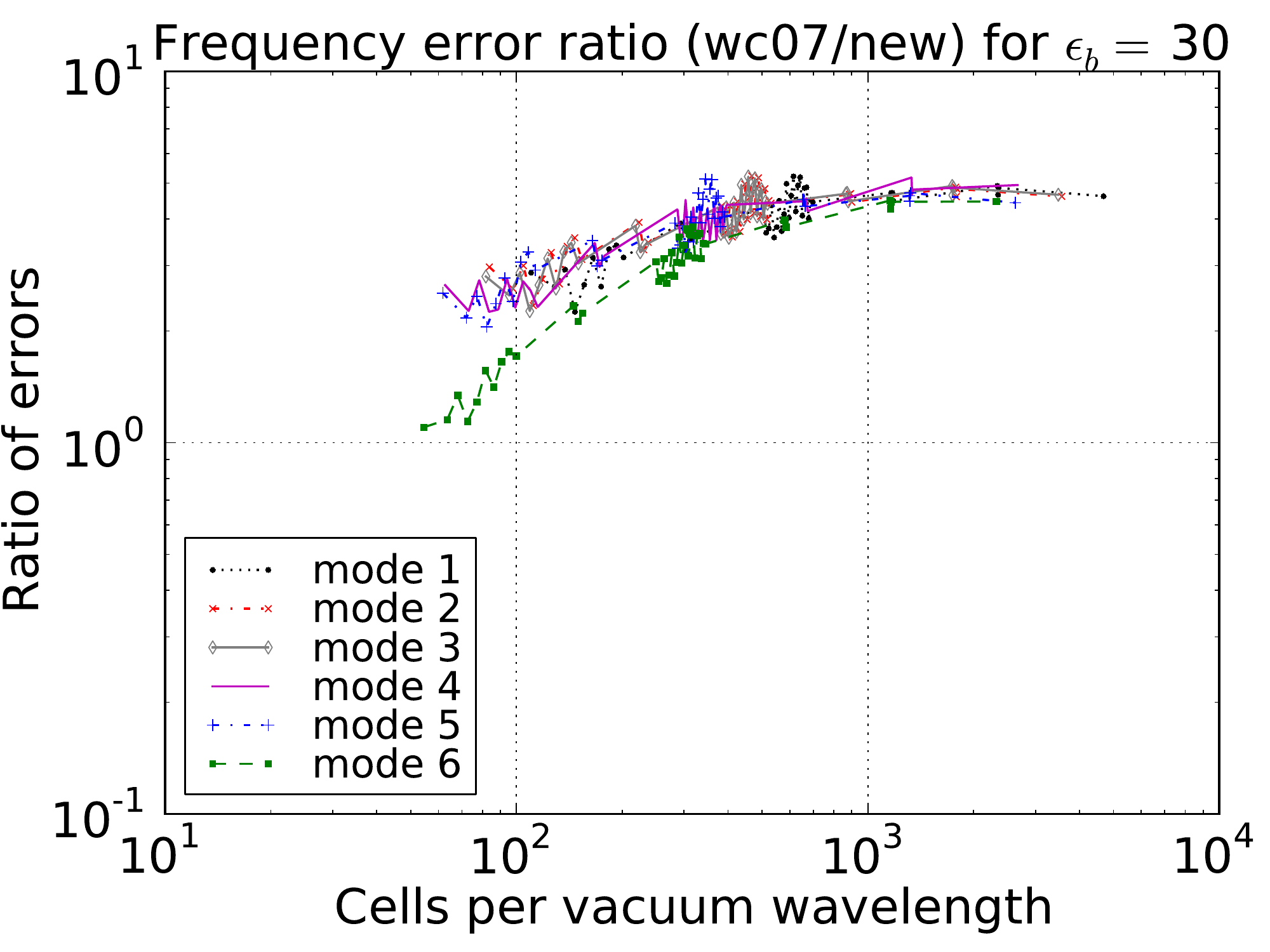}
\caption{(Color online.) For wc07/new, 2D, anisotropic:
the frequency error in wc07 over the error in the new algorithm,
for modes
of a 2D photonic crystal of $r/a=0.37$ anisotropic discs with
dielectric contrast $\epsilon_b$.
For $\epsilon_b \gtrsim 10$, the new method has lower frequency error;
however, Fig.~\ref{fig:newVsOldAnisoDiscSurfOutE} shows that
even for $\epsilon_b = 5$, the new method has lower field error.
\label{fig:newVsOldAnisoDisc}}
\end{figure}

Figure~\ref{fig:newVsOldAnisoDisc} shows the error for wc07
divided by the error for the new method, for the 2D square lattice
of anisotropic discs (as in Sec.~\ref{sec:2dAnisoDiscs}).
Not only is the new method guaranteed stable,
it performs better for medium and high dielectric contrast.

Occasionally, for a problem of low contrast,
one sees the error of wc07 plunge at some low resolution; the
same sometimes happens for other methods as well.  In this case,
it appears that
the frequency error has first- and second-order contributions:
$\alpha \Delta x + \beta \Delta x^2$.  When $\alpha$ and $\beta$
have opposite signs, there is a narrow range of $\Delta x$ for which
the error is nearly zero.  However, this drop in frequency error
is not reflected in the field error.  An example of this can
be seen for $\epsilon=5$ in Figs.~\ref{fig:newVsSimpleAnisoDisc} and
\ref{fig:newVsSimpleAnisoDiscSurfOutE}.

\begin{figure}[tp]
\centering
\includegraphics*[trim=0mm 0mm 0mm 0mm,width = 65mm]{%
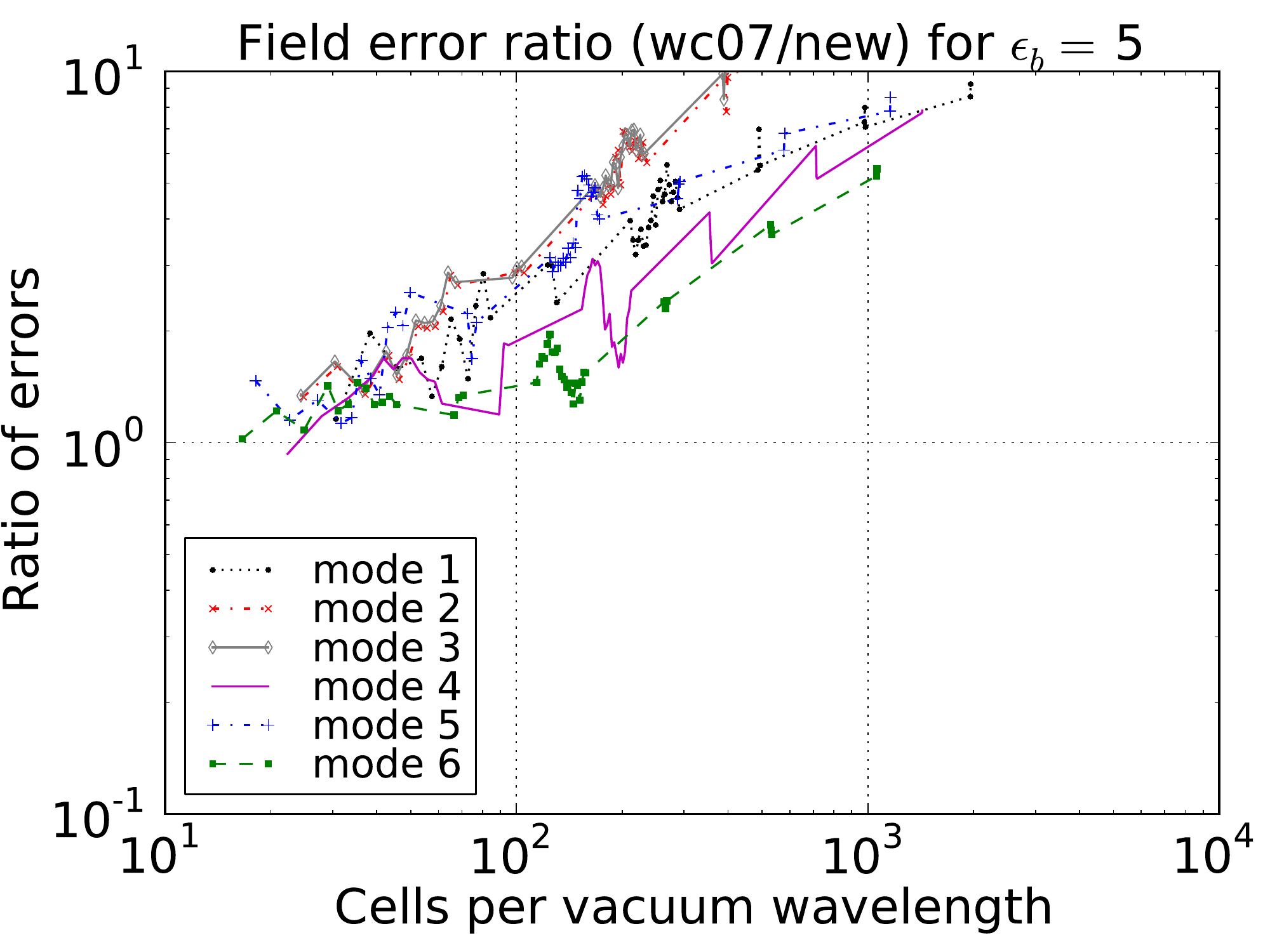}

\includegraphics*[trim=0mm 0mm 0mm 0mm,width = 65mm]{%
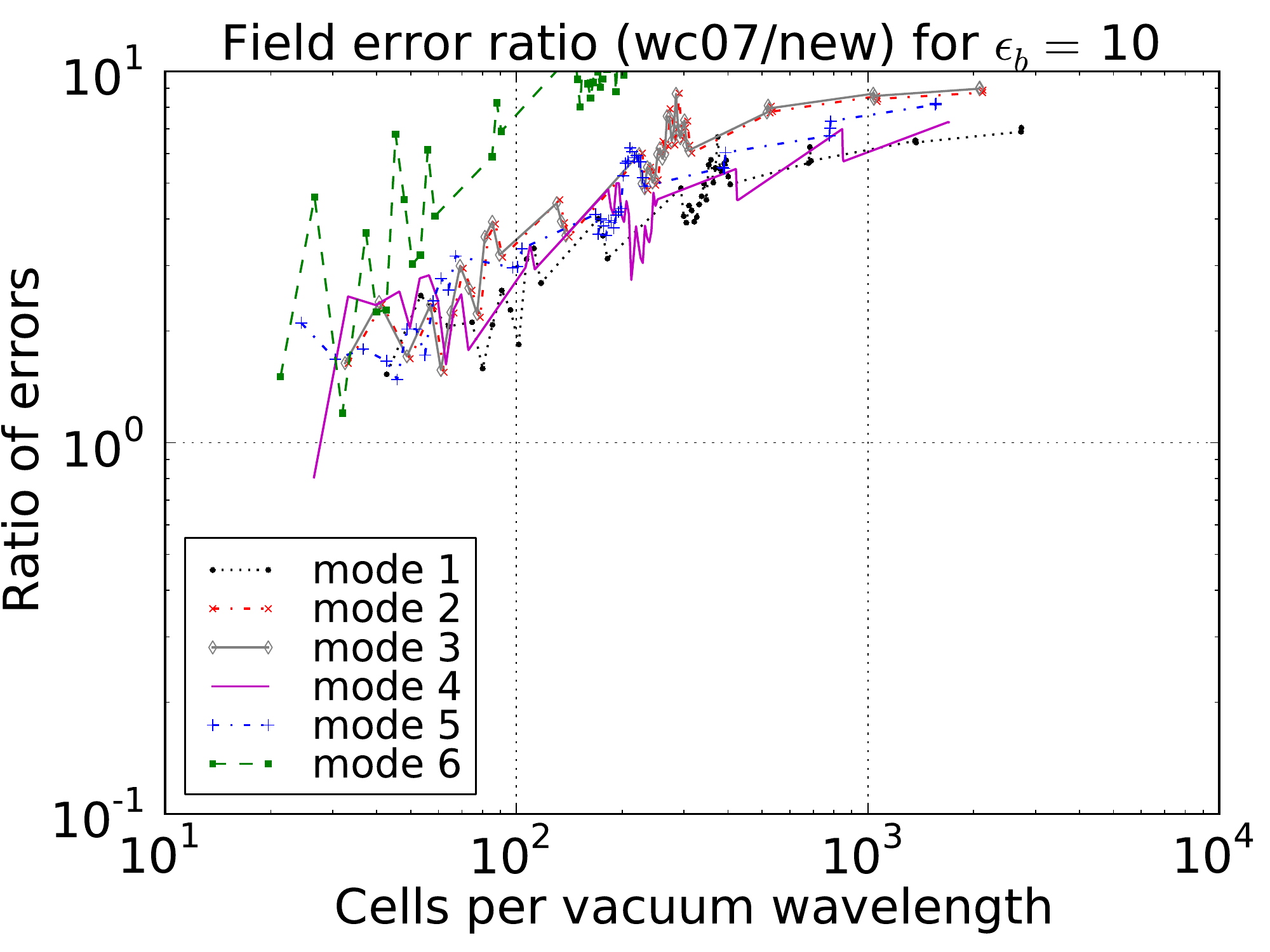}

\includegraphics*[trim=0mm 0mm 0mm 0mm,width = 65mm]{%
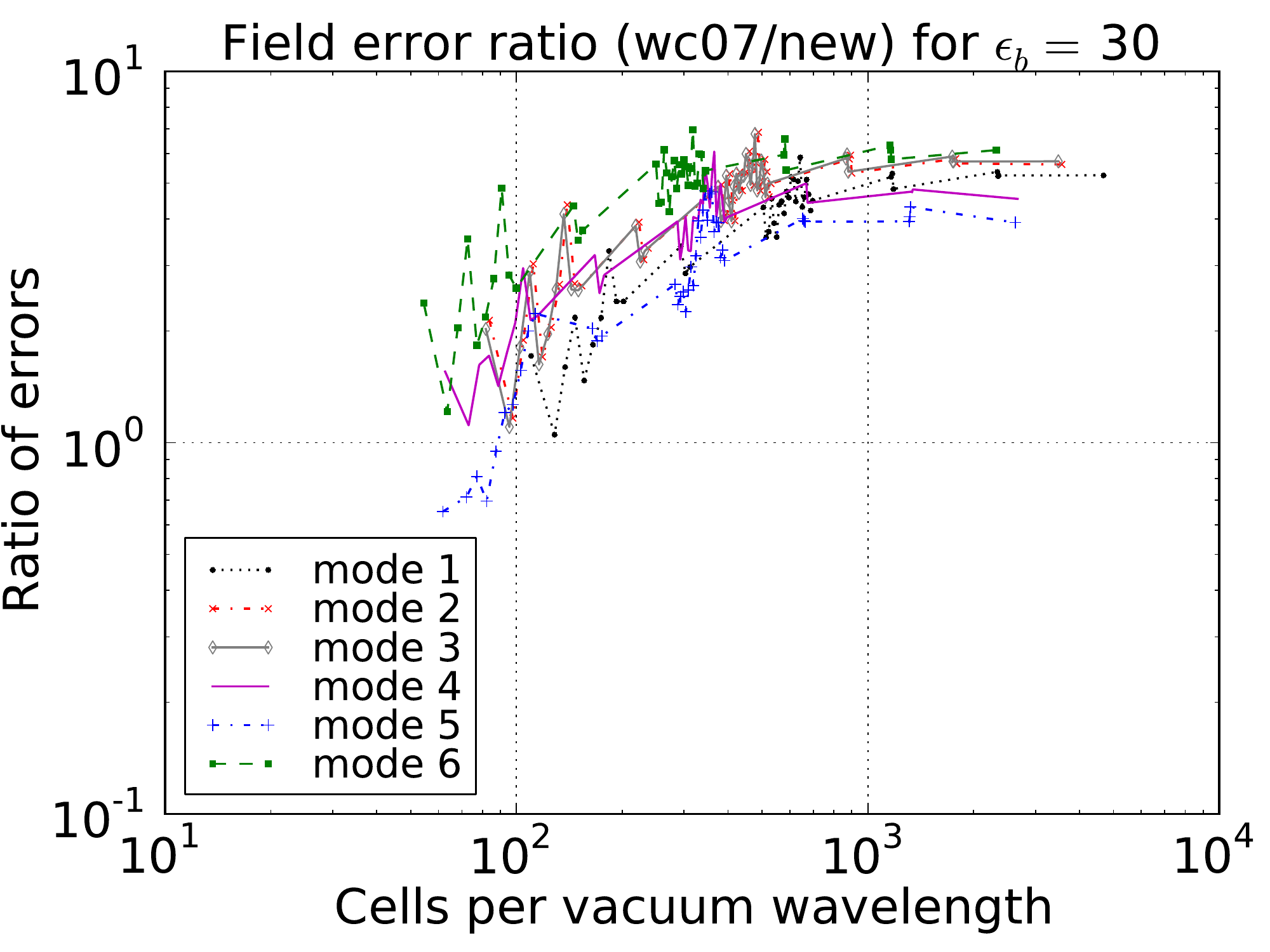}
\caption{(Color online.) For wc07/new, 2D, anisotropic:
the field error in wc07 over the error in the new algorithm,
for fields
on a circle at $r/a = 0.37 + 1/8$, a fixed distance outside the
dielectric disc.  The new method is better in almost every case,
even for $\epsilon_b=5$.  (The simulations are the same as in
Fig.~\ref{fig:newVsOldAnisoDisc}.)
\label{fig:newVsOldAnisoDiscSurfOutE}}
\end{figure}

Figure~\ref{fig:newVsOldAnisoDiscSurfOutE} shows the error of
algorithm wc07 divided by that of the new algorithm for surface
fields on the circle at $r/a = 0.37 + 1/8$.   In nearly all
cases, the wc07 algorithm has higher error.  The error ratio is
seen to increase with the number of cells per vacuum wavelength.

\begin{figure}[tp]
\centering
\includegraphics*[trim=0mm 0mm 0mm 0mm,width = 65mm]{%
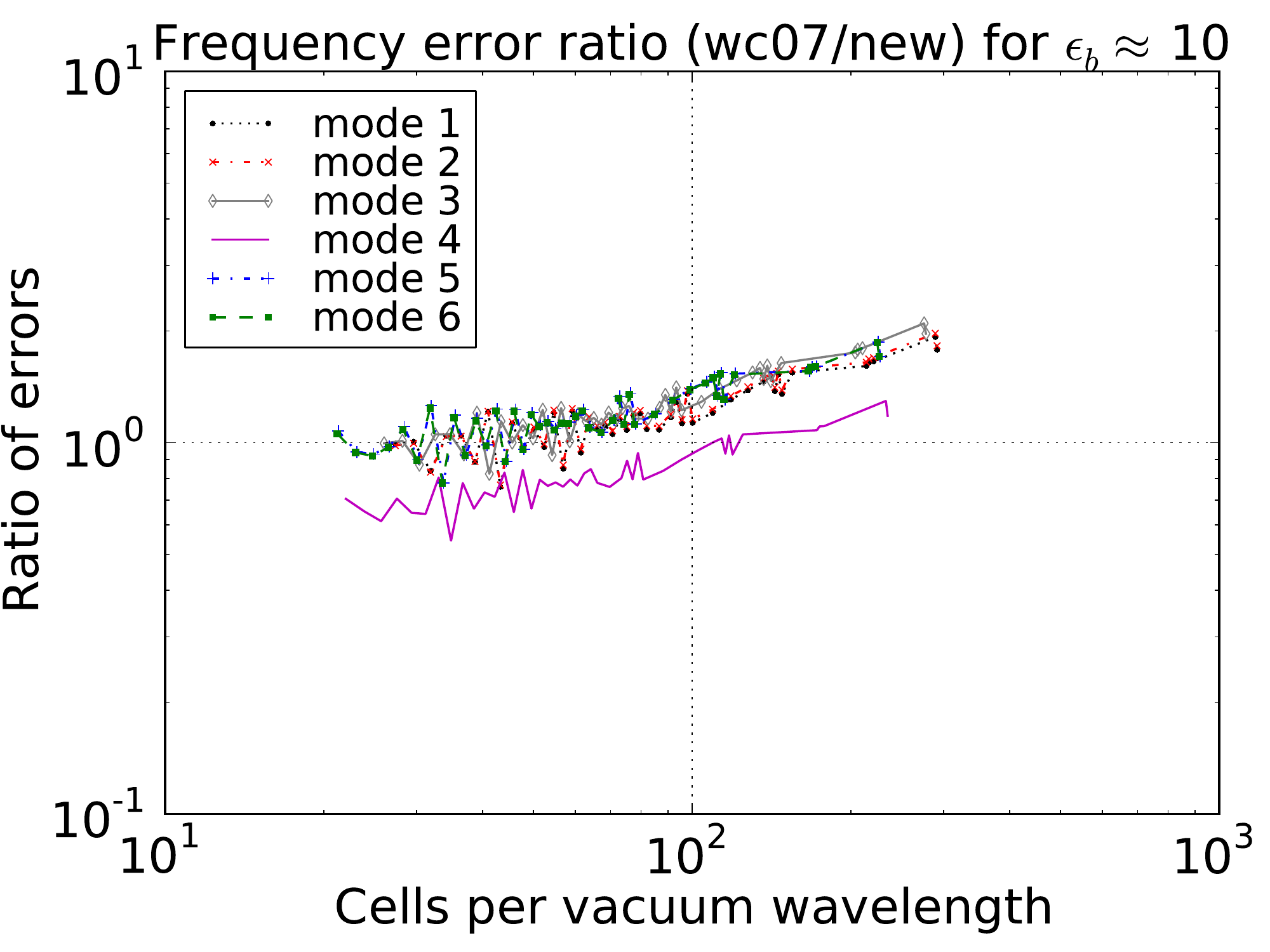}%
\includegraphics*[trim=0mm 0mm 0mm 0mm,width = 65mm]{%
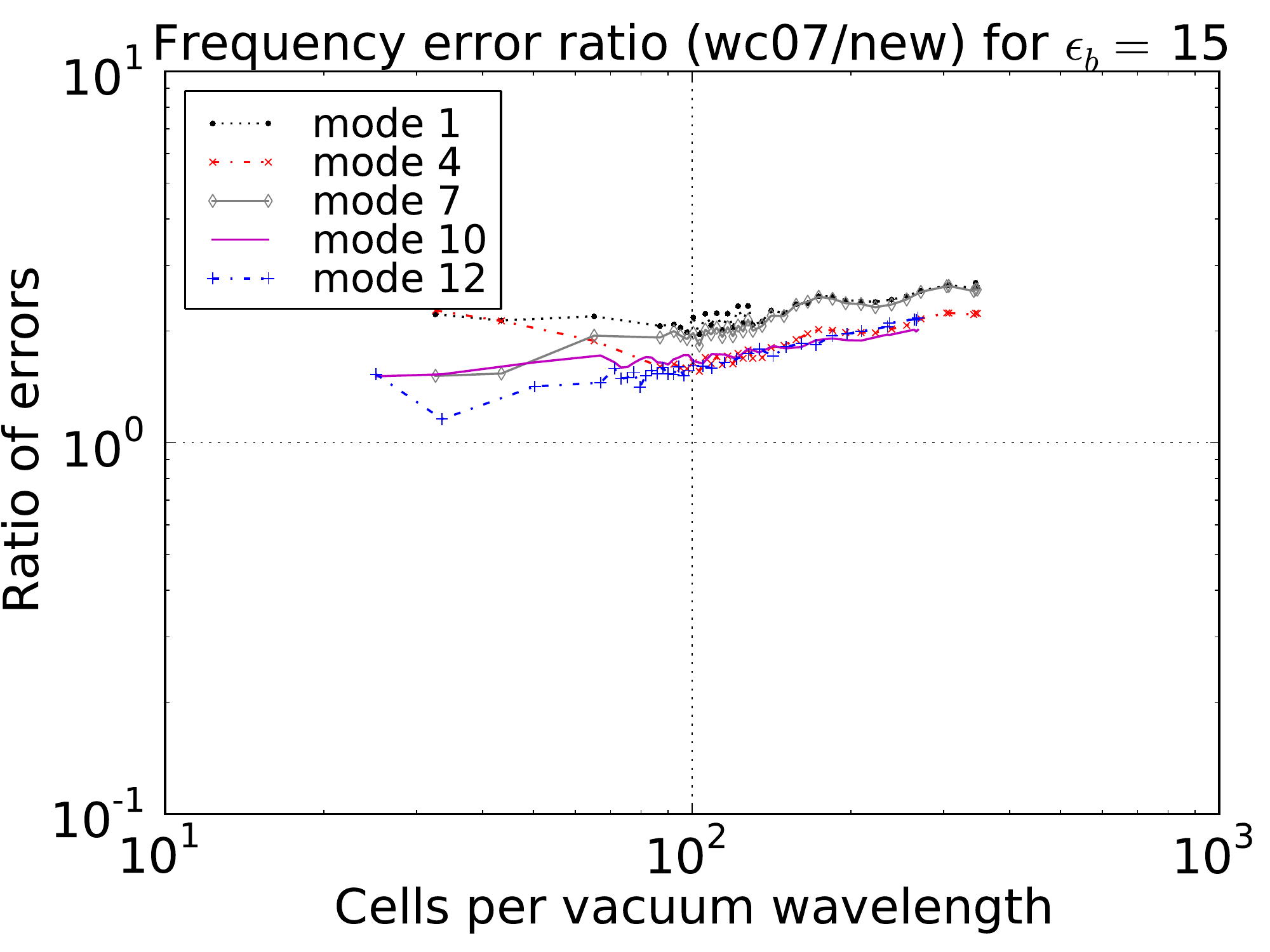}
\caption{(Color online.) For wc07/new, 3D, aniso- and iso-tropic:
the frequency error in wc07 over the error in the new algorithm,
for modes
of a 3D photonic crystal of $r/a=0.37$ spheres of sapphire (left)
and isotropic $\epsilon=15$ (right).
\label{fig:spheres}}
\end{figure}

A similar advantage for the new method is also apparent in 3D,
as shown in Fig.~\ref{fig:spheres},
for sapphire (anisotropic, $\epsilon_b\approx 10$)
and isotropic, $\epsilon_b=15$ spheres.
The dielectric tensor for the sapphire spheres was
\begin{eqnarray}
  \epsilon &=&
    \left( \begin{array}{ccc}
      10.225 & -0.825 & -0.55 \sqrt{3/2} \\
	-0.825 & 10.225 & 0.55 \sqrt{3/2} \\
	-0.55 \sqrt{3/2} & 0.55 \sqrt{3/2} & 9.95 \\
    \end{array} \right)
\end{eqnarray}
which has $\epsilon = 11.6$ along
its $c$-axis, and $9.4$ in the two perpendicular directions; we took the
$c$-axis along $y$, and then rotated the dielectric by $30^\circ$
about the $x$-axis, and then $45^\circ$ about the $z$-axis.

\begin{figure}[tp]
\centering
\includegraphics*[trim=0mm 0mm 0mm 0mm,width = 65mm]{%
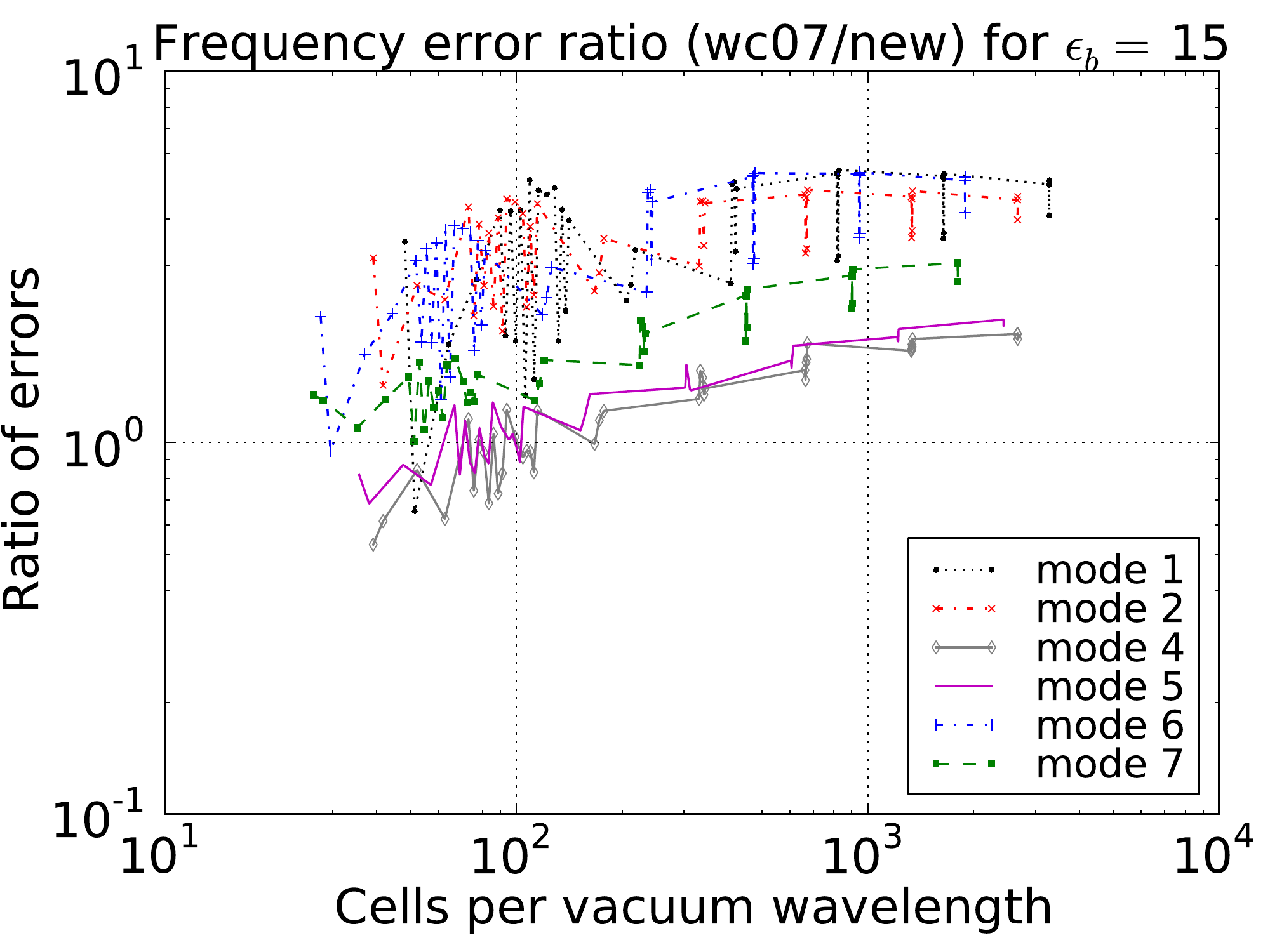}%
\caption{(Color online.) For wc07/new, 2D, isotropic (inverse):
the frequency error in wc07 over the error in the new algorithm,
for modes
of a 2D photonic crystal of $r/a=0.37$ vacuum discs (holes) inside an
isotropic background of $\epsilon=15$.
\label{fig:inverseDiscs}}
\end{figure}

The advantage of the new method remains when the concavity
of the dielectric interface changes: Figure~\ref{fig:inverseDiscs}
shows the advantage of the new method for vacuum discs in a background
of $\epsilon_b$.

\subsection{Comparison to wc07mod: 2D anisotropic discs}

Here we compare the new algorithm to
the wc07mod algorithm (Sec.~\ref{sec:simpleStable}), for
the modes of a square lattice of $r/a=0.37$ anisotropic discs,
as before.  The new
algorithm is better except (surprisingly, given how much worse
wc07mod is at medium contrast) for $\epsilon_b=100$.

\begin{figure}[tp]
\centering
\includegraphics*[trim=0mm 0mm 0mm 0mm,width = 65mm]{%
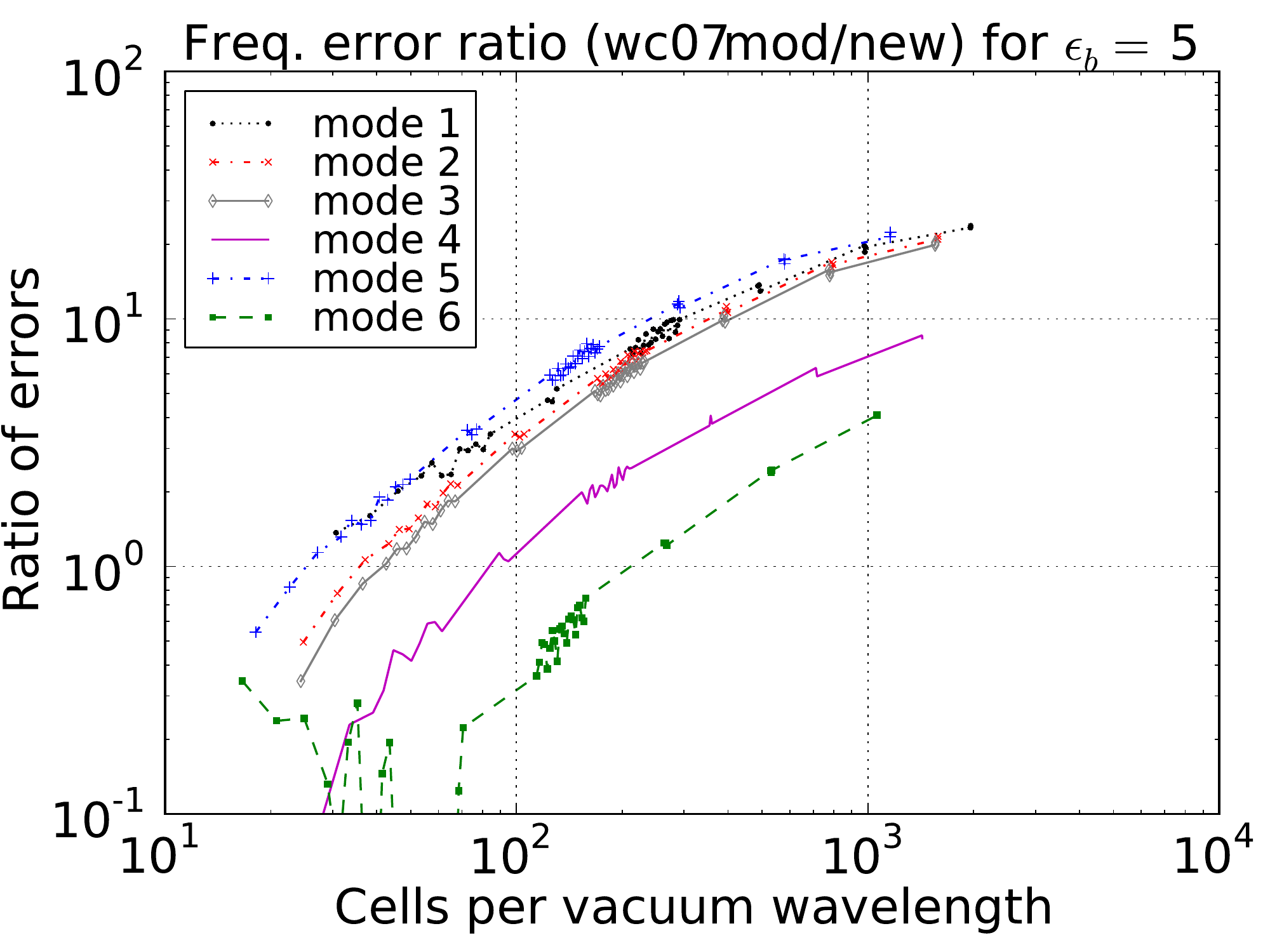}%
\includegraphics*[trim=0mm 0mm 0mm 0mm,width = 65mm]{%
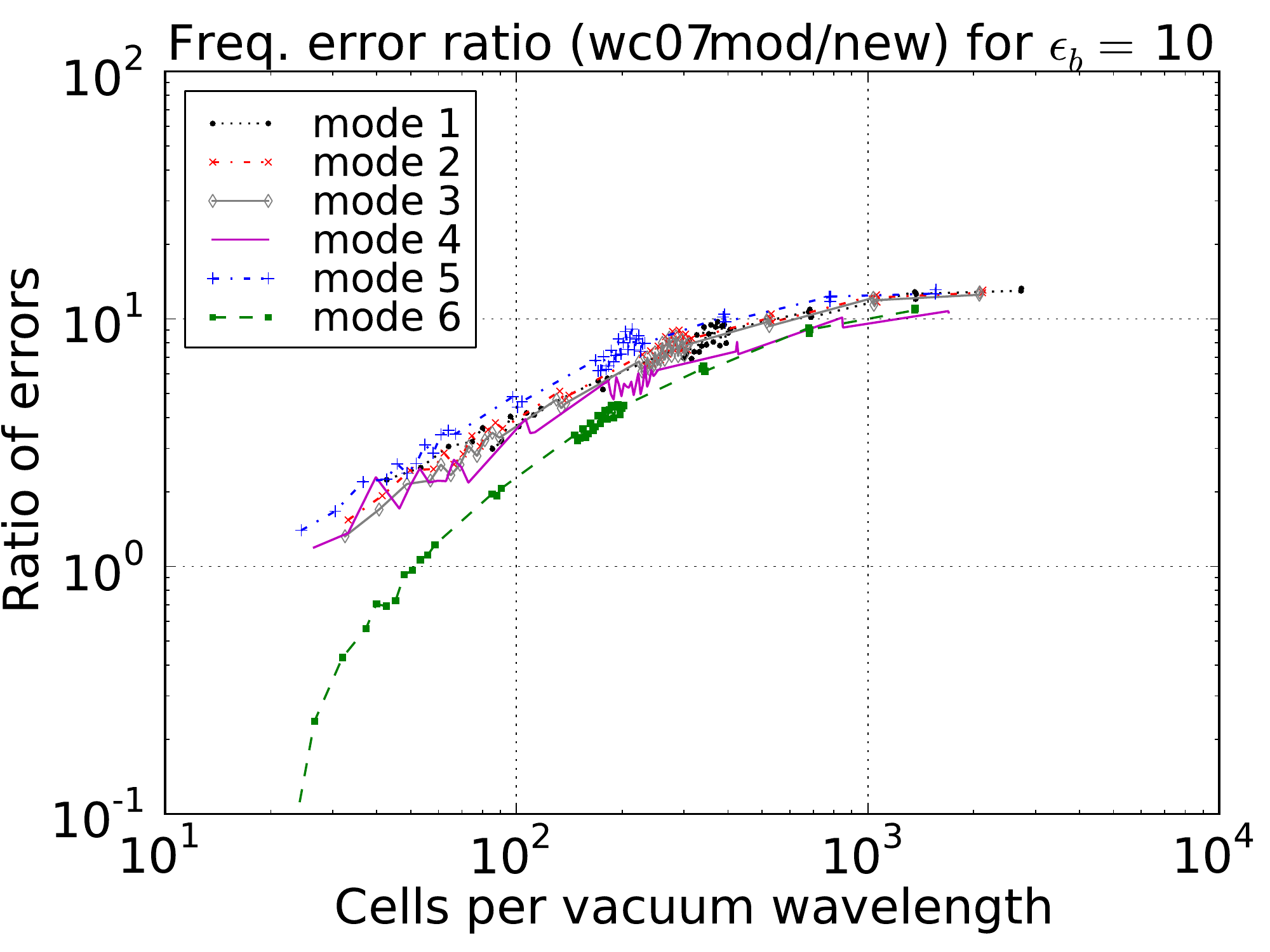}
\includegraphics*[trim=0mm 0mm 0mm 0mm,width = 65mm]{%
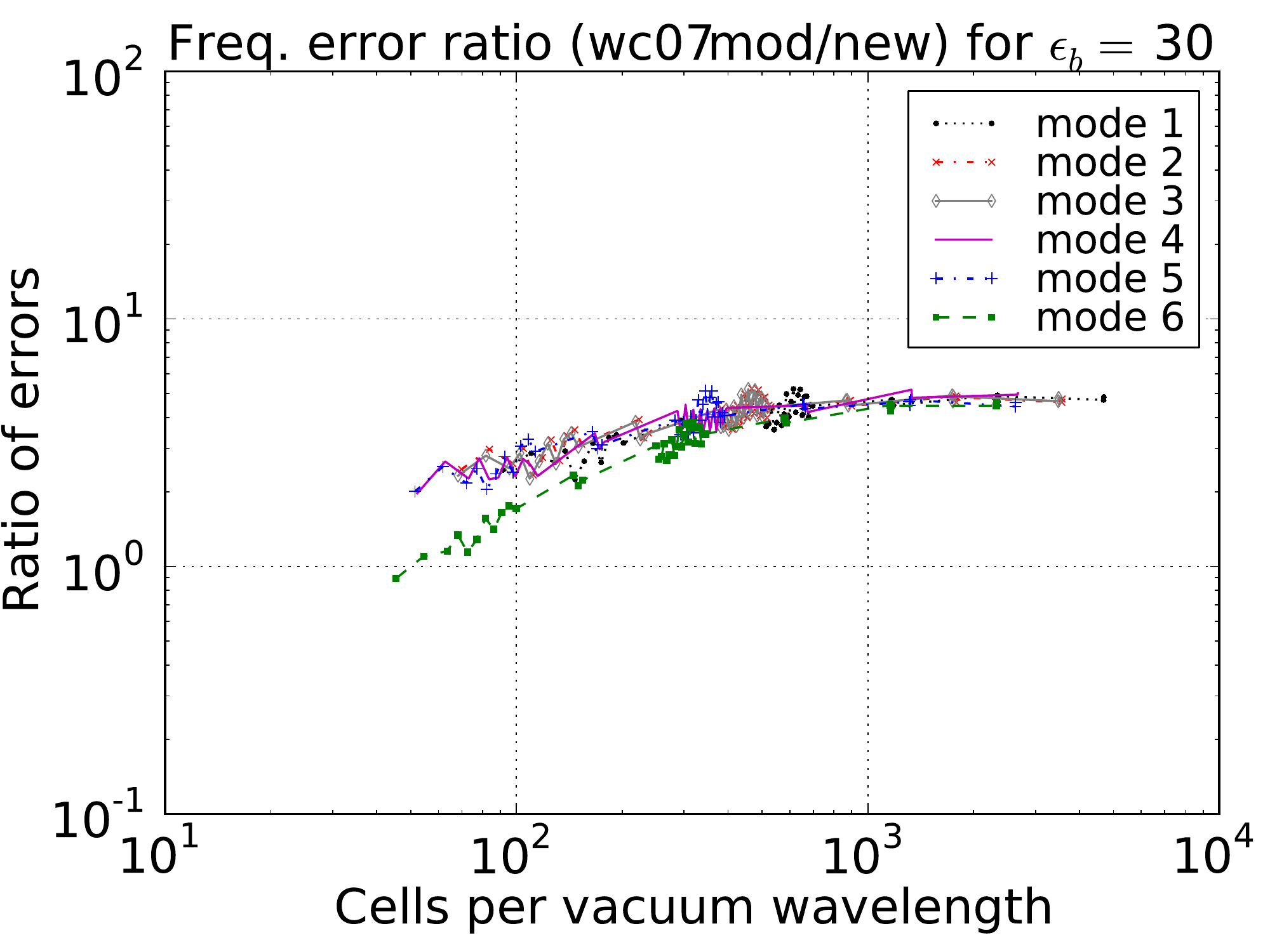}%
\includegraphics*[trim=0mm 0mm 0mm 0mm,width = 65mm]{%
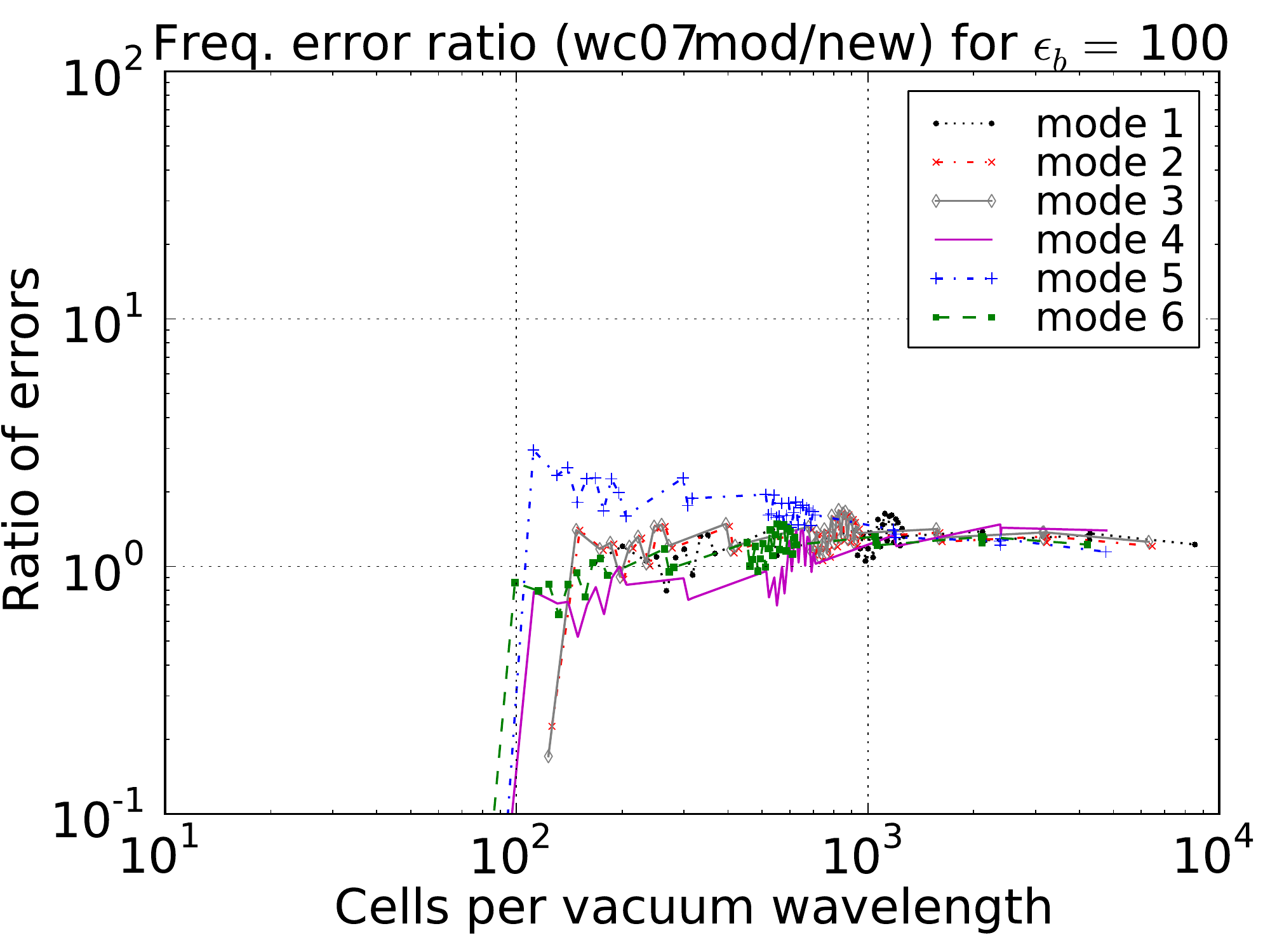}
\caption{(Color online.) For wc07mod/new, 2D, anisotropic:
the frequency error in wc07mod over the error in the new algorithm,
modes of a 2D photonic crystal of $r/a=0.37$ anisotropic discs with
dielectric contrast $\epsilon_b$.
\label{fig:newVsSimpleAnisoDisc}}
\end{figure}

Figure~\ref{fig:newVsSimpleAnisoDisc} shows the ratio in frequency
errors.
For low contrast, we see that the error for wc07mod
plunges within a narrow range of resolutions, due to fortuitous
cancellation of first- and second-order error; we note that one cannot
depend on this cancellation 
(it does not occur for all shapes, nor is the range
of resolutions predictable).  This is illustrated by the convergence
of field, where such fortuitous cancellation is much less likely;
there, wc07mod has no less error.

\begin{figure}[tp]
\centering
\includegraphics*[trim=0mm 0mm 0mm 0mm,width = 65mm]{%
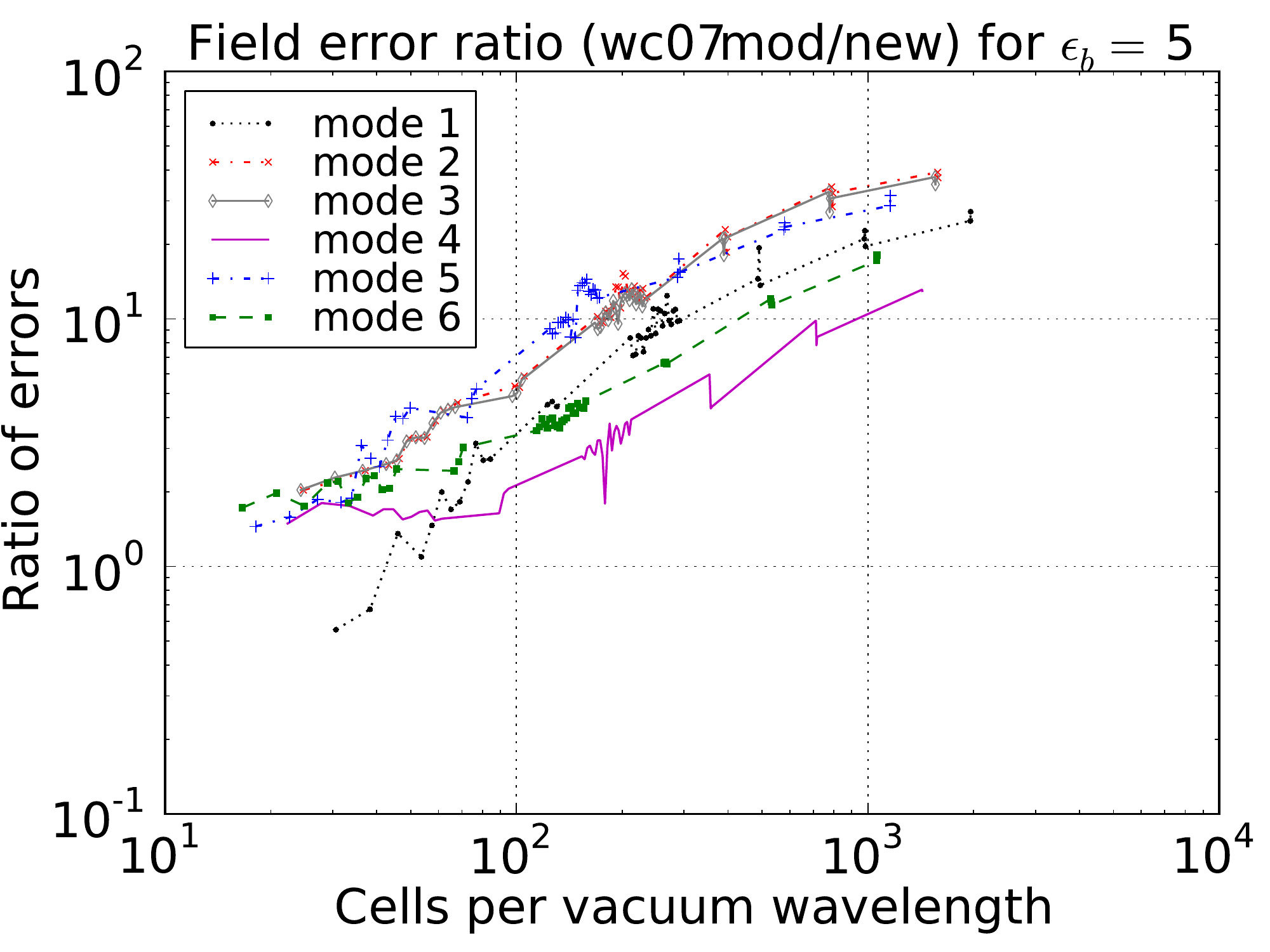}%
\includegraphics*[trim=0mm 0mm 0mm 0mm,width = 65mm]{%
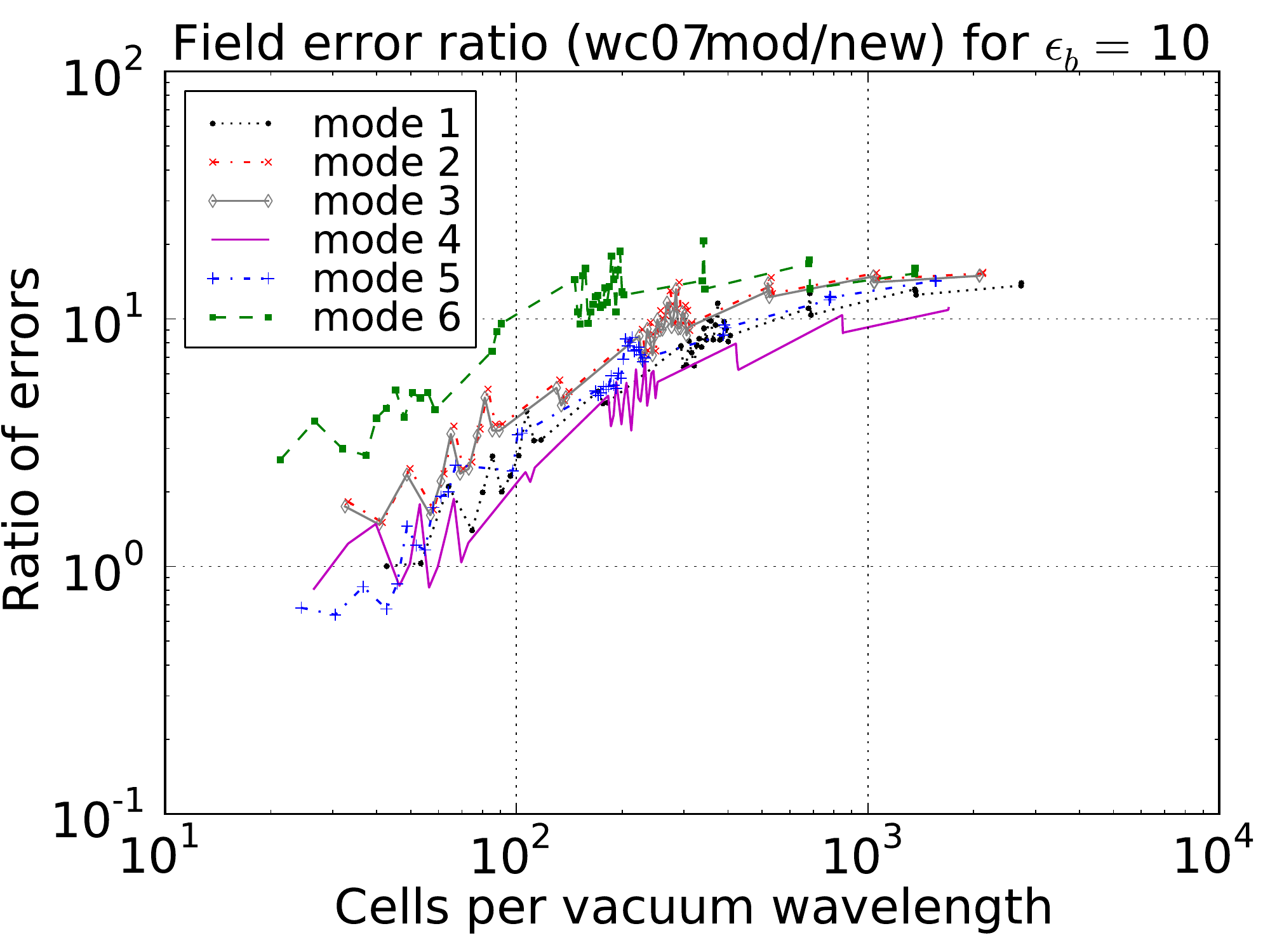}
\includegraphics*[trim=0mm 0mm 0mm 0mm,width = 65mm]{%
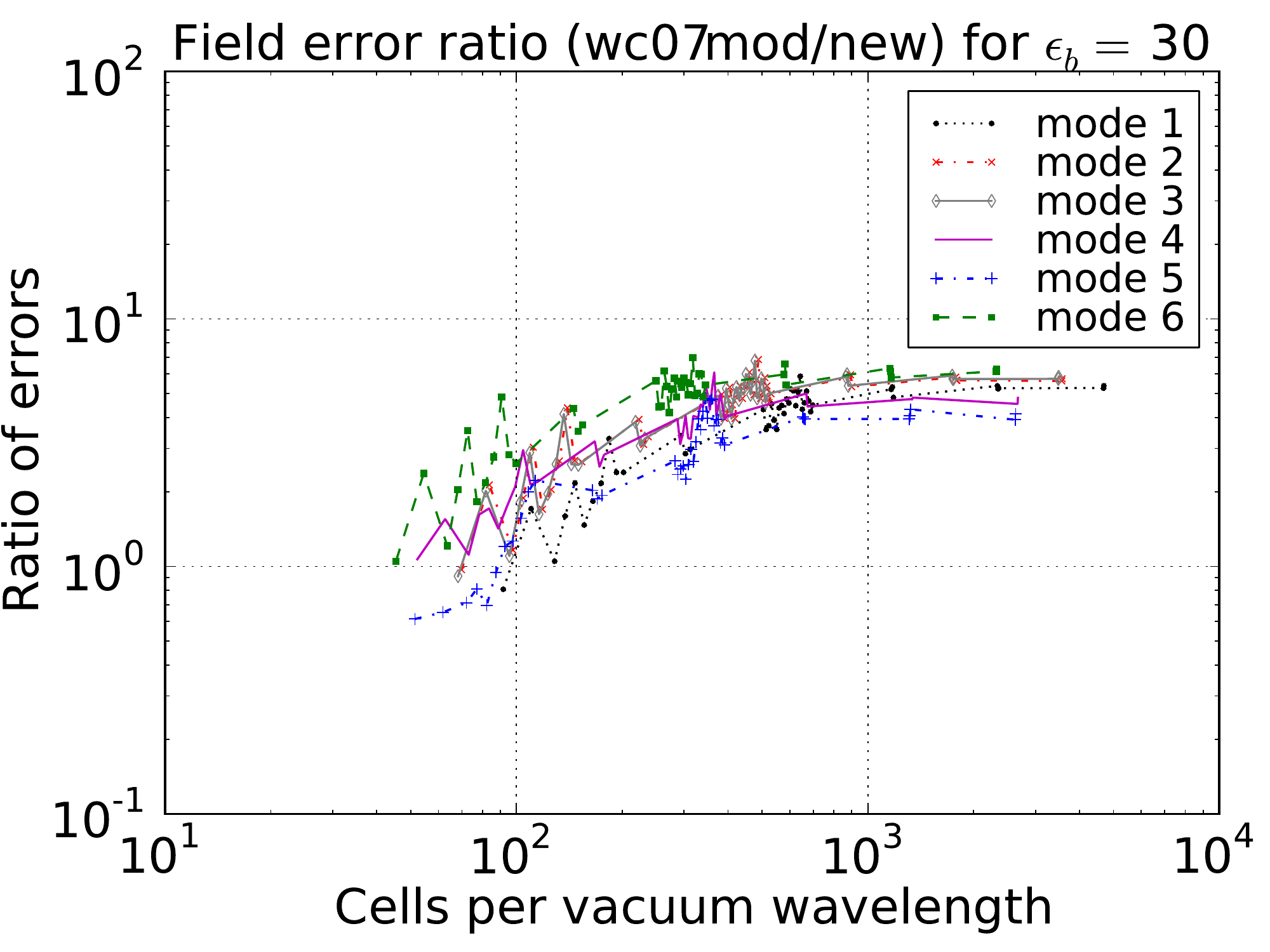}%
\includegraphics*[trim=0mm 0mm 0mm 0mm,width = 65mm]{%
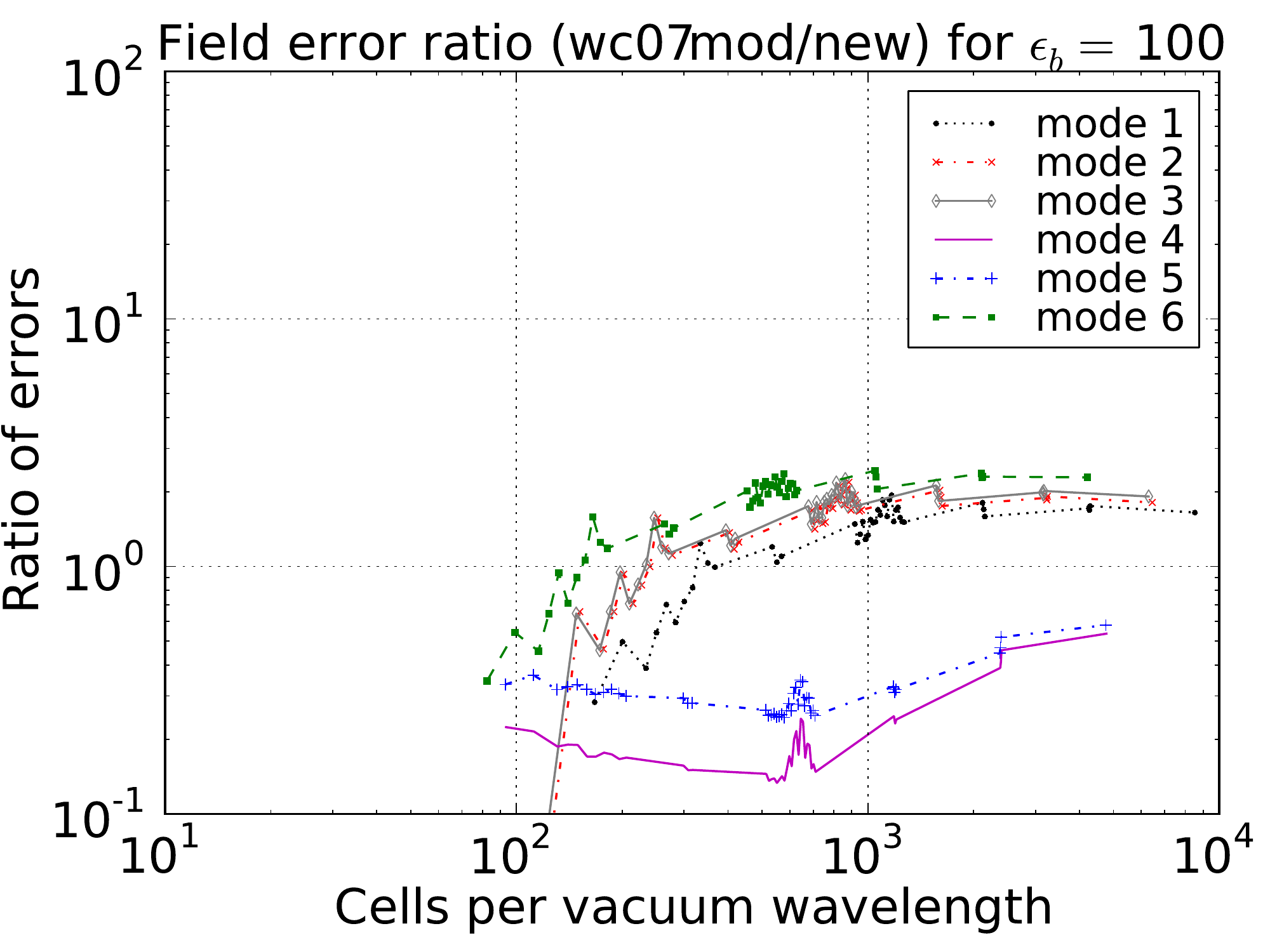}
\caption{(Color online.) For wc07mod/new, 2D, anisotropic: the field error in
wc07mod over the error in the new algorithm, for fields on a
circle at $r/a = 0.37 + 1/8$, a fixed distance outside the
dielectric disc.
\label{fig:newVsSimpleAnisoDiscSurfOutE}}
\end{figure}

Figure~\ref{fig:newVsSimpleAnisoDiscSurfOutE} shows the ratio of
the surface field error of the wc07mod algorithm to that of the
new algorithm on a circle at $r/a = 0.37 + 1/8$.  This shows
that the wc07mod algorithm generally has higher error, and that
the factor by which it is worse increases with the number of
cells per vacuum wavelength.

\section{Summary and Discussion}

We have demonstrated a new FDTD algorithm for simulating electromagnetics
in the presence of sharp dielectric transitions;
the new algorithm is generally as accurate or more accurate than
previous FDTD algorithms, and it is stable
at high dielectric contrast (unlike the algorithm of
\cite{Werner:2007,Oskooi:2009}).

We showed how to create a stable algorithm, given the ability to
form an ``effective'' (or ``average'') $3\times 3$ dielectric tensor
relating any three neighboring components of the $D$-field and
the $E$-field.  As long as each $3\times 3$ tensor is 
symmetric and positive definite (SPD),
the algorithm will be stable.  One can then
try to construct SPD effective dielectric tensors to achieve the
highest accuracy.

We compared three different ways to compute the effective dielectric:
a sometimes unstable method, wc07, described in \cite{Werner:2007,Oskooi:2009};
a small alteration of that method to achieve stability, wc07mod, but
still using the dielectric averaging of \cite{Kottke:2008};
and a new method based on
symmetrizing the asymmetric effective dielectric of \cite{Bauer:2011}.

All these algorithms (except \cite{Bauer:2011}, which is unstable
in the time domain)
have first-order error in the grid-cell-size $\Delta x$;
that is, the error in mode frequencies
and fields at given points decreases ultimately as $O(\Delta x)$.
However, at coarse resolutions the error decreases 
as $O(\Delta x^2)$, before transitioning to $O(\Delta x)$.  For
low dielectric contrast (less than about 10), the transition point
occurs at fairly high resolution, so that for many simulations,
these methods may be practically considered to have second-order error.

By examining the convergence of mode fields, we showed that the fields
at fixed points (fixed as $\Delta x$ varies) converge at the same rate
as the mode frequencies.  However, when we look at the fields at a fixed
cell-distance away from a dielectric interface (so that the points
move closer to the interface as $\Delta x$ decreases), we find that
the error is $O(1)$---it does not decrease with $\Delta x$.

This demonstrates two important points.  First,
the global error is $O(\Delta x)$: the local error in
the bulk material is $O(\Delta x^2)$ (due to centered differencing),
is (eventually) eclipsed by the $O(1)$ local error at the interface
(due to the discontinuity in fields).  However, the interface
cuts a fraction of cells scaling as $O(\Delta x)$, so the ultimate
global error (e.g., in mode frequency) in $O(\Delta x)$.  Heuristically,
we can write ``$\textrm{Error} = a \Delta x + b \Delta x^2$,'' where
$a$ comes from the interface, and $b$ from the bulk (as well higher
order terms from the interface).  Our results show that $a$ can be
relatively small for low dielectric contrast, and therefore
the error at low resolutions (large $\Delta x$) appears second-order
for a while.  Sometimes $a$ and $b$ may be of opposite signs, in which
case the error may plunge briefly for a small range of $\Delta x$ at
the transition between second- and first-order (but only for frequency
error; field error is not a one-dimensional quantity like frequency,
and so the first- and second-order parts cannot fortuitously cancel).

Although we cannot prove that the symmetric effective dielectrics
used in FDTD
simulations must yield (ultimate) first-order error, the
logic of \cite{Bauer:2011} strongly hints that should be the case.
Reference~\cite{Bauer:2011} achieved second-order error by demanding
local first-order, or $O(\Delta x)$, error at the dielectric
interface---accomplished
by finding an effective dielectric that exactly
maps $\mathbf{D}$ to $\mathbf{E}$ in the limit
of infinite wavelength (and planar interface).
Unfortunately, that effective dielectric is
asymmetric, and unusable in FDTD codes due to the instability it creates.
Equally unfortunate, the
symmetric effective dielectrics do not appear to be able to satisfy this
property of mapping $\mathbf{D}$ to $\mathbf{E}$ exactly in the
infinite wavelength limit.

The $O(1)$ convergence of error at a fixed cell-distance from the
surface could be a problem for finding surface fields with
arbitrary accuracy.  However, the error drops rapidly by several
cells away from the surface, and we believe that it will be possible
to obtain surface fields with arbitrary accuracy by extrapolating
from the fields several cells away from the surface.  Of course,
that extrapolation introduces some error, but we currently believe that
the field error plus the extrapolation error can be made to vanish
as $\Delta x \rightarrow 0$.  However, a much more thorough study will
be needed in the future to determine this issue.

It would be great if we could somehow form a symmetric effective dielectric
with the accuracy of the asymmetric effective dielectric of
\cite{Bauer:2011}.  Indeed, there are some degrees of freedom one could
exploit.  For example, we used a symmetric average of the $\Xi^{\pm\pm\pm}$
matrices is Sec.~\ref{sec:stability} to form $\Xi$.  We could have
have used different non-negative coefficients (not $1/8$) that add up to one,
without affecting stability; however, the accuracy within uniform,
anisotropic dielectric would be degraded.  In principle, one can
vary those coefficients in space; however, varying them can destroy
the symmetry of $\Xi$, so one would need some sort of global solution
to attain accuracy and symmetry (not to mention positive definiteness).
We have made some attempts to do this, without complete success.
There may be a way; if not, it may still be possible to achieve greater
accuracy, though not as high as \cite{Bauer:2011}, while maintaining
stability.

\section{Acknowledgments}

This work was supported by the U.S. Department of Energy grant
DE-FG02-04ER41317.

Most of the simulations described in this work were performed with the
\textsc{Vorpal} framework; we
would like to acknowledge the efforts of the \textsc{Vorpal} team:
D.~Alexander,
K.~Amyx,
E.~Angle,
T.~Austin,
G.~I.~Bell,
D.~L.~Bruhwiler,
E.~Cormier-Michel,
Y.~Choi,
B.~M.~Cowan,
R.~K.~Crockett
D.~A.~Dimitrov,
M.~Durant,
B.~Jamroz,
M.~Koch,
S.~E.~Kruger,
A.~Likhanskii,
M.~C.~Lin,
M.~Loh,
J.~Loverich,
S.~Mahalingam,
P.~J.~Mullowney,
C.~Nieter,
K.~Paul,
I.~Pogorelov,
C.~Roark,
B.~T.~Schwartz,
S.~W.~Sides,
D.~N.~Smithe,
P.~H.~Stoltz,
S.~A.~Veitzer,
D.~J.~Wade-Stein,
N.~Xiang,
C.~D.~Zhou.

\newcommand{\SortNoop}[1]{}

\appendix

\section{The error in the asymmetric second-order method}
\label{sec:secondOrderConvergence}

\begin{figure}[tp]
\centering
\underline{Frequency Error}
\includegraphics*[trim=0mm 0mm 0mm 0mm,width = 65mm]{%
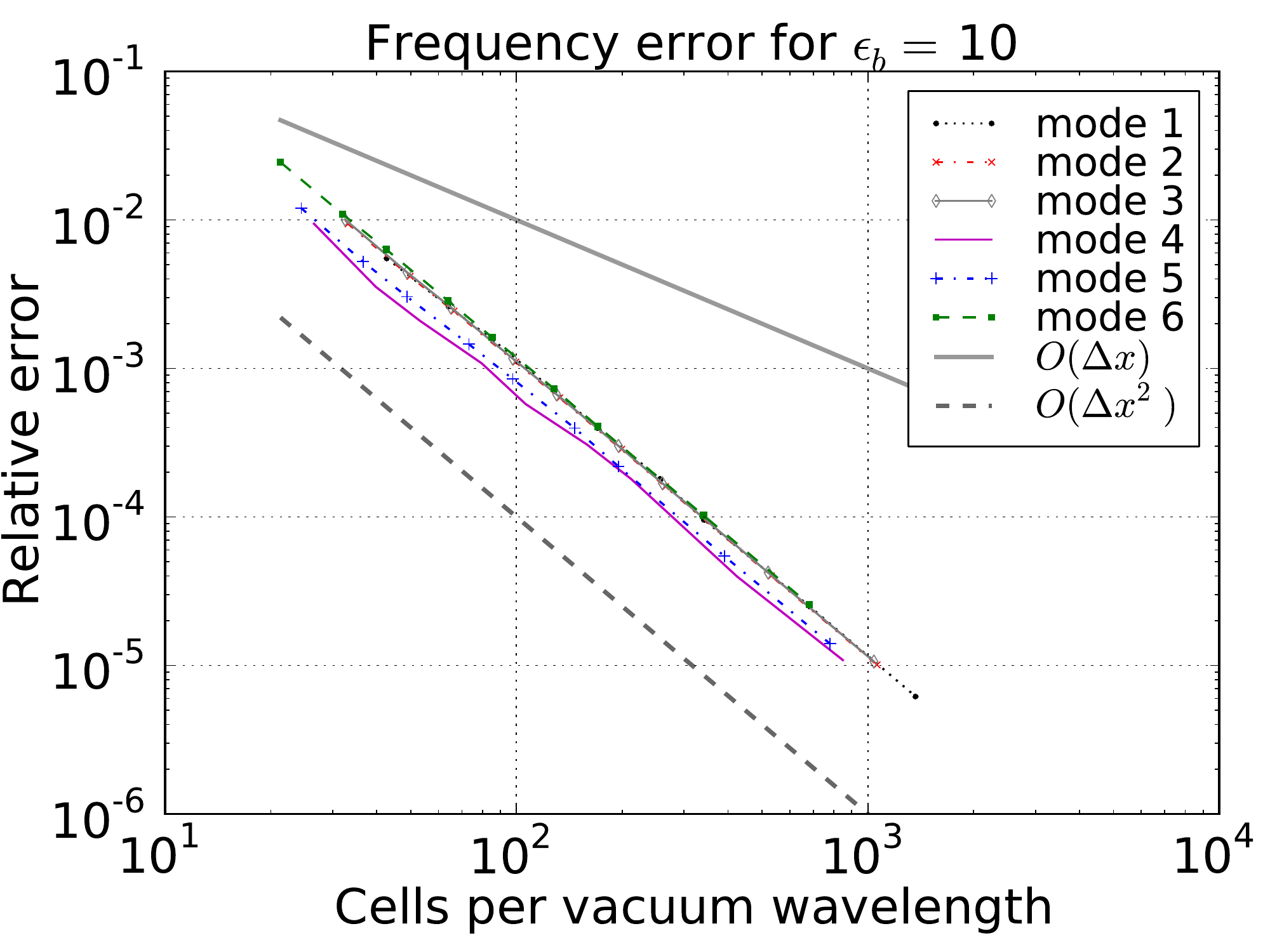}%
\includegraphics*[trim=0mm 0mm 0mm 0mm,width = 65mm]{%
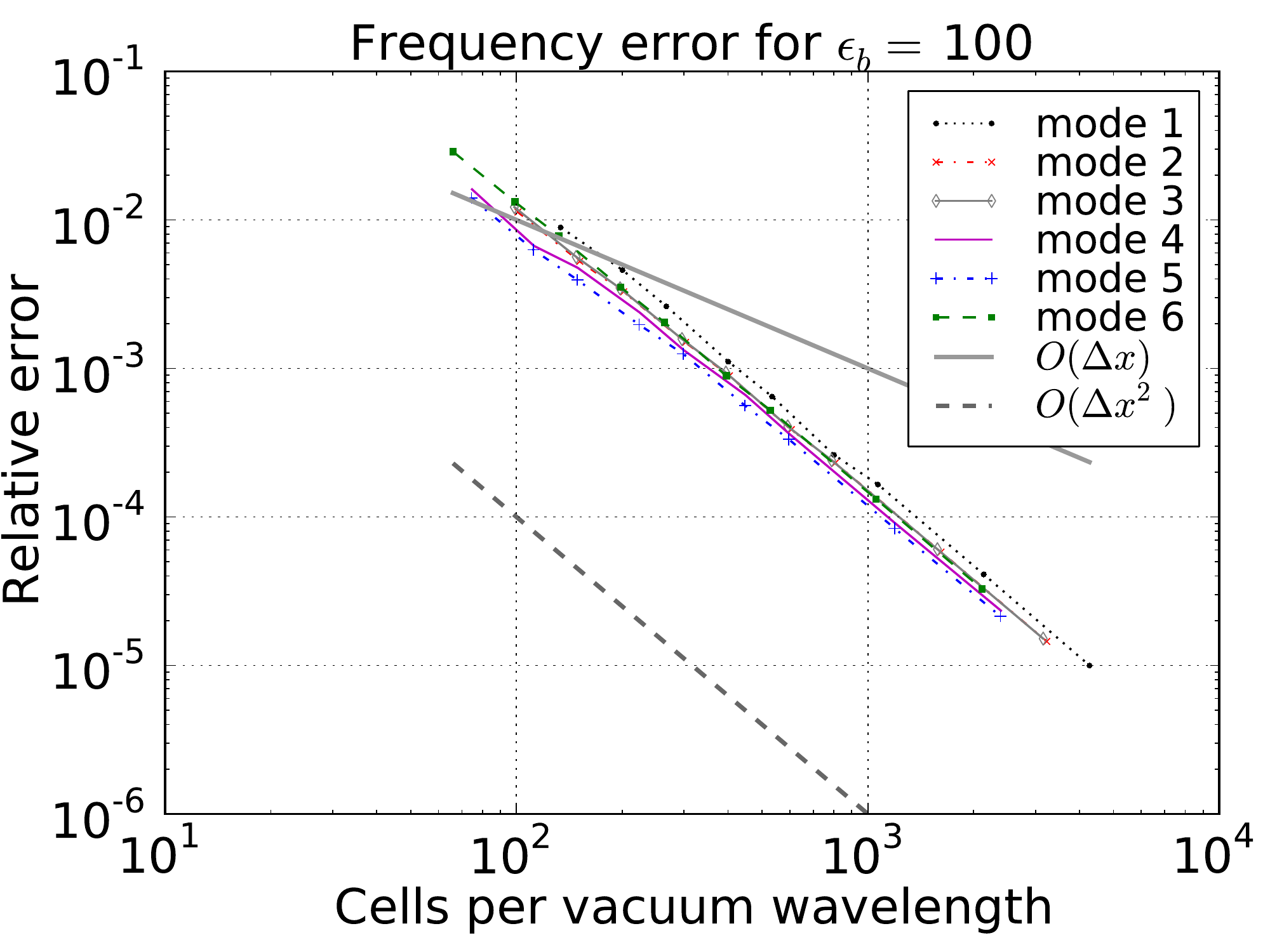}
\underline{Field Error ($a/8$ outside interface)}
\includegraphics*[trim=0mm 0mm 0mm 0mm,width = 65mm]{%
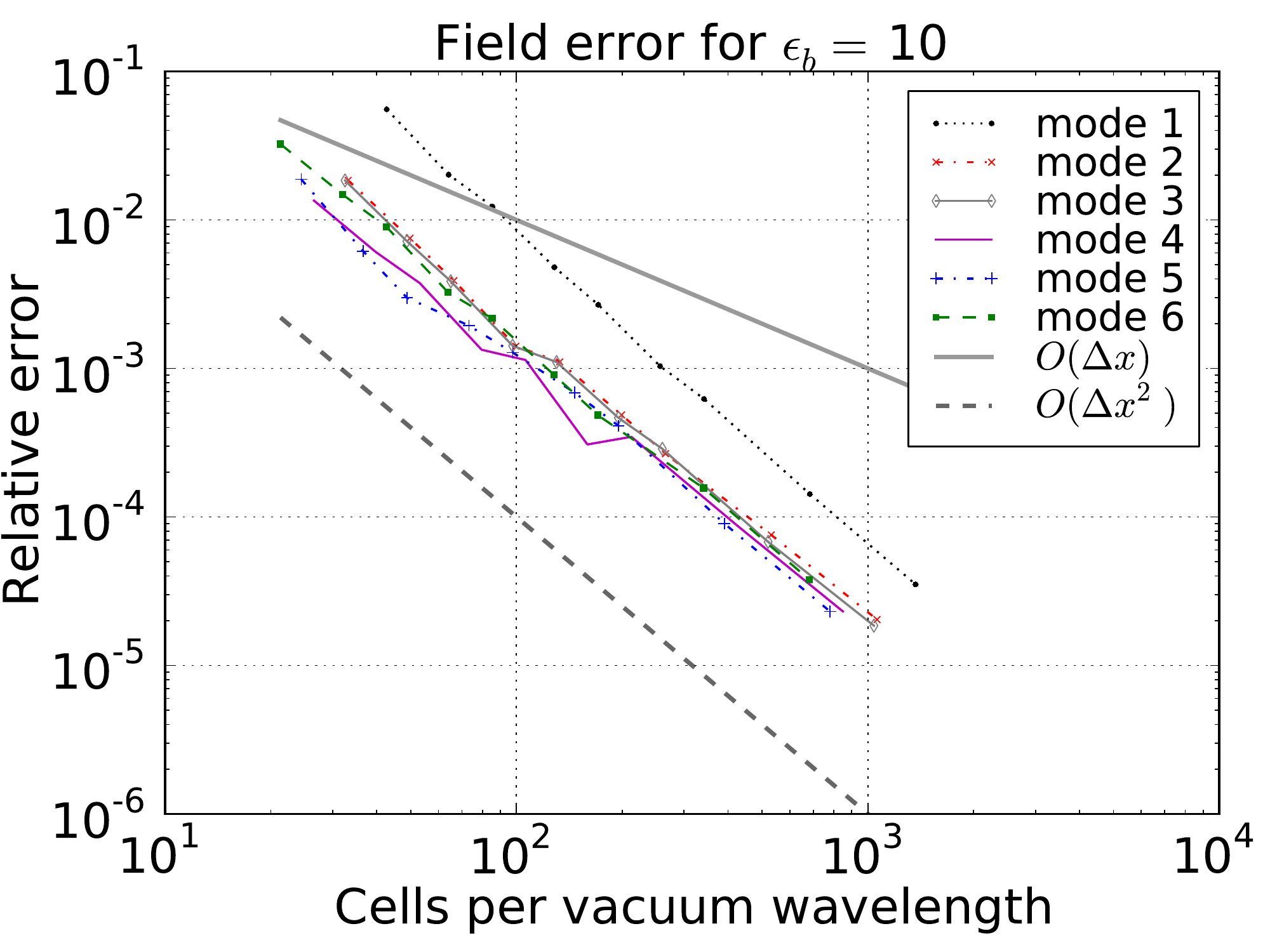}%
\includegraphics*[trim=0mm 0mm 0mm 0mm,width = 65mm]{%
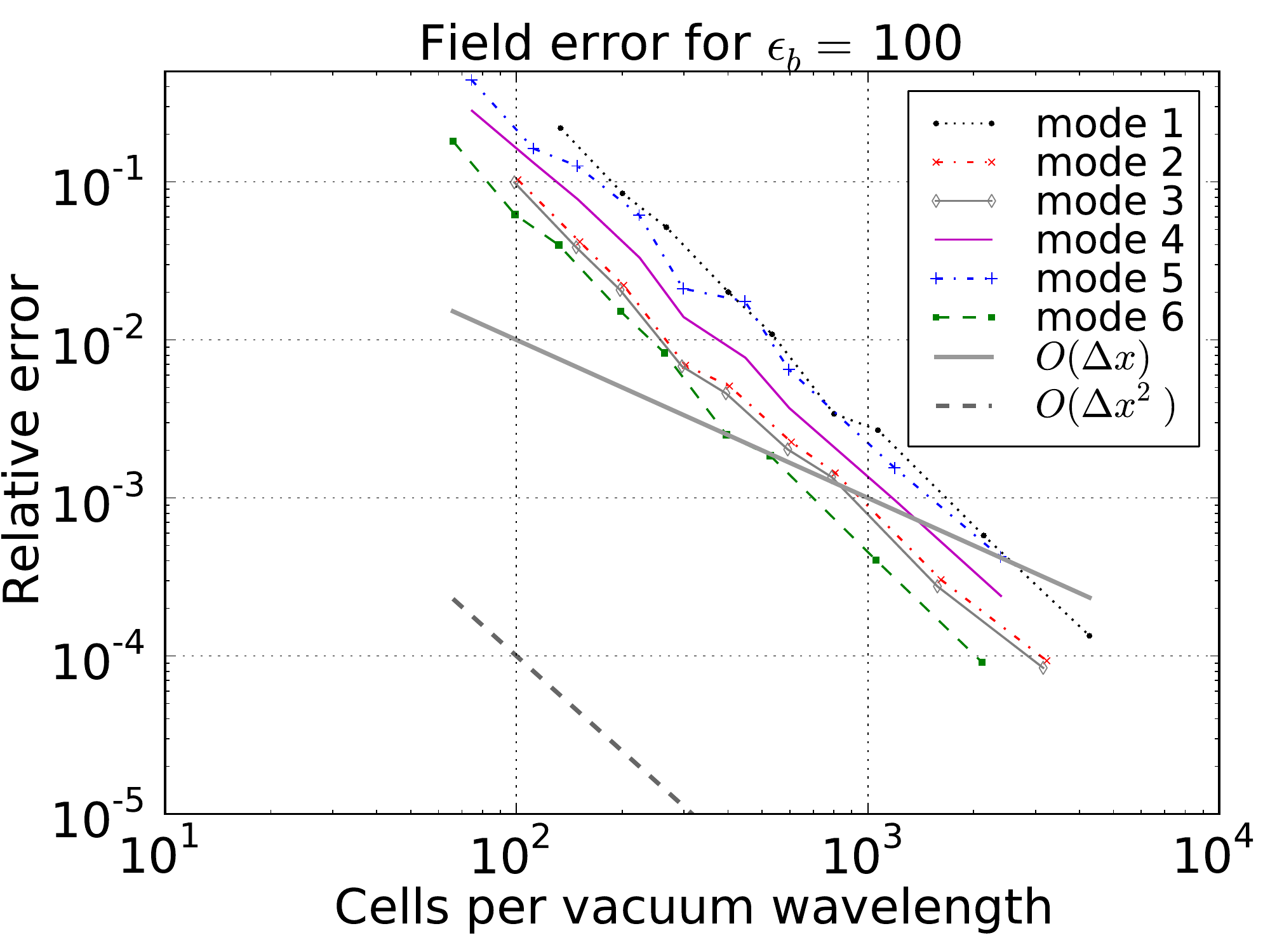}
\underline{Field Error ($3\Delta x$ outside interface)}
\includegraphics*[trim=0mm 0mm 0mm 0mm,width = 65mm]{%
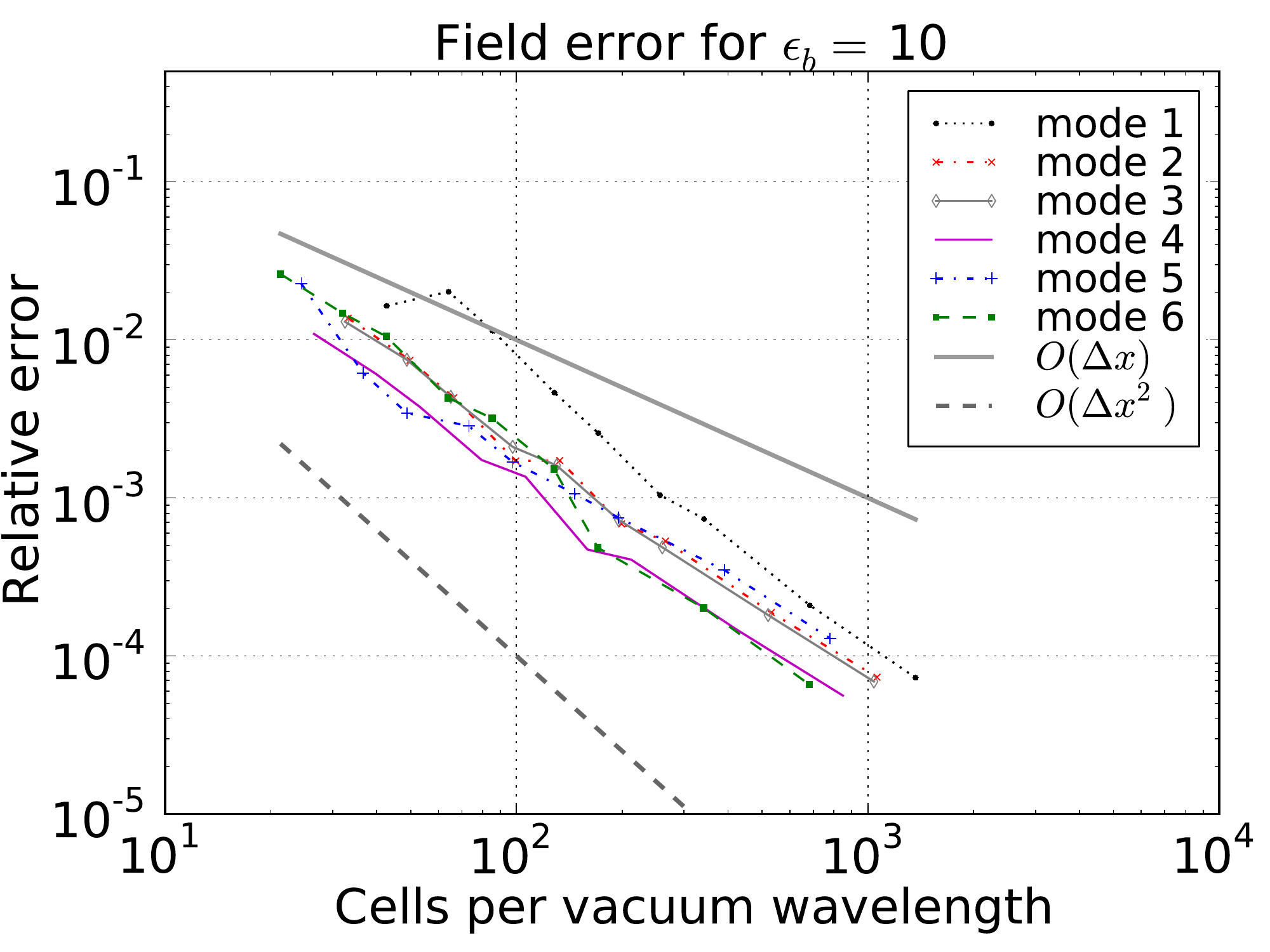}%
\includegraphics*[trim=0mm 0mm 0mm 0mm,width = 65mm]{%
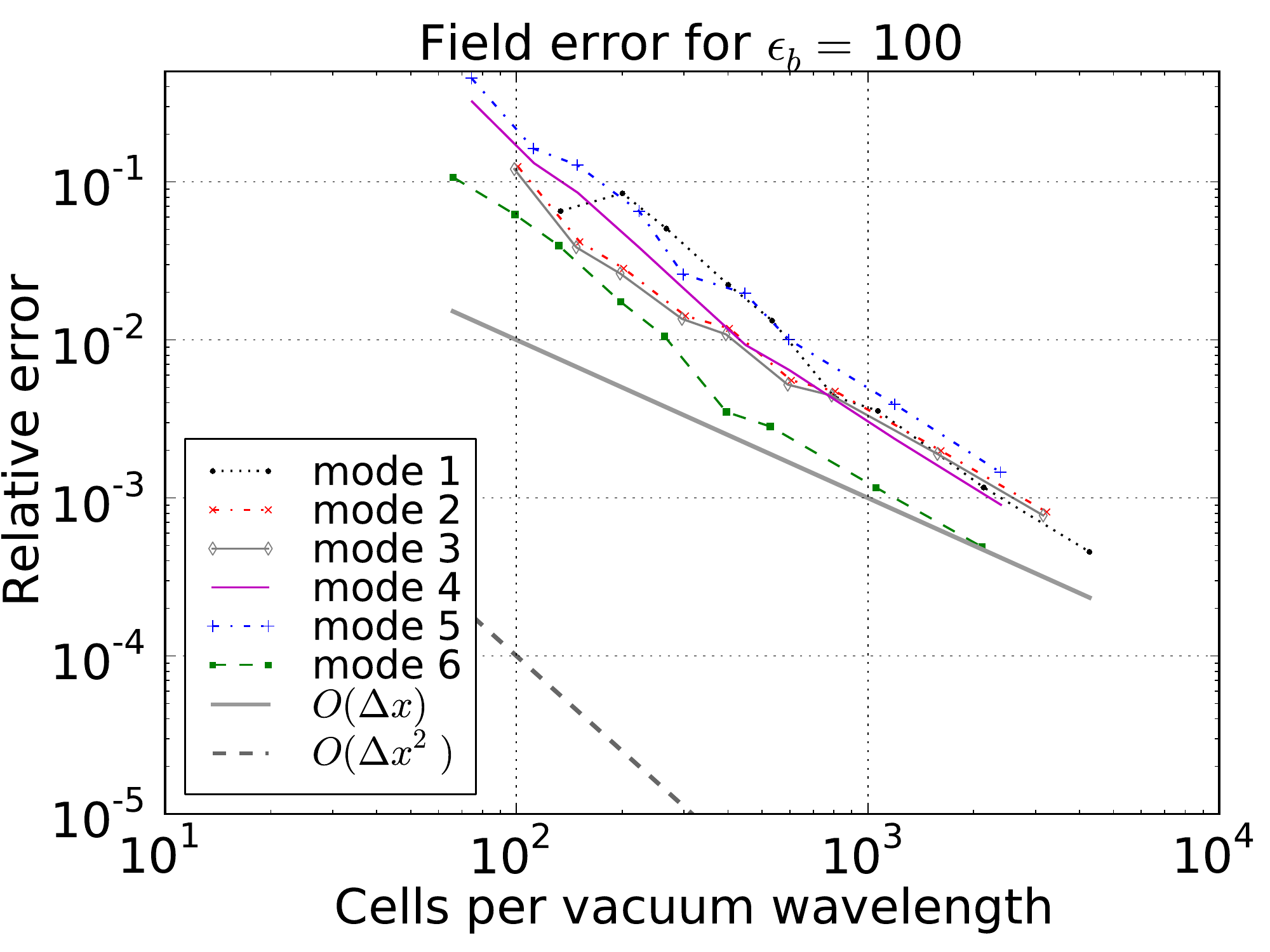}
\caption{(Color online.) For the second-order method, 2D, anisotropic:
relative error vs. resolution for the second-order method
of \cite{Bauer:2011} (which would be unstable in the time-domain),
for the anisotropic 2D photonic crystal with
$\epsilon_b=10$ (left) and $\epsilon_b=100$ (right).
Errors shown are in frequency (top), and
in $\mathbf{E}$ (middle) for a circle at
radius $(0.37 + 1/8)a$, and (bottom) for a circle of
radius $(0.37 + 3\Delta x)a$.  As expected, global errors (top and middle)
are
$O(\Delta x^2)$, but local field errors at the interface are
ultimately $O(\Delta x)$ (as clearly seen in $\epsilon_b=100$).
\label{fig:secondOrderConvergence}}
\end{figure}

To demonstrate what second-order convergence looks like, we show
convergence using the results for the 2D anisotropic dielectric
problem for $\epsilon_b=10$ and $\epsilon_b = 100$
using the algorithm of \cite{Bauer:2011}
in Fig.~\ref{fig:secondOrderConvergence}.
Here we compare to the Richardson-extrapolated values from simulations
at resolutions $N_a=512$ and $N_a=1024$ (and therefore, we do not
plot the values for $N_a=1024$).

The code used to produce these graphs is open-sourced at
\url{https://github.com/bauerca/maxwell}.

\section{The effective dielectric tensor of (Kottke, 2008)
is positive semi-definite}
\label{sec:posDefEffDiel}

If two dielectrics, $\varepsilon^1$ and $\varepsilon^2$, coexist in a
cell in which the normal to the interface between dielectrics is
$\hat{\bf n}$, then we compute the effective dielectric as follows,
according to \cite{Kottke:2008}.

Rotating into a coordinate system where the first coordinate is
$\hat{\bf n}$ (the other two directions, perpendicular to $\hat{\bf n}$,
are labeled 2 and 3),
we calculate $\tau^1$ and $\tau^2$:
\begin{equation}
  \tau^{i} \equiv \tau(\varepsilon^i) \equiv 
    \left( \begin{array}{c@{\quad}c@{\quad}c}
   \displaystyle
    -\frac{1}{\varepsilon^i_{nn}}
    & \displaystyle
    \frac{\varepsilon^i_{n2}}{\varepsilon^i_{nn}}
    & \displaystyle
    \frac{\varepsilon^i_{n3}}{\varepsilon^i_{nn}}
    \\
   \displaystyle
    \frac{\varepsilon^i_{2n}}{\varepsilon^i_{nn}}
    & \displaystyle
    \varepsilon^i_{22} - \frac{\varepsilon^i_{2n} \varepsilon^i_{n2}}{\varepsilon^i_{nn}}
    & \displaystyle
    \varepsilon^i_{23} - \frac{\varepsilon^i_{2n} \varepsilon^i_{n3}}{\varepsilon^i_{nn}}
    \phantom{\frac{{|^|}^|}{{|_|}_|}} 
    \\
   \displaystyle
    \frac{\varepsilon^i_{3n}}{\varepsilon^i_{nn}}
    & \displaystyle
    \varepsilon^i_{32} - \frac{\varepsilon^i_{3n} \varepsilon^i_{n2}}{\varepsilon^i_{nn}}
    & \displaystyle
    \varepsilon^i_{33} - \frac{\varepsilon^i_{3n} \varepsilon^i_{n3}}{\varepsilon^i_{nn}}
    \\
    \end{array} \right)
.\end{equation}
According to \cite{Kottke:2008}, $\tau$ (not, e.g., $\epsilon$) is the 
quantity that should be volume-averaged.
We perform a simple average on $\tau^1$ and $\tau^2$ to get
the effective $\tilde{\tau}$, and then transform back to get the effective
$\tilde{\varepsilon}$.
\begin{equation}
  \varepsilon(\tau ) = \left( \begin{array}{c@{\quad}c@{\quad}c}
   \displaystyle
    -\frac{1}{\tau_{nn}}
    & \displaystyle
    -\frac{\tau_{n2}}{\tau_{nn}}
    & \displaystyle
    -\frac{\tau_{n3}}{\tau_{nn}}
    \\
   \displaystyle
    -\frac{\tau_{2n}}{\tau_{nn}}
    & \displaystyle
    \tau_{22} - \frac{\tau_{2n} \tau_{n2}}{\tau_{nn}}
    & \displaystyle
    \tau_{23} - \frac{\tau_{2n} \tau_{n3}}{\tau_{nn}}
    \phantom{\frac{{|^|}^|}{{|_|}_|}} 
    \\
   \displaystyle
    -\frac{\tau_{3n}}{\tau_{nn}}
    & \displaystyle
    \tau_{32} - \frac{\tau_{3n} \tau_{n2}}{\tau_{nn}}
    & \displaystyle
    \tau_{33} - \frac{\tau_{3n} \tau_{n3}}{\tau_{nn}}
    \\
    \end{array} \right)
.\end{equation}
For example, if the volume fractions for $\varepsilon^1$ and
$\varepsilon^2$ are $V_1$ and $V_2$, respectively ($V_1 + V_2 = 1$),
then the effective dielectric is
\begin{equation}
  \tilde{\varepsilon} = \varepsilon(V_1 \tau^1 + V_2 \tau^2)
.\end{equation}

We can show that, if $\varepsilon^1$ and $\varepsilon^2$ are
symmetric and positive semi-definite (SPSD), then $\tilde{\varepsilon}$
is also SPSD.

We remember that a real, symmetric matrix $A$ is positive semi-definite
when, for all $x$, $x^T A x \geq 0$.  It follows that
the sum of SPSD matrices is again SPSD.

Whenever $A$ is real and SPSD,
there exists a matrix $P$ such that
$A=P^T P$.\footnote{
Since $A$ is symmetric, it has an eigendecomposition $A = V^T D V$;
since $A$ is positive semi-definite, the eigenvalues are non-negative,
and $D$ has a real square root; therefore, $A = (\sqrt{D} V)^T (\sqrt{D} V)$.}
The reverse also holds: if $A=P^T P$, then for all $x$,
$x^T A x = x^T P^T P x = \| Px \|^2 \geq 0$.
From this we can conclude that $Q^T A Q = (PQ)^T (PQ)$
is SPSD for any matrix $Q$.

We will prove that
$\tilde{\varepsilon} = \varepsilon(\tilde{\tau}) =
 \varepsilon(V_1 \tau^1 + V_2 \tau^2)$
is SPSD, by showing that there is an
invertible matrix $\Gamma$ such that
$\Gamma^T \tilde{\varepsilon} \Gamma$ is SPSD.

The matrix $\Gamma$
\begin{equation}
  \Gamma(\varepsilon) = \left( \begin{array}{c@{\quad}c@{\quad}c}
    \displaystyle \frac{1}{\varepsilon_{nn}}
    & \displaystyle
    -\frac{\varepsilon_{n2}}{\varepsilon_{nn}}
    & \displaystyle
    -\frac{\varepsilon_{n3}}{\varepsilon_{nn}}
    \\
   \displaystyle
    0 & 1 & 0
    \\
    0 & 0 & 1 \\
    \end{array} \right)
\end{equation}
is invertible, since it is diagonalizable and none of
its eigenvalues ($1/\epsilon_{nn}$, $1$, and $1$) are zero.
The matrix product
\begin{equation}
  \Gamma^T \varepsilon \Gamma =
  \left( \begin{array}{c@{\quad}c@{\quad}c}
    \displaystyle \frac{1}{\varepsilon_{nn}}
    & 0 & 0 \\
    0
    & \displaystyle
    \varepsilon_{22} - \frac{\varepsilon_{2n} \varepsilon_{n2}}{\varepsilon_{nn}}
    & \displaystyle
    \varepsilon_{23} - \frac{\varepsilon_{2n} \varepsilon_{n3}}{\varepsilon_{nn}}
    \phantom{\frac{{|^|}^|}{{|_|}_|}} 
    \\
    0
    & \displaystyle
    \varepsilon_{32} - \frac{\varepsilon_{3n} \varepsilon_{n2}}{\varepsilon_{nn}}
    & \displaystyle
    \varepsilon_{33} - \frac{\varepsilon_{3n} \varepsilon_{n3}}{\varepsilon_{nn}}
    \\
    \end{array} \right)
    =
  \left( \begin{array}{c@{\quad}c@{\quad}c}
    \displaystyle -\tau_{nn}
    & 0 & 0 \\
    0
    & \displaystyle \tau_{22}
    & \displaystyle \tau_{23}
    \\
    0
    & \displaystyle \tau_{32}
    & \displaystyle \tau_{33}
    \\
    \end{array} \right)
\end{equation}
is SPSD whenever $\varepsilon$ is
(since
$x^T \Gamma^T \varepsilon \Gamma x = (\Gamma x)^T \varepsilon (\Gamma x)$).
Since $\Gamma$ is invertible
the converse also holds: if $\Gamma^T \varepsilon \Gamma$ is positive
semi-definite, then so is $\varepsilon$.

All this means that because the $\varepsilon^i$ are SPSD,
$\Gamma(\varepsilon^i)^T \varepsilon^i \Gamma(\varepsilon^i)$ are
SPSD.
Since $\tilde{\tau} = V_1 \tau^1 + V_2 \tau^2$,
the matrix
\begin{eqnarray}
     \Gamma(\tilde{\varepsilon})^T
	\tilde{\varepsilon}
	\Gamma(\tilde{\varepsilon})
	&=&
  \left( \begin{array}{c@{\quad}c@{\quad}c}
    \displaystyle -\tilde{\tau}_{nn}
    & 0 & 0 \\
    0
    & \displaystyle \tilde{\tau}_{22}
    & \displaystyle \tilde{\tau}_{23}
    \\
    0
    & \displaystyle \tilde{\tau}_{32}
    & \displaystyle \tilde{\tau}_{33}
    \\
    \end{array} \right)
    =
      V_1
      \Gamma(\varepsilon^1)^T
	\varepsilon^1
	\Gamma(\varepsilon^1)
	+
      V_2
      \Gamma(\varepsilon^2)^T
	\varepsilon^2
	\Gamma(\varepsilon^2)
	\nonumber \\
\end{eqnarray}
is the sum of two positive semi-definite matrices, which is itself
positive semi-definite.
Therefore the effective dielectric tensor,
$\tilde{\varepsilon}$ is positive semi-definite.

Incidentally, the matrix $\Gamma$ plays a well-known role.  If
$\mathbf{F} = (D_n, E_2, E_3)$ is the 3-tuple of continuous field
components (at the dielectric interface), then
$\mathbf{F} = \Gamma(\varepsilon) \mathbf{E}$.  In fact, the
effective dielectric is derived (see \cite{Kottke:2008}) by taking
the perturbative formula for change in eigenvalue:
\begin{equation}
	\Delta \omega^2 \propto
	\langle E | \varepsilon(x) - \varepsilon'(x) | E' \rangle
\end{equation}
where $E$ is the eigenmode corresponding to $\varepsilon(x)$ and
$E'$ is the (exact) eigenmode corresponding to $\varepsilon'(x)$.  The
above expression contains no approximation; the problem is that one
typically doesn't know $E'$---usually a first-order approximation is
realized by approximating $E' \approx E$.  In this case, the authors of
\cite{Kottke:2008} argued that because $E$ is discontinuous, it jumps
when $\varepsilon$ changes by a finite amount, and so $E'\approx E$
is a bad approximation (it has zeroth-order error).  It is better to
approximate $F' \approx F$.  Therefore, we write
\begin{eqnarray}
	\Delta \omega^2 & \propto  &
	\langle E | \varepsilon(x) - \varepsilon'(x) | E' \rangle
	=
	\langle F |
	\Gamma(\varepsilon)^T [\varepsilon(x) - \varepsilon'(x)]
	\Gamma(\varepsilon')
	| F' \rangle \nonumber \\
	& \approx &
	\langle F |
	\Gamma(\varepsilon)^T [\varepsilon(x) - \varepsilon'(x)]
	\Gamma(\varepsilon')
	| F \rangle
.\end{eqnarray}
To obtain approximately zero frequency shift when we
substitute $\epsilon'$ for $\epsilon$,
we demand that
$\int_V \Gamma(\varepsilon)^T [\varepsilon(x) - \varepsilon'(x)] \Gamma(\varepsilon')$
vanish, which leads directly to the condition that
$\int_V \tau(\epsilon(x)) = \int_V \tau(\epsilon'(x))$.

\end{document}